\documentclass[12pt]{article}
\usepackage{eurosym}
\usepackage[left=2.2cm, right=2.2cm, top=2.2cm, bottom=2.2cm]{geometry}
\usepackage{amsfonts}
\usepackage{amssymb}
\usepackage{amsmath}
\usepackage{lscape}
\usepackage{graphicx}
\usepackage{pgfplots}
\usepackage{tikz}
\pgfplotsset{compat=newest}
\pgfplotsset{plot coordinates/math parser=false}
\newlength\figureheight
\newlength\figurewidth
\usepackage{mathtools}
\usepackage{multirow}
\usepackage{float}

\def\tvi{\vrule width 0pt height 15pt depth 5pt}

\DeclareMathOperator*{\plim}{plim}
\newtheorem{theorem}{Theorem}

\newtheorem{result}[theorem]{Result}

\newcommand{\Keywords}[1]{\par\noindent{{\em \large{Keywords}\/}: #1}}

\newcommand{\dprod}{\displaystyle\prod}

\def\1{1\!{\rm l}}

\DeclareMathOperator*{\argmax}{arg\,max}

\begin{document}
	
	\title{Approximate Maximum Likelihood for Complex Structural Models}
	\author{Veronika Czellar\thanks{Department of Data Science, Economics and Finance, EDHEC Business School, France}, David T. Frazier\thanks{Department of Econometrics and Business Statistics, Monash University,
			Melbourne, Australia. Corresponding author: \texttt{david.frazier@monash.edu}} 
		and Eric Renault\thanks{Department of Economics, University of Warwick and Department of Econometrics and Business Statistics, Monash University.}}
	\maketitle

	\begin{abstract}
Indirect Inference (I-I) is a popular technique for estimating complex parametric models
whose likelihood function is intractable, however, the statistical efficiency of I-I estimation is
questionable. While the efficient method of moments, Gallant and
Tauchen (1996), promises efficiency, the price to pay for this efficiency is a loss of parsimony and thereby a potential lack of robustness to model misspecification. This stands in contrast to simpler I-I estimation strategies, which are known to display less sensitivity to model misspecification precisely due to their focus on specific elements of the underlying structural model. In this research, we propose a new simulation-based approach that maintains the parsimony of I-I estimation, which is often critical in empirical applications, but can also deliver estimators that are nearly as efficient as maximum likelihood. This new approach is based on using a constrained approximation to the structural model, which ensures identification and can deliver estimators that are nearly efficient. We demonstrate this approach through several examples, and show that this approach can deliver estimators that are nearly as efficient as maximum likelihood, when feasible, but can be employed in many situations where maximum likelihood is infeasible. 
	\end{abstract}
	\hspace{3cm}
	

\section{Introduction}

Indirect inference (hereafter, I-I), as proposed by Smith (1993) and
Gourieroux, et al. (1993), is a simulation-based estimation
method often used when the underlying likelihood for the model of interest
is computationally challenging, or intractable. The key idea underpinning
I-I is that, regardless how complicated the structural model, it is often
feasible to simulate artificial data from this fully parametric model. As a
result, statistics based on the observed data and data simulated from the model
can be compared, with the resulting difference minimized in a given
norm to produce an estimator of the structural parameters.

The implementation of I-I is most often carried out using an auxiliary model
that represents an incorrect, {but tractable} version of the
structural model under analysis. User-friendly estimators {for the} parameters of
this auxiliary model provide the statistics, based on the observed and
simulated data, respectively, that are used to conduct inference on
the underlying structural parameters. However, by definition the
information encapsulated in the
auxiliary parameter estimates is less than the information carried in the
likelihood for the structural parameters. As such, {in any}
implementation of I-I there is a fundamental trade-off between the
statistical efficiency of the resulting estimators and their computational feasibility.

The main contribution of this paper is to propose an alternative to I-I that
 produces structural parameter estimates that, albeit also
simulation-based, are arguably closer to {reaching} the Cramer-Rao efficiency
bound {for the parametric structural model.} The new
method proposed {herein}, dubbed ``Approximate Maximum Likelihood" (hereafter, AML), maintains the standard philosophy of I-I that one can resort to
a possibly biased approximation of the structural model, insofar as matching {statistics calculated from this approximation using both simulated and observed data will
	allow us to erase} the misspecification bias. {In contrast to standard I-I,} instead of matching
estimators of auxiliary parameters, we directly match a proxy/approximation to the score vector of the intractable log-likelihood. {These proxies are} indexed by the
vector of structural parameters, for which a preliminary plug-in estimator (based on
observed data) must be used.

{However, as we later demonstrate, the dependence of this approach on the preliminary plug-in estimator differs from standard I-I estimation: as far as the asymptotic
	distribution of our AML estimator is concerned, the asymptotic distribution of the preliminary estimator is immaterial, and only its probability limit (a pseudo-true value possibly different from the true unknown value) will impact the information conveyed by the approximate score. This is in stark contrast to I-I estimation, where the key feature in determining the asymptotic efficiency of I-I is the efficiency of the auxiliary parameter estimates. As such, since it is only the probability limits of the plug-in estimators that matters, our new AML approach can not be directly placed in the standard I-I framework.}

{While this new approach is based on matching types of scores,  it} should not be confused with the score-based version
of I-I proposed by Gallant and Tauchen (1996). As shown by
Gourieroux, Monfort and Renault (1993) (see ``The Third Version of the
Indirect Estimator'' in their Appendix 1), Gallant and Tauchen's (1996)
estimator is actually tantamount to match estimators of auxiliary
parameters. In particular, when fishing for efficiency, Gallant and Tauchen
(1996) (see the proof of their theorem 2) ultimately import the efficiency
for the estimator of auxiliary parameters to reach the Cramer-Rao efficiency
bound for the structural parameters, {with this efficiency claim ultimately requiring that the auxiliary model ``smoothly embeds'' the structural model.}

In short, ``efficient method of moments", Gallant and
Tauchen (1996), must {resort to a} semi-nonparametric score generator
as an auxiliary model. Thanks to its steadily increasing dimension, the
score of this auxiliary model may asymptotically span the score of the
structural model, and {thereby} deliver efficient estimators of the resulting
structural parameters. However, the price to pay for this efficiency is a
highly-parametrized auxiliary model that may be ill-behaved (due to the
non-parsimonious nature of the auxiliary model) when there are
deviations from the underlying model structure, i.e., when the structural
model may be partly misspecified. This is in contrast to
standard I-I estimation, which has been shown to be somewhat robust to
deviations from the underlying modelling assumptions (see, e.g., Dridi et
al., 2007), precisely because it is based on calibrating a limited
number of structural parameters. Our new method remains true to this
parsimony principle since we match proxies for the actual score vector, whose
dimension is the same as the structural parameters.

In our AML approach, (approximate) {efficiency of structural
parameter estimates} does not rest upon high-dimensional inference {or the near-efficiency
	of auxiliary parameter estimates}, but on the conjunction of two
properties. 
\begin{itemize}
	\item First, the {efficiency gap between our estimates and the MLE is tightly related to
		the difference between the asymptotic value of our plug-in estimator for the structural parameters (i.e., the pseudo-true value that will asymptotically feature in our proxy/ approximation for the true limiting score function) and the true unknown value of the structural parameters.}
	\item Second, {the fact that the} Cramer-Rao efficiency bound can be (nearly) reached if the
	information identity is (nearly) maintained. More precisely, the question is
	to assess the difference between the curvature of the log-likelihood at the true value of the structural parameters (as measured by the slope of the
	expected score vector as a function of the structural parameters) and the slope
	of the score vector when the structural parameters enter the score
	through data simulated at a specific parameter value. Satisfaction of the information identity in this context requires a type of multiplicative separability
	of the score vector, which we later demonstrate is satisfied for exponential models.	
\end{itemize}

The motivation for our AML approach is the observation {that
	there are many cases} of interest where the intractability of the assumed
model, and its likelihood, is entirely due to a sub-vector of structural
parameters. Examples include, for instance, dynamic discrete choice models
with ARMA errors (Robinson, 1982, Gourieroux et al., 1985, Poirier and Ruud,
1988), spatial discrete choice models (see, e.g., Pinske and Slade, 1998),
and many dynamic equilibrium models. In such models, a few well-chosen
restrictions would allow us to alleviate the intractability of the likelihood due to the presence of certain latent variables.

More generally, many complex economic models are such that imposing a
(potentially false) constraint on the structural
model yields a simpler auxiliary {model with} a computationally
tractable likelihood. This is precisely the reason why score/LM tests are
popular in econometrics: estimation and testing \textquotedblleft under the
null\textquotedblright is feasible even in very
complicated models. Unfortunately, imposition of this constraint, and
subsequent optimization of the constrained log-likelihood, will not deliver
consistent estimates of the structural parameters if the constraint is not satisfied at the truth.

As recently pointed out by Calvet and Czellar (2015), imposing potentially
false equality constraints on a given structural model can be an attractive
method for obtaining simple and rich auxiliary models for the purposes of
I-I. For instance, in the context of a long-run risk model (Bansal and
Yaron, 2004), Calvet and Czellar (2015) demonstrate that imposing specific
equality constraints on certain parameters produces a simple auxiliary model
for use in I-I (with a computationally tractable likelihood function) that
closely resemble the structural model. The fact that this resulting
auxiliary model may not deliver consistent estimates {of the true structural parameters} is
immaterial insofar as matching a {simulation-based approximation against the
	observation-based version will allow us} to erase the misspecification bias. The
benefits of such an approach are two-fold: one, by using
constraints to define the auxiliary model, we sketch a
systematic strategy for the choice of an auxiliary model; two, this auxiliary
model closely matches the structural model and so for issues of robustness
and efficiency this auxiliary model is very useful.

However, while highly-useful, the suggestion of Calvet and Czellar (2015) is
incomplete, and does not allow for consistent estimation of the structural
parameters on its own. That is, since we impose a number of constraints on the
auxiliary model, by definition the auxiliary model can not consistently
estimate all the structural parameters, except in the unlikely case where
the constraints are satisfied at the true value of the structural
parameters. To circumvent this issue, Calvet and Czellar (2015) propose to
add to the statistics obtained from the auxiliary model {additional
	statistics so that, when considered jointly, this new vector can
	jointly identify} the structural parameters {when estimated by I-I}.

Motivated by the above ideas and the approach to handling constraints within
I-I proposed in Calzolari et al. (2004) and Frazier and Renault (2019), we
propose a novel inference approach based on constraining the structural model
parameters to create a simple, but highly
informative, {proxy for the score vector that can be used} to estimate the
structural parameters. However, unlike the strategy put forward by
Calvet and Czellar (2015), {our approach provides an automatic, and
	nearly-efficient, method to identify the}
structural parameters.

In addition, we {demonstrate that this AML strategy can be based on a proxy for the score
	vector which entails additional layers of approximation} beyond simply plugging in
a (wrongly) constrained estimation of the structural parameters. {For example,} in the
context of stable probability distributions, the likelihood function
is known in closed-form only {at certain specific values of the parameters; as an example, a unit shape parameter ($a=1$) and a zero value of the asymmetry parameter ($b=0$) yield a Cauchy likelihood, however, even then the partial derivatives of the likelihood function with,
	respect to $a$ and $b$, is not available in closed-form. In such settings, our AML strategy can be implemented by invoking an additional layer of approximation and replacing the directions of our score vector proxy that can not be obtained in closed-form by a finite-difference approximation. Approximating certain directions of the score vector by finite-differences is obviously even more useful when some structural parameters are only defined on the integers. We demonstrate our methodology in such cases using the example of Markov-Switching multifractal (MSM) volatility processes,  Calvet and Fisher (2004, 2008), which are especially well-suited to capture volatility dynamics through an unknown, but finite, number of multiplicative components.}

While we apply our AML methodology within the confines of a MSM volatility model, we note here that the use of MSM models are not exclusive to the analysis of volatility. Indeed, Chen, Diebold and Schorfheide (2013) propose a novel Markov-switching multifractal duration (MSMD) model to analyze inter-trade duration data in financial markets, and demonstrate its superiority over competing duration models. While we exemplify the AML procedure within a MSM volatility model, we note here that AML can be equivalently applied to the MSMD model of Chen et al. (2013) using precisely the same approach detailed in this paper. 

The remainder of the paper is organized as follows. In Section 2, we give
the general setup, discuss several
interesting examples where equality constraints on the structural model
yield a tractable score vector that can be used for inference through score
matching, and discuss our AML estimation strategy. We also demonstrate that, in contrast to standard I-I, the
choice of an auxiliary estimator is immaterial, beyond the pseudo-true value
of structural parameters that it defines.

In Section 3, we provide the asymptotic theory of AML. Further, we demonstrate that, in
the case of an exponential model, a sufficient (but not necessary) condition for  AML
estimators to achieve the Cramer-Rao efficiency bound is that the pseudo-true value used in AML coincides with the true one. Section 4 provides Monte Carlo evidence on the finite-sample performance of AML in two leading examples: one based on false equality constraints, and one where we are required to define some of the pseudo-score components using a finite-difference approximation, with the later example containing an empirical application to financial returns data using a multifractal stochastic volatility model. Monte Carlo evidence on the application to stable distribution is provided in Appendix D. Section 5 concludes with 
suggestions for future research on extensions of I-I where
not only the two vectors to match both depend on the
observed data, as in this paper, but even the simulator itself may depend on the observed
data. Mathematical details for the proofs of main results and developments of theoretical examples are provided in Appendices A, B and C.  

\section{Approximate Maximum Likelihood vs Indirect Inference}

\subsection{Model Setup: Nonlinear State Space Models}
Following Gourieroux, et al. (1993) (hereafter, GMR), {our goal is inference on the unknown parameters of a dynamic structural model that has a nonlinear state space representation.} The structural model is specified through a transition, or state, equation and a measurement equation. The transition equation is of the following form
\begin{equation*}
u_{t}=\varphi \left( u_{t-1},\varepsilon _{t},\theta \right) ;\theta \in
\Theta \subset 
\mathbb{R}
^{p}  \label{tr},
\end{equation*}
where $\varphi $ is a known function, $\left(u_{t},\varepsilon _{t}\right)_{t=1}^{T}$ are
latent processes and $\varepsilon _{t}$ is a strong white noise {process} with a
known distribution; and the measurement equation satisfies
\begin{equation*}
{y}_{t}=r\left( {y}_{t-1},x_{t},u_{t},\varepsilon _{t},\theta
\right) ;\theta \in \Theta \subset 
\mathbb{R}
^{p}  \label{meas},
\end{equation*}
where $r$ is a known function and $\left(x_{t},y_{t}\right)_{t=1}^{T}$ are observed
processes.
In the two equations, known functions $\varphi $\ and $r$ are indexed by a
$p$-dimensional vector of unknown parameters $\theta \in \Theta $. We assume that $%
\left(x_{t}\right)_{t\le T} $ {is a homogenous Markov process of order 1, and is}
independent of the process $\left(\varepsilon _{t}\right)_{t\le T} $ (and $\left(
u_{t}\right)_{t\le T} $). Then the process $\left( x_{t}\right) $ is exogenous and
the process $\left( x_{t},y_{t}\right)_{t\le T} $ is stationary. It is worth
recalling that, by standard arguments, the fact that the Markov process is
of order 1 and the probability distribution of the white noise $\varepsilon_t $
is known are not restrictive assumptions.

Under the above conditions, assuming absolute continuity with respect to
some dominating measure, for a given initial condition $z_{0}=\left(
y_{0},u_{0}\right) ,$ it should be possible to write down the joint
conditional probability density function
\begin{equation}
l^{\ast }\left\{ \left( y_{t}\right) _{1\leq t\leq T},\left( u_{t}\right)
_{1\leq t\leq T}\left\vert (x_{t}\right) _{1\leq t\leq T},z_{0};\theta
\right\}  \label{likelat}.
\end{equation}
{The density of the observed sequence $(y_t)_{ t\le T}$, conditional on $(x_t)_{t\le T}$, is obtained by integrating out the latent variables $\left( u_{t}\right)
_{1\leq t\leq T}$ from the density \eqref{likelat} and can generally be stated as}
\begin{equation}
l\left\{ \left( y_{t}\right) _{1\leq t\leq T}\left\vert (x_{t}\right)
_{1\leq t\leq T};\theta \right\} =\dprod\limits_{_{1\leq t\leq
		T}}l\{y_{t}\left\vert \left( y_{\tau }\right) _{1\leq \tau \leq
	t-1},x_{t},z_{0};\theta \right\}  \label{likeobs},
\end{equation}
where the last equality comes from {the Markovianity and exogeneity of
the process $\left( x_{t}\right) $. This density function allows us to construct the log-likelihood function}
\begin{equation}
L_{T}\left( \theta \right) =\frac{1}{T}\sum_{_{1\leq t\leq T}}\log \left(
l\{y_{t}\left\vert \left( y_{\tau }\right) _{1\leq \tau \leq
	t-1},x_{t},z_{0};\theta \right\} \right).  \label{loglike}
\end{equation}

A maintained assumption in this paper will be that the log-likelihood
{asymptotically identifies} some true unknown value, $\theta ^{0}$, of the unknown
parameters, $\theta $, and is the unique maximizer of the population criterion
\begin{eqnarray*}
	\theta ^{0} &=&\arg \max_{\theta \in \Theta }L_{\infty }\left( \theta \right),
	\text{ where }
	L_{\infty }\left( \theta \right) =\plim_{T\rightarrow \infty }L_{T}\left(
	\theta \right).
\end{eqnarray*}
It is important to realize that more often than not, this assumption is
neither testable nor associated to a feasible estimator of $\theta ^{0}$.
{The likelihood function in equation (\ref{likeobs}) does not have an analytically tractable form: it is constructed via the latent likelihood in \eqref{likelat} through an integration step that is infeasible to carry out, integration with
respect to the $T$ variables $(u_{t})_{t\le T}$, with $T$ going to infinity.}\footnote{Clearly, such examples are exclusive of cases where the integration, or filtering, can be performed analytically, such as cases where the Kalman filter can be performed, as in linear Gaussian state space models, or as in certain qualitative Markov switching
models. The focus of this paper is nonlinear state space models, where the above simplifications are not generally applicable.}

Even though direct inference on $\theta^0$ associated with $L_T(\theta)$ may be infeasible, it is well-known that inference can be carried out using simulation-based filtering and inference approaches. Under the assumed model, it is possible to simulate
values of $y_{1},...,y_{T}$, for a given initial condition $z_{0}=\left(
y_{0},u_{0}\right) $ and a given value $\theta $ of the parameters,
conditionally on the observed path of the exogenous variables $%
x_{1},...,x_{T} $. This is done by independently drawing simulated values $
\tilde{\varepsilon}_{1},...,\tilde{\varepsilon}_{T}$ from the assumed
distribution of the strong white noise $\left( \varepsilon _{t}\right) $
(the simulated values are also independent of the realized values $%
\varepsilon _{1},...,\varepsilon _{T}$ that underpin the observations) and
by computing
\begin{equation*}
\tilde{y}_{t}\left( \theta ,z_{0}\right),\text{ for }t=0,1,\dots,T,
\end{equation*}
with $\tilde{y}_{0}\left( \theta ,z_{0}\right) =y_{0}$ and where
\begin{eqnarray*}
		\tilde{y}_{t}\left( \theta ,z_{0}\right) &=&r\left[ \tilde{y}_{t-1}\left(
		\theta ,z_{0}\right) ,x_{t},\tilde{u}_{t}\left( \theta ,u_{0}\right)
		,\tilde\varepsilon _{t},\theta \right] \\
		\tilde{u}_{t}\left( \theta ,u_{0}\right) &=&\varphi \left[ \tilde{u}_{t-1}\left( \theta ,u_{0}\right) ,\tilde{\varepsilon}_{t},\theta \right].
\end{eqnarray*}
	
While simulation is the most prevalent mechanism for inference in such settings, we note that in many cases inference could be based directly on $L_T(\theta)$ if we were to instead consider sub-models defined by restricting the parameters $\theta$ to lie in a given set $\Theta_0\subset\Theta$. Indeed, it will often be that case that the sub-models could be chosen by imposing $\theta\in\Theta_{0}$ so that we obtain a convenient factorization of the probability density
function, which ensures that integration of the $T$ latent variables,
$\left( u_{t}\right) _{t\leq T}$, no longer requires solving a $T$-dimensional integral, and consequently inference (over the sub-models) could be based directly on the log-likelihood function (\ref{loglike}). However, in general the sub-models specified by this constraint will not be correctly specified and the resulting estimates will be asymptotically biased for the parameter of interest $\theta^0$. However, as we will later see, following the intuition of I-I, this misspecification bias can be corrected by matching these estimators against a simulated counterpart.

The following section demonstrates that there are many interesting cases where restricting the parameters $\theta$ to lie in some set $\Theta_0\subset\Theta$ results in log-likelihood functions that are easily tractable.

\subsection{Illustrative Examples}\label{sec:examps1}

\subsubsection{Example 1: \textit{Autoregressive Discrete Choice Models}}

We observe the sample $\{y_{t},x_{t}\}_{t=1}^{T}$ generated from 
\begin{eqnarray*}
	y_{t} &=&\bigg{\{} 
	\begin{array}{lr}
		1 & \quad\text{ if }y_{t}^{\ast }>0 \\ 
		0 & \quad\text{ if }y_{t}^{\ast }\leq 0%
	\end{array},\;\;%
	y_{t}^{\ast } =x_{t}^{\prime }\theta _{1}+u_{t},\;\;u_{t}=\theta
	_{2}u_{t-1}+\nu _{t},
\end{eqnarray*}%
where $x_{t}$ is a vector of explanatory variables, $\nu _{t}$ is a Gaussian
white noise and the $AR(1)$ process $(u_t)_{t\le T}$ is stationary ( $-1<\theta _{2}<1$), $%
\theta =\left( \theta _{1}^{\prime },\theta _{2}\right)^{\prime }.$
Following the standard normalization practice for a Probit error term, we set $\nu
_{t}\sim \aleph \left( 0,1\right) $. In what follows, panel data can easily
be accommodated at the cost of more involved notations, and so we omit this
extension for simplicity.

Unlike the standard Probit model, the autoregressive nature of $u_{t}$ means
that the data density can only be stated as the $T$-dimensional integral: Let $A_t=[0,+\infty)$ if $y_t=1$ and $A_t=(-\infty,0)$ if $y_t=0$,  
\begin{flalign*}
l\left\{ \left( y_{t}\right) _{t\leq T}\left\vert (x_{t}\right)
_{t\leq T};\theta \right\} &=\int_{A_{1}}\cdots \int_{A_{T}}l^{\ast
}\left\{ \left( y_{t}^{\ast }\right) _{ t\leq T}\left\vert
(x_{t}\right) _{t\leq T},z_{0};\theta \right\} dy_{1}^{\ast }\cdots
dy_{T}^{\ast },\\
l^{\ast }\left\{ \left( y_{t}^{\ast }\right) _{t\leq T}\left\vert
(x_{t}\right) _{t\leq T},z_{0};\theta \right\} &=(2\pi)^{-T/2}R(\theta _{2})^{-1/2}\exp \left( -\frac{1}{2R(\theta _{2})}%
u_{1}^{2}(\theta _{1})\right) \prod_{t=2}^{T}\exp \left( -\frac{\left[
	u_{t}(\theta _{1})-\theta _{2}u_{t-1}(\theta _{1})\right] ^{2}}{2}\right)
\end{flalign*}%
where $R\left( \theta _{2}\right) =1/(1-\theta _{2}^{2})$ and $u_{t}(\theta
_{1})=y_{t}^{\ast }-x_{t}^{\prime }\theta _{1}$. However, note that if one
were to impose the constraint $\theta _{2}=0$ in $l^{\ast }\left\{ \left(
y_{t}^{\ast }\right) _{t\leq T}\left\vert (x_{t}\right) _{ t\leq
	T},z_{0};\theta \right\} $, {the integral that defines this density can be factorized into
a product of $T$ univariate integrals, which ultimately yields the usual Probit likelihood function.} As such,  a convenient parametric sub-model is given by
\begin{equation*}
l\left\{ \left( y_{t}\right) _{t\leq T}\left\vert (x_{t}\right)
_{t\leq T};\theta \right\} ;\theta \in \Theta _{0}=\left\{ \theta \in
\Theta ,\theta =\left( \theta _{1}^{\prime },0\right) ^{\prime }\right\}
\end{equation*}

A similar finding to the above can also be applied, albeit with different
notations, to spatially correlated Probit models, instead of the
autoregressive Probit model.

\subsubsection{Example 2: \textit{GARCH-like Stochastic Volatility Model}}\label{sec:g-sv}
Observed log-returns are assumed to evolve according to
\begin{equation*}
r_{t+1}=\mu +\varepsilon _{t+1},\;E[\varepsilon _{t+1}\left\vert I_{t}\right]=0,
\end{equation*}where the error term $\varepsilon _{t+1}$ is a martingale difference sequence
(hereafter, mds). We are interested in the volatility dynamics of the process
\begin{equation*}
\sigma _{t}^{2}=E[\varepsilon _{t+1}^{2}\left\vert I_{t}\right],
\end{equation*}
As usual, the observed counterpart of volatility dynamics is given by the
dynamics of the squared return process. We assume that $\varepsilon _{t}^{2}$
is a weak $ARMA(p,p):$%
\begin{equation}
\varepsilon _{t+1}^{2}-\omega -\sum_{j=1}^{p}\gamma _{j}\varepsilon
_{t+1-j}^{2}=\xi _{t+1}-\sum_{j=1}^{p}\beta _{j}\xi _{t+1-j}  \label{ARMA}
\end{equation}
where $\xi _{t+1}$ is a weak white noise that defines the innovation process
of $\varepsilon _{t}^{2}$. In other words, the ARMA representation (\ref{ARMA}) is causal and invertible.

It is known (see e.g. Meddahi and Renault (2004)) that $\varepsilon _{t}$ is
a (semi-strong) $GARCH(p,q)$\ with $q\leq p$ if and only if $\xi _{t}$ is a
mds. Inspired by Franses et al. (2008), albeit with a different model, we
want to relax this restriction about the white noise $\xi _{t+1}$, so that
we define of family of stochastic volatility models, which contains the $%
GARCH(p,q)$\ with $q\leq p$ as a particular case, but, beyond this
particular case, belong to the realm of nonlinear state space models. For
this purpose, it is worth setting the focus on the difference between the
innovation process $\xi _{t+1}$\ and the mds $\nu _{t+1}=\varepsilon_{t+1}-\sigma^2_t$.

By definition (see equation (\ref{ARMA})), the difference $\left( \xi
_{t+1}-\varepsilon _{t+1}^{2}\right) $\ is $I_{t}$-measurable, so that we
are allowed to introduce the notation: 
\begin{equation*}
\xi _{t+1}-\nu _{t+1}=\eta _{t}=\sigma _{t}^{2}-k_{t}
\end{equation*}so that
\begin{equation*}
\xi _{t+1}-\varepsilon _{t+1}^{2}=-\sigma _{t}^{2}+\eta _{t}=-k_{t},
\end{equation*}which allows us to rewrite the volatility dynamics in equation (\ref{ARMA}) as
\begin{equation*}
\varepsilon _{t+1}^{2}-\omega -\sum_{j=1}^{p}\gamma _{j}\varepsilon
_{t+1-j}^{2}=\varepsilon _{t+1}^{2}-k_{t}-\sum_{j=1}^{p}\beta _{j}\left[
\varepsilon _{t+1-j}^{2}-k_{t-j}\right]
\end{equation*}
so that
\begin{eqnarray}
k_{t} &=&\omega +\sum_{j=1}^{p}\alpha _{j}\varepsilon
_{t+1-j}^{2}+\sum_{j=1}^{p}\beta _{j}k_{t-j}  \label{GARCHlike} \\
\alpha _{j} &=&\gamma _{j}-\beta _{j}
\end{eqnarray}

In other words, we see that, without any additional assumption, the $%
ARMA(p,q)$ representation for $\varepsilon _{t+1}^{2}$ in equation (\ref{ARMA}) can be
characterized by a GARCH-like equation (\ref{GARCHlike}) with
\begin{equation}
\sigma _{t}^{2}=k_{t}+\eta _{t},\eta _{t}=\xi _{t+1}-\nu _{t+1},\nu
_{t+1}=\varepsilon _{t+1}^{2}-\sigma _{t}^{2}  \label{nomds}
\end{equation}

Note that, since since $\nu _{t+1}$\ is a mds, we deduce from (\ref{nomds})
that
\begin{equation*}
\eta _{t}=E[\eta _{t}\left\vert I_{t}\right] =E[\xi _{t+1}\left\vert I_{t}%
\right]
\end{equation*}
and thus
\begin{eqnarray*}
	E[\xi _{t+1}\left\vert I_{t}\right] &=&0\Longleftrightarrow \sigma
	_{t}^{2}=k_{t} \\
	&\Longleftrightarrow &\sigma _{t}^{2}=\omega +\sum_{j=1}^{p}\alpha
	_{j}\varepsilon _{t+1-j}^{2}+\sum_{j=1}^{p}\beta _{j}\sigma _{t-j}^{2}.
\end{eqnarray*}That is, we again find that the GARCH case is tantamount to the mds property for the
noise process $\xi _{t+1}$, which implies that the process $\eta _{t}$ is identically zero.

Now, beyond the GARCH case, it is worth questioning whether a non-zero
process $\eta _{t}$\ is just a white noise or encapsulates some additional
dynamic features of conditional variance. It is then natural to consider the
following model for $\eta_t$:
\begin{equation}
\eta _{t}=\rho \eta _{t-1}+\varpi \chi _{t},\;\;\left\vert \rho \right\vert <1
\label{transitvolat}
\end{equation}
where $\chi _{t}$ is i.i.d. with a known distribution with zero mean. Such a model for $\eta_t$ leads to a nonlinear state space model with the measurement
equation
\begin{equation*}
r_{t+1}=\mu +\left[ \omega +\sum_{j=1}^{p}\alpha _{j}\varepsilon
_{t+1-j}^{2}+\sum_{j=1}^{p}\beta _{j}\left( \sigma _{t-j}^{2}-\eta
_{t-j}\right) +\eta _{t}\right] ^{1/2}u_{t+1}
\end{equation*}
for $u_{t+1}$\ and $\chi
_{t}$\ i.i.d. with known distributions, and where the transition equation is given by (\ref{transitvolat}).

Similar to the general case treated in equation \eqref{likeobs}, the likelihood function of this model is only expressible as a $T$-dimensional integral (due to the dynamics in \eqref{transitvolat}). However, as we have already seen in the autoregressive Probit example, Example 1, imposing the constraint $\rho =0$ in this state space model means that the $T$-dimensional integral can be factorized into the product of $T$ univariate integrals. As a consequence, stable numerical
procedures can be used to compute these univariate integrals and the resulting likelihood can then be maximized. More precisely, since
\begin{eqnarray*}
	\sigma _{t}^{2} &=&k\left[\{r_{\tau }\}_{\tau \leq t}\right] +\eta _{t} \\
	k\left[ \{r_{\tau }\}_{\tau \le t}\right] &=&k_{t}=\omega +\sum_{j=1}^{p}\alpha
	_{j}\varepsilon _{t+1-j}^{2}+\sum_{j=1}^{p}\beta _{j}k_{t-j}
\end{eqnarray*}%
$k_{t}$ can be computed recursively as a function of past observed returns $%
\{ r_{\tau }\}_{\tau \le t} $, as is standard in GARCH models. Therefore, when $\rho =0$, the overall likelihood is the product of the increments $l[r_{t+1}\left\vert \{r_{\tau }\}_{\tau \leq t};\theta \right]$, where for $t\ge1$, 
\begin{equation*}
l[r_{t+1}\left\vert \{r_{\tau }\}_{\tau \leq t};\theta \right] =\int_{-\infty
}^{+\infty }\frac{1}{\left[ k\left[ \{r_{\tau }\}_{\tau \leq t}\right] +\eta _{t}%
	\right] ^{1/2}}f_{u}\left[ \frac{r_{t+1}-\mu }{k\left[ \{r_{\tau }\}_{\tau \leq t}
	\right] +\eta _{t}}\right] \frac{1}{\varpi }f_{\chi }\left[ \frac{\eta _{t}}{%
	\varpi }\right] d\eta _{t}
\end{equation*}
and where $f_u(.)$\ (resp. $f_{\chi }(.)$) denote the probability density function of
the standardized log-return $u_{t+1}$ (resp. of the noise $\chi _{t}$)

\bigskip

\subsubsection{Example 3: \textit{Generalized Tobit Model}}\label{sec:gtm}
Amemiya (1985) defines the generalized Tobit Model of Type 2 by the
following observation scheme for the outcome variable $y_{i}:$ 
\begin{equation}
y_{i}=%
\begin{cases}
y_{1i}^{\ast } & {\text{ if }}y_{2i}^{\ast }\geq 0 \\ 
{\text{missing}} & {\text{ if }}y_{2i}^{\ast }<0%
\end{cases}
\label{struct1},
\end{equation}%
with 
\begin{equation}
y_{1i}^{\ast }=x_{i}^{\prime }\theta _{1}+\sigma \varepsilon _{i}
\label{struct2},
\end{equation}%
where $x_{i}$ is a vector of exogenous explanatory variables, $(\theta
_{1}^{\prime },\sigma )^{\prime }$\ a vector of unknown parameters and $%
\varepsilon _{i}$ is a standardized Gaussian error $\varepsilon _{i}\sim
\aleph \left( 0,1\right) .$ {A complete specification for the likelihood
	function requires specifying the conditional probability of missingness in the data:}
\begin{equation*}
\Pr [y_{2i}^{\ast }<0\left\vert y_{1i}^{\ast },{z}_{i},\theta
_{2},\theta _{3}\right],
\end{equation*}%
where $z_{i}$ is a vector of exogenous explanatory variables and $%
(\theta _{2}^{\prime },\theta _{3}^{\prime })^{\prime }$\ is a vector of
unknown parameters. {The parameter $\theta_2$ govern the relationship between ${z}_i$ and the missingness mechanism, and the parameter $\theta
	_{3}$ characterizes the dependence between the two latent endogenous
	variables $y_{1i}^{\ast }$ and $y_{2i}^{\ast }$.} Then, if $I_{1}$ (resp. $%
I_{0}$) stands for the subset of indices for which $\left( y_{2i}^{\ast
}\geq 0\right) $ (resp. $y_{2i}^{\ast }<0$), the likelihood function can be
written as
\begin{eqnarray*}
	l\left\{ \left( y_{i}\right) _{1\leq i\leq T}\left\vert (x_{i},{z}%
	_{i}\right) _{1\leq i\leq T};\theta \right\}
	=\dprod\limits_{i\in I_{1}}\frac{1}{\sigma }\varphi \left( \frac{%
		y_{i}-x_{i}^{\prime }\theta _{1}}{\sigma }\right) \Pr [y_{2i}^{\ast }\geq
	0\left\vert y_{i},{z}_{i},\theta _{2},\theta _{3}\right]
	\dprod\limits_{i\in I_{0}}\Pr [y_{2i}^{\ast }<0\left\vert {z}%
	_{i},\theta \right],
\end{eqnarray*}
with
\begin{equation*}
\Pr [y_{2i}^{\ast }<0\left\vert {z}_{i},\theta \right] =\int \Pr
[y_{2i}^{\ast }<0\left\vert y_{1i}^{\ast },z_{i},\theta _{2},\theta
_{3}\right] \frac{1}{\sigma }\varphi \left( \frac{y_{1i}^{\ast
	}-x_{i}^{\prime }\theta _{1}}{\sigma }\right) dy_{1i}^{\ast },
\end{equation*}
where the function $\varphi (.)$\ stands for the probability density
function of the standard normal distribution and $$\theta =(\theta
_{1}^{\prime },\theta _{2}^{\prime },\theta _{3}^{\prime },\sigma )^{\prime
}\text{ where }\theta _{1}\in \mathbb{R}^{p_{1}}\;, \theta _{2}\in \mathbb{R}%
^{p_{2}},\;\theta _{3}\in \mathbb{R},\;\sigma >0$$ denotes the vector of
unknown structural parameters. Estimation of $\theta $ may be challenging
because the likelihood function involves an integral that may be
necessary to compute numerically. {However, imposing the (possibly
	false) equality constraint $\theta _{3}=0$  implies that $y_{1i}^{\ast }$ and 
	$y_{2i}^{\ast }$ are conditionally independent, given ${z}_{i}$, and
	the likelihood function under the} constraint $\theta _{3}=0$ becomes
	\begin{eqnarray*}
		l\left\{ \left( y_{i}\right) _{1\leq i\leq T}\left\vert (x_{i},{z}%
		_{i}\right) _{1\leq i\leq T};\theta \right\}
		&=&\dprod\limits_{i\in I_{1}}\frac{1}{\sigma }\varphi \left( \frac{%
			y_{i}-x_{i}^{\prime }\theta _{1}}{\sigma }\right) \Pr [y_{2i}^{\ast }\geq
		0\left\vert {z}_{i},\theta _{2},0\right] \dprod\limits_{i\in I_{0}}\Pr
		[y_{2i}^{\ast }<0\left\vert {z}_{i},\theta _{2},0\right].
	\end{eqnarray*}
	
	Amemiya (1985) notes that the ``special case of independence" makes the
	likelihood function almost as simple as a standard Tobit when the
	probability distribution of $y_{2i}^{\ast }$\ given ${z}_{i}$ is also
	Gaussian. However, by reference to an empirical paper (Dudley and
	Montmarquette (1976) about the foreign aid from United States to a
	particular country), Amemiya (1985) notes that "it makes their model
	computationally advantageous. However, it seems unrealistic to assume that
	the potential amount of aid, $y_{1}^{\ast }$ is independent of the variable
	that determines whether or not aid is given, $y_{2}^{\ast }$". More
	generally, Amemiya (1985) considers that the joint conditional distribution
	of $\left( y_{1i}^{\ast },y_{2i}^{\ast }\right) ^{\prime }$ given $\left(
	x_{i},{z}_{i}\right) $ is Gaussian and $\theta _{3}$\ stands for the
	correlation coefficient between $\ y_{1i}^{\ast }$ and $y_{2i}^{\ast }$. 
	
	{However, an alternative, and often computationally more convenient choice, is to assume that the
	conditional probability distribution of $y_{2i}^{\ast }$\ given $%
	(y_{1i}^{\ast },x_{i},{z}_{i})$ is logistic, which yields}
	\begin{equation}\label{eq:logist}
	\Pr [y_{2i}^{\ast }\geq 0\left\vert y_{1i}^{\ast },{z}%
	_{i},x_{i},\theta _{2},\theta _{3}\right] =[1+\exp (-{z}_{i}^{\prime
	}\theta _{2}-\theta _{3}y_{1i}^{\ast })]^{-1}.
	\end{equation}
		{In this case, imposing the (potentially false) equality constraint  $\theta _{3}=0,$ leads to a ``computationally advantageous" model with log-likelihood function, when
	evaluated at $\theta =(\theta _{1}^{\prime },\theta _{2}^{\prime },0,\sigma
	)^{\prime }$, with a particularly simple form}
	\begin{eqnarray*}
		&&L_{T}\left[ (\theta _{1}^{\prime },\theta _{2}^{\prime },0,\sigma )\right]
		\\
		&=&\frac{1}{T}\sum_{i\in I_{1}}\left\{ -\frac{1}{2}\log \left( 2\pi \sigma
		^{2}\right) -\frac{1}{2\sigma ^{2}}\left( y_{i}-x_{i}^{\prime }\theta
		_{1}\right) ^{2}-\log \left( 1+e^{-{z}_{i}^{\prime }\theta
			_{2}}\right) \right\} -\frac{1}{T}\sum_{i\in I_{0}}\log \left( 1+e^{{z}%
			_{i}^{\prime }\theta _{2}}\right).
	\end{eqnarray*}
	
	\bigskip

\subsubsection{Example 4: \textit{Markov-Switching Multifractal (MSM) Model}}\label{sec:msm}
Similarly to {Example 2, consider that observed asset returns evolve according to}
\begin{equation*}
r_{t+1}=\mu +\varepsilon _{t+1},\;E[\varepsilon _{t+1}\left\vert I_{t}\right]
=0,
\end{equation*}where the error process $\varepsilon_{t}$ is assumed to follow 
$$
\varepsilon _{t+1}=\sigma _{t}u_{t+1},\;\;E[u_{t+1}^{2}\left\vert I_{t}\right]
=1$$with $\sigma_t$ denoting the volatility process. 
{Our goal remains the analysis of the volatility process, however, in this example we use the Binomial MSM model proposed in Calvet and
Fisher (2001, 2004, 2008), and consider that the volatility process is defined as the product of several volatility components}
\begin{flalign*}
	\sigma _{t}^{2} &=\sigma ^{2}\dprod\limits_{k=1}^{\overline{k}}M_{k,t}.
\end{flalign*}{The components $M_{k,t}$ are unobservable (i.e., latent) variables that are often referred to as multipliers or volatility components, and the overall number of components, $\overline{k}$, is unknown.}

We will assume that the standardized return $u_{t+1}$\ is i.i.d with a
probability density function $f_u\left( .\right) $. The latent state variables 
$M_{k,t},k=1,...,\overline{k}$,\ {are assumed to be stationary Markov processes with common}
marginal distribution, denoted by $M$. Given a value $M_{k,t}$ for the $k^{th}$\
component at time $t$, the next-period multiplier {is assumed to evolve according to}
\begin{equation*}
M_{k,t+1}=\left\{ 
\begin{array}{cc}
\sim M & \text{with probability }\gamma _{k} \\ 
M_{k,t} & \text{with probability }(1-\gamma _{k})%
\end{array}%
\right.
\end{equation*}%
where the notation $\left( \sim M\right) $\ stands for ``drawn in the
distribution $M$" and $M_0$ is generated from the stationary distribution $\pi_0$, where
\begin{equation*}
\pi^j_0=\Pr [M_{0}=m^{j}]=1/d,\;\forall j=1,...,d,
\end{equation*}and where $d=2^{\overline{k}}$. 

The switching events (with transition probabilities $%
\gamma _{k},k=1,...,\overline{k}$) and new draws from $M$ are assumed to be
independent across $k$\ and $t$. {To ensure a non-negative and stationary volatility}
process ($E\left( \sigma _{t}^{2}\right) =\sigma ^{2}$), we assume
\begin{equation*}
E(M)=1,\;M\geq 0
\end{equation*}
For sake of parsimony, we introduce an unknown parameter $m_{0}\in (1,2)$\
such that:%
\begin{equation*}
\Pr \left[ M=m_{0}\right] =\Pr \left[ M=2-m_{0}\right] =\frac{1}{2}.
\end{equation*}

{Then the state vector $M_{t}=\left( M_{1,t},...,M_{\overline{k}%
	,t}\right) ^{\prime }$ can take $d$ possible values $m^{j}$, $j=1,...,d $, so that at each date the squared volatility process
takes $d$ possible values}
\begin{equation*}
\sigma ^{2}g\left( m^{j}\right),\text{ where }
g\left[ \left( M_{1,t},...,M_{\overline{k},t}\right) \right] =\dprod%
\limits_{k=1}^{\overline{k}}M_{k,t}.
\end{equation*}{Furthermore, we parametrize the transition probabilities $\gamma _{k},k=1,...,\bar{k%
},$ such that the first components (small $k$) are the most persistent}
\begin{equation*}
\gamma _{k}=\bar{\gamma}b^{k-\overline{k}},\bar{\gamma}\in (0,1],b>1,k=1,...,\bar{%
	k},
\end{equation*} {and where a possibly higher ``volatility of volatility'' can be accommodated by increasing $\overline{k}$.}

For this model, the structural parameter vector is
\begin{equation*}
\theta =\left( m_{0},\bar{\gamma},b,\sigma ,\overline{k}\right) ^{\prime }
\end{equation*}
and the log-likelihood associated with {observed returns} $(r_{t+1})_{t\leq T}$ is given by:%
\begin{equation}
L_{T}\left( \theta \right) =\frac{1}{T}\sum_{t=1}^{T}\log \left(
\sum_{j=1}^{d}\frac{1}{\sigma \sqrt{g\left( m^{j}\right) }}f_u\left( \frac{%
	r_{t+1}-\mu }{\sigma \sqrt{g\left( m^{j}\right) }}\right) \Pr
[M_{t}=m^{j}\left\vert r_{\tau },\tau \leq t\right] \right)  \label{MSM}
\end{equation}
where the conditional probabilities $\pi _{t}^{j}=\Pr [M_{t}=m^{j}\left\vert
r_{\tau },\tau \leq t\right] $ are computed recursively. {By Bayes' rule, the probability $\pi _{t}^{j}$ can be expressed as a function of the previous probabilities 
$\pi _{t-1}=\left( \pi _{t-1}^{1},...,\pi _{t-1}^{d}\right) :$}
\begin{eqnarray*}
	\pi _{t}^{j} &\propto &\sum_{i=1}^{d}\frac{1}{\sigma \sqrt{g\left(
			m^{i}\right) }}f_u\left( \frac{r_{t}-\mu }{\sigma \sqrt{g\left( m^{i}\right) }}%
	\right) \pi _{t-1}^{i}a_{i,j} \\
	a_{i,j} &=&\Pr [M_{t}=j\left\vert M_{t-1}=i\right] =\dprod\limits_{k=1}^{%
		\overline{k}}\left[ (1-\gamma _{k})1_{\left[ m_{k}^{i}=m_{k}^{j}\right] }+\frac{\gamma
	_{k}}{2} \right].
\end{eqnarray*}

Hence, unlike {continuous stochastic volatility models, such as in Example 2,} the Markov-switching multifractal model has a closed-form
likelihood, precisely because the filtering techniques a la Hamilton can be
applied. However, the price to pay for a volatility process with a discrete
state space is that, for sake of goodness\ of fit, it often takes a state
space with many elements, {which implies a
large number of multipliers $\overline{k}$}. Calvet and Fisher (2004) documents
that for exchange rate data, the multifractal model ``works better for larger
values of $\overline{k}$'' and choose to set the focus on the case $\overline{k}=10$
for all currencies. 

{While the log-likelihood is available in closed-form, a single evaluation requires $O\left( 2^{2\overline{k}}T\right) $ computations, where $%
O\left( .\right) $ denotes the order of the evaluation. Therefore, if the upper bound on the parameter space for $\overline{k}$ is too large, estimation
via maximum likelihood becomes prohibitively expensive.}

{Given the potentially prohibitive computational requirements associated with a large value of $\overline{k}$, it is worth revisiting the likelihood function with the false equality
constraint $\overline{k}=2$, which} is the smallest possible value of $\overline{k}$
allowing to identify all the other parameters. {Under the constraint $\overline{k}=2$, a single likelihood evaluation requires only $16\cdot T$, i.e., $2^{4}T$, computations. Therefore, such a constraint could easily be imposed, and the resulting estimation procedure implemented, to alleviate the computational burden associated with searching over the entire parameter space for $\overline{k}$.}

\subsubsection{Example 5: \textit{Stable Distribution}}

Consider i.i.d. observations $y_{1},\dots ,y_{T}$ generated from a stable
distribution with stability parameter $a\in (0,2]$, skewness parameter $b\in
\lbrack -1,1]$, scale parameter $c>0$ and location parameter $\mu \in 
\mathbb{R}$. The structural parameter vector is given by:%
\begin{equation}
\theta =(a,b,c,\mu )^{\prime }  \label{paramstable}
\end{equation}

The practical problem for maximum likelihood inference in this context\ does
not come from a non-linear state space where the likelihood function would
involve integrals over the state variables. However, it is known that the
log-likelihood function $L_{T}(\theta )$ is not available in general, except
for some specific values of the parameters $a$ and $b$. {As such, maximum likelihood inference can only be implemented by the time-consuming task of numerical inverting the characteristic function, which is known in closed-form, to obtain the resulting (numerical approximation to) the stable density.}

However, for $a=1$ and $b=0$, the stable distribution coincides with the
Cauchy distribution which has a closed-form log-likelihood function $%
L_{T}(1,0,c,\mu )$. \ Moreover, the stable model also allows to simulate
sample paths, for instance with the method of Chambers, Mallows and Stuck
(1976). This will pave the way again for an AML strategy.

\subsection{Pseudo-Score Vector}\label{sec:psv}
The common feature of all the previously discussed examples is that for all values of $\theta$ {in some subset $\Theta_0\subset\Theta$, obtained by imposing some (possibly false)
equality constraints, the log-likelihood function $L_T(\theta)$ in \eqref{loglike} is available in closed form (up to the evaluation of univariate integrals).} {Moreover, we can also show that for all five examples considered in Section \ref{sec:examps1}, considering $\theta \in \Theta _{0}$ allows us to compute, in closed-form,
a pseudo-score vector
\begin{equation}
{\Delta_\theta L_{T}(\theta)}{ };\;\theta \in
\Theta _{0}  \label{pseudoscore}
\end{equation}that can be used as the basis for inference on the unknown $\theta^0$.}

{The notation} $\Delta_\theta L_{T}(\theta)$ is used since certain components
of the pseudo-score vector may not be computed as exact partial derivatives. Of course such an
approximation will be required when some components of $\theta $\ are
integers, {such as  $\overline{k}$ in the multifractal case (Example 4).} Moreover, this approximation will also be relevant
in the case of stable distributions (Example 5), where genuine
partial derivatives with respect to parameters $a$ and $b$ cannot always be
computed.

{Importantly, we note that the pseudo-score vector in (\ref{pseudoscore}) is of the same dimension as the unknown parameters, i.e., it is a $p$-dimensional vector. That is, the
partial derivatives for the pseudo-score are computed with respect to all components of $\theta $, including those {dimensions} whose values are fixed when $\theta \in \Theta _{0}$.
In the following, we demonstrate that, in the examples considered above, constraining
$\theta \in \Theta _{0}$ allows us to compute the pseudo-score in closed-form, at least up to the evaluation of univariate integrals.}

\paragraph{Example 1: (\textit{Autoregressive Discrete Choice Models})}
The dynamic Probit model is a striking example of the fact
that, while the complete likelihood function $l\left\{ \left( y_{t}\right)
_{t\leq T}\left\vert (x_{t}\right) _{ t\leq T};\theta \right\} $
can only be stated as a $T$-dimensional integral, the sub-model defined by $%
\theta _{2}=0$ is much simpler, since it coincides with the usual Probit
likelihood. Not only does the (possibly false) equality
constraint $\theta _{2}=0$ lead to a closed-form likelihood, but the results of Gourieroux, et al. (1985) demonstrate that the partial derivatives of the likelihood function are also available in closed-form. 

Under the restriction $\theta_2=0$, for
	\begin{equation*}
	\tilde{u}_{t}(\theta _{1},0)=\frac{\varphi \left( x_{t}^{\prime }\theta
		_{1}\right) }{\Phi \left( x_{t}^{\prime }\theta _{1}\right) \left[ 1-\Phi
		\left( x_{t}^{\prime }\theta _{1}\right) \right] }\left[ y_{t}-\Phi \left(
	x_{t}^{\prime }\theta _{1}\right) \right],
	\end{equation*}where $\varphi $\ (resp. $\Phi $ ) denotes the probability density
	function (resp. the cumulative distribution function) of the standard
	normal, the computations in Gourieroux et al. (1985) yield
	\begin{flalign*}
	\frac{\partial L_{T}\left( \theta _{1},0\right)}{\partial \theta _{1}} =%
	\frac{1}{T}\sum_{t=1}^{T}x_{t}\tilde{u}_{t}(\theta _{1},0),\quad \frac{\partial L_{T}\left( \theta _{1},\theta_2\right)}{\partial \theta _{2}}\bigg{|}_{\theta_2=0} =%
	\frac{1}{T}\sum_{t=2}^{T}\tilde{u}_{t-1}(\theta _{1},0)\tilde{u}_{t}(\theta
	_{1},0)
	\end{flalign*}
	
The term $\tilde{u}_{t}(\theta_1,0)$ is the generalized residual under the restriction $\theta_2=0$. Gourieroux et al. (1987) \ show that $\tilde{u}_{t}(\theta _{1},0)$
can be interpreted as the conditional expectation of the error term $u_{t}$\
given $y_{t}$\ when the true value of $\theta $ is $(\theta _{1}^{\prime},0)^{\prime }$.



\paragraph{Example 2: (\textit{GARCH-like Stochastic Volatility Model})}
In the case of an $ARCH(1)$-like stochastic volatility model, observed returns are assumed to evolve according to
\begin{eqnarray*}
	r_{t+1} &=&\mu +\varepsilon _{t+1},\varepsilon _{t+1}=\sigma _{t}u_{t+1}, \\
	k_{t} &=&\omega +\alpha \varepsilon _{t}^{2},\sigma _{t}^{2}=k_{t}+\eta _{t},
	\\
	\eta _{t} &=&\rho \eta _{t-1}+\varpi \chi _{t},
\end{eqnarray*}we now demonstrate that the derivatives of the log-likelihood are also available in closed-form. We treat the case of an ARCH(1)-like model for the sake of expositional simplicity, and note that the result extends to other members of this class but require more lengthy derivations. Furthermore, we assume that standardized asset (log)return $u_{t+1}$\ is
Gaussian white noise. For this model, the structural parameter vector is
given by:%
\begin{equation*}
\theta =(\zeta ^{\prime },\rho )^{\prime },\zeta =\left( \mu ,\omega ,\alpha
,\varpi \right) ^{\prime },
\end{equation*}
and the likelihood function (calculated from observed returns $(r_{t+1})_{t\le T}$) is
\begin{equation*}
l[\left\{ r_{t+1}\right\} _{t=1}^{T}\left\vert \theta \right] =\int_{-\infty
}^{+\infty }...\int_{-\infty }^{+\infty }l^{\ast }[\left\{ r_{t+1},\eta
_{t}\right\} _{t=1}^{T}\left\vert \theta \right] d\eta _{1}...d\eta _{T},
\end{equation*}
where $l^{\ast }[\left\{ r_{t+1},\eta _{t}\right\} _{t=1}^{T}\left\vert
\theta \right] $ is the latent likelihood:%
\begin{eqnarray*}
l^{\ast }[\left\{ r_{t+1},\eta _{t}\right\} _{t=1}^{T}\left\vert \theta 
\right] &=&\dprod\limits_{t=1}^{T}\frac{1}{\sqrt{2\pi }}\frac{1}{\sqrt{
	\omega +\alpha \varepsilon _{t}^{2}+\eta _{t}}}\exp \left( -%
\frac{1}{2}\left[ \frac{r_{t+1}-\mu }{\sqrt{\omega +\alpha \varepsilon
	_{t}^{2}+\eta _{t}}}\right] ^{2}\right) f_{\eta }[\eta _{1},...,\eta
_{T}\left\vert \eta _{0},\varpi ,\rho \right],  \\
f_{\eta }[\eta _{1},...,\eta _{T}\left\vert \eta _{0},\varpi ,\rho \right]
&=&\dprod\limits_{t=1}^{T}\frac{1}{\varpi }f_{\chi }\left( \frac{\eta
	_{t}-\rho \eta _{t-1}}{\varpi }\right).  \notag
\end{eqnarray*}

As already announced, imposing the equality constraint $\rho =0$\
will greatly simplify the computation of the observed likelihood and
corresponding score vector. The main reason for that is the implied additive
structure for the latent and observed log-likelihood functions that can be
written:%
\begin{eqnarray*}
	L_{T}^{\ast }\left( \zeta ,0\right) &=&\frac{1}{T}\sum_{t=1}^{T}\log \left(
	l^{\ast }[r_{t+1},\eta _{t}\left\vert r_{\tau },\tau \leq t;\left( \zeta
	,0\right) \right] \right), \\
	L_{T}\left( \zeta ,0\right) &=&\frac{1}{T}\sum_{t=1}^{T}\log \left(
	l[r_{t+1}\left\vert r_{\tau },\tau \leq t;\left( \zeta ,0\right) \right]
	\right) ,\\
	l[r_{t+1}\left\vert r_{\tau },\tau \leq t;\left( \zeta ,0\right) \right]
	&=&\int_{-\infty }^{+\infty }l^{\ast }[r_{t+1},\eta _{t}\left\vert r_{\tau
	},\tau \leq t;\left( \zeta ,0\right) \right] d\eta _{t}.
\end{eqnarray*}

This additive structure is very convenient, not only for its computational
advantages, but also because it allows us to resort to a formula provided by
Gourieroux et al (1987) to compute the observed score vector from the latent score. While this formula had been established by Gourieroux et al. (1987) (as
a generalization of Louis (1982) ) for i.i.d. data, it obviously allows us
to write (the algebra for proving it is perfectly similar):
\begin{equation}
\frac{\partial \log \left( l[r_{t+1}\left\vert r_{\tau },\tau \leq t;\left(
	\zeta ,0\right) \right] \right) }{\partial \zeta }=E\left[ \frac{\partial
	\log \left( l^{\ast }[r_{t+1},\eta _{t}\left\vert r_{\tau },\tau \leq
	t;\left( \zeta ,0\right) \right] \right) }{\partial \zeta }\bigg{|}\{r_{\tau }\}_{\tau \leq t+1} \right]  \label{louis2}.
\end{equation}
Hence, we can compute
\begin{equation}
\frac{\partial L_{T}\left( \zeta ,0\right) }{\partial \zeta }=\frac{1}{T}%
\sum_{t=1}^{T}E\left[ \frac{\partial \log \left( l^{\ast }[r_{t+1},\eta
	_{t}\left\vert r_{\tau },\tau \leq t;\left( \zeta ,0\right) \right] \right) 
}{\partial \zeta }\bigg{|} \{r_{\tau }\}_{\tau \leq t+1}  \right].
\label{obssco}
\end{equation}
Two remarks are in order. First, and by contrast with Gourieroux et al. (1987), due to dynamic
conditional information, (\ref{obssco}) does not give the observed score as
the conditional expectation of the latent score given the observed data.
However, we will see below that it allows a recursive extension of the
concept of generalized residual. Second, it is worth keeping in mind that formulas (\ref{louis2}) and (\ref%
{obssco}) are written by assuming that $\left( \zeta ,0\right) $ is the true
unknown value of the structural parameters that defines the probability
distribution used in the computation of the conditional expectations. Since
in our case, the constraint $\rho =0$ is likely to be a false equality
constraint, the application of (\ref{louis2}) and (\ref{obssco}) will only
provide us with proxies of the true score that we dub pseudo-scores.

Thanks to equation \eqref{louis2}, we can compute the pseudo-score in closed-form. We summarize this result in the following result, and place the derivation of the result in Appendix \ref{sec:glsv}. 
\begin{result}\label{result:g-sv} For $k\in\{-1,1,2\}$, let $[ 1/\left(\sigma_{t}^{2}\right)^{k}]_{F,t}=E[1/\left(\sigma_{t}^{2}\right)^{k}\left\vert r_{\tau
	},\tau \leq t\right] $ denote the filtered function of volatility, computed under the assumed model (and under the parameter restriction $\rho=0$). Then, a closed-form pseudo-score can be obtained with the corresponding components
\begin{eqnarray*}
	\frac{\partial {L}_{T}\left( \zeta ,0\right) }{\partial \mu } &=&\frac{%
		1}{T}\sum_{t=1}^{T}\left[ \frac{1}{\sigma _{t}^{2}}\right] _{F,t}\left(
	r_{t+1}-\mu \right)  \\
	\frac{\partial L_{T}\left( \zeta ,0\right) }{\partial \omega } &=&%
	\frac{1}{2T}\sum_{t=1}^{T}\left[ \frac{1}{\sigma _{t}^{2}}\right] _{F,t}-%
	\frac{1}{2T}\sum_{t=1}^{T}\left[ \frac{1}{\sigma _{t}^{4}}\right]
	_{F,t}\left( r_{t+1}-\mu \right) ^{2} \\
	\frac{\partial {L}_{T}\left( \zeta ,0\right) }{\partial \alpha } &=&%
	\frac{1}{2T}\sum_{t=1}^{T}\left[ \frac{1}{\sigma _{t}^{2}}\right]
	_{F,t}\varepsilon _{t}^{2}-\frac{1}{2T}\sum_{t=1}^{T}\left[ \frac{1}{\sigma
		_{t}^{4}}\right] _{F,t}\left( r_{t+1}-\mu \right) ^{2}\varepsilon _{t}^{2} \\
	\frac{\partial {L}_{T}\left( \zeta ,0\right) }{\partial \varpi } &=&-%
	\frac{1}{\varpi }+\frac{1}{\varpi ^{3}}\frac{1}{T}\sum_{t=1}^{T}\left[ \left[
	\sigma _{t}^{2}\right] _{F,t}-\omega -\alpha \varepsilon _{t}^{2}\right] 
\end{eqnarray*}In addition, a pseudo-score for $\rho$, i.e., ${\partial {L}_{T}\left( \zeta ,0\right) }/{\partial \rho }$,  can be based on the approximation  
	\begin{equation*}
	\frac{1}{\varpi ^{2}}\frac{1}{T}\sum_{t=2}^{T}\left( \left[{\sigma}_{t}^{2}\right]_{F,t}-\omega -\alpha \varepsilon _{t}^{2}\right) \left( \left[{\sigma}_{t-1}^{2}\right]_{F,t-1}-\omega -\alpha \varepsilon _{t-1}^{2}\right). 
	\end{equation*}\hfill\(\Box\)
\end{result}

\paragraph{Example 3: (\textit{Generalized Tobit Model})}
Recall that the log-likelihood for the generalized Tobit model is given by
\begin{eqnarray*}
	L_{T}(\theta ) &=&\frac{1}{T}\sum_{i\in I_{1}}\log \left[ \frac{1}{\sigma }%
	\varphi \left( \frac{y_{i}-x_{i}^{\prime }\theta _{1}}{\sigma }\right) \Pr
	[y_{2i}^{\ast }\geq 0\left\vert y_{i},{z}_{i},\theta _{2},\theta _{3}%
	\right] \right] +\frac{1}{T}\sum_{i\in I_{0}}\log \left[ \Pr [y_{2i}^{\ast }<0\left\vert 
	{z}_{i},\theta \right] \right]\\&=&L_{1,T}\left( \theta \right)
	+L_{2,T}\left( \theta \right),
\end{eqnarray*}where 
\begin{eqnarray*}
	\Pr [y_{2i}^{\ast } &<&0\left\vert {z}_{i},\theta \right] =\int \Pr
	[y_{2i}^{\ast }<0\left\vert y_{1i}^{\ast },{z}_{i},\theta _{2},\theta
	_{3}\right] \frac{1}{\sigma }\varphi \left( \frac{y_{1i}^{\ast
		}-x_{i}^{\prime }\theta _{1}}{\sigma }\right) dy_{1i}^{\ast }, \\
	\Pr [y_{2i}^{\ast } &<&0\left\vert y_{1i}^{\ast },{z}_{i},\theta
	_{2},\theta _{3}\right] =\left[ 1+\exp \left( {z}_{i}^{\prime }\theta
	_{2}+\theta _{3}y_{1i}^{\ast }\right) \right] ^{-1}.
\end{eqnarray*}
As was noted previously, under the restrictions $\theta _{3}=0$, the above log-likelihood has a simple closed-form. 

The score of this likelihood under the restriction $\theta_3=0$ can also be obtained in closed-form. First, we can compute
\begin{eqnarray*}
	\frac{\partial L_{1,T}\left( \theta _{1},\theta _{2},0,\sigma \right) }{\partial \theta _{1}} &=&-%
	\frac{1}{T}\sum_{i\in I_{1}}x_{i}\left[ \frac{y_{i}-x_{i}^{\prime }\theta
		_{1}}{\sigma ^{2}}\right], \;\;
	\frac{\partial L_{1,T}\left( \theta _{1},\theta _{2},0,\sigma \right) }{%
		\partial \theta _{2}} =\frac{1}{T}\sum_{i\in I_{1}}{z}_{i}\left[
	1+e^{{z}_{i}^{\prime }\theta _{2}}\right] ^{-1} \\
	\frac{\partial L_{1,T}\left( \theta _{1},\theta _{2},0,\sigma \right) }{%
		\partial \theta _{3}} &=&\frac{1}{T}\sum_{i\in I_{1}}y_{i}\left[ 1+e^{\tilde{%
			x}_{i}^{\prime }\theta _{2}}\right] ^{-1}, \;\;
	\frac{\partial L_{1,T}\left( \theta _{1},\theta _{2},0,\sigma \right) }{\partial \sigma } =\frac{1}{T%
	}\sum_{i\in I_{1}}\left[ -\frac{1}{\sigma }+\frac{(y_{i}-x_{i}^{\prime
		}\theta _{1})^{2}}{\sigma ^{3}}\right]
\end{eqnarray*}
While we can also check that 
\begin{eqnarray*}
	\frac{\partial L_{2,T}\left( \theta _{1},\theta _{2},0,\sigma \right) }{%
		\partial \theta _{1}} &=&0,\;\;\frac{\partial L_{2,T}\left( \theta _{1},\theta
		_{2},0,\sigma \right) }{\partial \sigma }=0, \\
	\frac{\partial L_{2,T}\left( \theta _{1},\theta _{2},0,\sigma \right) }{%
		\partial \theta _{2}} &=&-\frac{1}{T}\sum_{i\in I_{0}}z_{i}\left[
	1+e^{-z_{i}^{\prime }\theta _{2}}\right] ^{-1}, \\
	\frac{\partial L_{2,T}\left( \theta _{1},\theta _{2},0,\sigma \right) }{%
		\partial \theta _{3}} &=&-\frac{1}{T}\sum_{i\in I_{0}}x_{i}^{\prime }\theta
	_{1}\left[ 1+e^{-z_{i}^{\prime }\theta _{2}}\right] ^{-1}.
\end{eqnarray*}
The pseudo-score can then be the above derivatives, computed under the restriction $\theta_3=0$, i.e.,
$$
\Delta_\theta L_T(\theta)=\frac{\partial L_{1,T}(\theta_1,\theta_2,0,\sigma)}{\partial\theta}+\frac{\partial L_{2,T}(\theta_1,\theta_2,0,\sigma)}{\partial\theta}.
$$

\paragraph{Example 4: (\textit{Markov-Switching Multifractal  (MSM) Model})}

For this model, the structural parameter vector is given by:%
\begin{equation*}
\theta =\left( \zeta ^{\prime },\overline{k}\right) ^{\prime },\zeta =\left(
m_{0},\bar{\gamma},b,\sigma \right) ^{\prime }.
\end{equation*}As already announced, if we consider this model under the false equality
constraint
\begin{equation*}
\overline{k}=2,
\end{equation*}
the log-likelihood associated with observed data $\left\{ r_{t+1}\right\}
_{t=1}^{T}$ is given by
\begin{equation*}
L_{T}\left( \zeta ,2\right) =\frac{1}{T}\sum_{t=1}^{T}\log \left(
\sum_{j=1}^{4}\frac{1}{\sigma \sqrt{g\left( m^{j}\right) }}f_u\left( \frac{%
	r_{t+1}-\mu }{\sigma \sqrt{g\left( m^{j}\right) }}\right) \Pr
[M_{t}=m^{j}\left\vert r_{\tau },\tau \leq t\right] \right).
\end{equation*}

We can then define a pseudo-score vector by
\begin{equation*}
{\Delta_\theta L_{T}}\left( \zeta ,2\right) =\left( \frac{%
	\partial L_{T}\left( \zeta ,2\right) }{\partial \zeta ^{\prime }}%
,L_{T}\left( \zeta ,3\right) -L_{T}\left( \zeta ,2\right) \right) ^{\prime }.
\end{equation*}
Note that filtered $\Pr [M_{t}=m^{j}\left\vert r_{\tau },\tau \leq t\right] $%
\ probabilities depend on all structural parameters as explained above
through in particular two transition probabilities:%
\begin{equation*}
\gamma _{1}=\frac{\bar{\gamma}}{b},\gamma _{2}=\bar{\gamma}.
\end{equation*}

\subsection{Pseudo-Score Matching and AML Estimation}\label{sec:cons1}
{In the previous section, we have exemplified} the computation of pseudo-score vectors
\begin{equation*}
\Delta _{\theta }L_{T}\left( \theta \right) ;\theta \in \Theta _{0},
\text{ where }
L_{T}\left( \theta \right) =\frac{1}{T}\sum_{t=2}^{T}\log \left(
l\{y_{t}\left\vert \left( y_{\tau }\right) _{1\leq \tau \leq
	t-1},x_{t},z_{0};\theta \right\} \right),
\end{equation*}{from which we can compute estimators of the unknown $\theta^0\in\Theta$. While feasible, these estimators do not in general deliver a consistent estimator of $\theta^0$. We now demonstrate how these pseudo-scores can be used to conduct inference on $\theta^0$. Throughout the remainder, we  maintain the following assumption on the parameters and $\Delta _{\theta }L_{T}(\theta)$. }\medskip 

\noindent\textbf{Assumption A1(\textit{False Equality Constraints}):} {The parameter space can be partitioned as}
\begin{eqnarray*}
	\Theta &=&\Theta ^{1}\times \Theta ^{2}, \quad
	\Theta ^{1} \subset 
	\mathbb{R}
	^{p_{1}},\Theta ^{2}\subset 
	\mathbb{R}
	^{p_{2}},\quad p=p_{1}+p_{2} \\
	\Theta _{0} &=&\Theta ^{1}\times \left\{ \left( \beta _{j}^{0}\right)
	_{p_{1}<j\leq p}\right\} =\Theta ^{1}\times \left\{ \beta ^{2,0}\right\}
\end{eqnarray*}
and the application
\begin{equation*}
\beta ^{1}=\left( \theta _{j}\right) _{1\leq j\leq p_{1}}\longrightarrow
\Delta _{\theta }L_{T}\left[ (\beta ^{1^{\prime }},\beta ^{2,0^{\prime
}})^{\prime }\right]
\end{equation*}
is continuously differentiable on the interior of $\Theta ^{1}$.\medskip

{We highlight that this} assumption is fulfilled in the five examples considered
above. We {also require the components of the derivative map in \textbf{Assumption A1} to satisfy the following regularity condition.}\medskip

\noindent\textbf{Assumption A2:} (\textit{Hessian matrix}) Uniformly on the interior
of $\Theta ^{1}$, {for some $(p\times p_1)$-dimensional matrix $K^0$},
\begin{equation*}
\plim_{T\rightarrow \infty }\frac{\partial \Delta _{\theta }L_{T}\left[
	(\beta ^{1^{\prime }},\beta ^{2,0^{\prime }})^{\prime }\right] }{\partial
	\beta ^{1^{\prime }}}=-K^{0}\left[ (\beta ^{1^{\prime }},\beta ^{2,0^{\prime
}})^{\prime }\right],
\end{equation*}{and where $-K^{0}\left[ (\beta ^{1^{\prime }},\beta ^{2,0^{\prime
}})^{\prime }\right]$ has full column-rank.} \medskip

Consider the log-likelihood function computed for a simulated path $%
\{ \tilde{y}_{t}^{(h)}\left( \theta ,z_{0}\right)\} _{t=1}^{T}$ (for $h=1,\dots,H$) and at a value $\beta $ of the structural parameters:\footnote{For the sake of notational simplicity, we have not made explicit the
	dependence of the {likelihood function on the initial value $z_0$ of the simulated data}. Since we are confining ourselves to standard settings, the dependence of $L_T^{(h)}$ on $z_{0}$ will be immaterial asymptotically.}
\begin{equation}
L_{T}^{(h)}\left( \theta ,\beta \right) =\frac{1}{T}\sum_{t=2}^{T}\log \left( l\left\{\tilde{y}_{t}^{(h)}\left( \theta \right) \big{|} \left( 
\tilde{y}_{\tau }^{(h)}\left( \theta \right) \right) _{1\leq \tau \leq
	t-1},x_{t};\beta  \right\} \right)  \label{simsco}.
\end{equation}
{Associated to $L_{T}^{(h)}\left( \theta ,\beta \right)$ is} the simulated pseudo-score vector
\begin{equation*}
\Delta _{\beta }L_{T}^{(h)}\left( \theta,\beta \right) ;\beta \in \Theta
_{0},
\end{equation*}
where the (pseudo) derivative $\Delta _{\beta }$ is computed with respect to
the vector $\beta\in\Theta_0 $ of parameters in (\ref{simsco}), and not with respect to the set of structural parameters, $\theta\in\Theta$, used to simulate $\tilde{y}^{(h)}_t(\theta)$. 

{As is standard, we require regularity on the behavior of the Hessian matrix associated with $\Delta _{\beta }L_{T}^{(h)}\left( \theta,\beta \right)$. 
}\medskip 

\noindent\textbf{Assumption A3 (\textit{Cross-Derivative})}: For all $\beta $\ $\in
\Theta _{0}$, the application
\begin{equation*}
\theta \longrightarrow \Delta _{\beta }L_{T}^{(h)}\left( \theta,\beta
\right)
\end{equation*}
is continuously differentiable on the interior of $\Theta $\ and
\begin{equation*}
\plim_{T\rightarrow \infty }\frac{\partial \Delta _{\beta }L_{T}^{(h)}\left(
	\theta,\beta \right) }{\partial \theta ^{\prime }}=-J^{0}\left( \theta
,\beta \right),
\end{equation*}
{for $J^{0}\left( \theta,\beta \right) $\ a $(p\times p)$-dimensional matrix, with $J^{0}\left( \theta^0 ;\beta^0 \right) $ non-singular.} \medskip 


{Our estimation approach for $\theta^0$ will be based on matching a pseudo-score at a preliminary estimator $\hat{\beta}_{T}$ ($%
\hat{\beta}_{T}\in \Theta _{0}$) of $\beta^0 $.} {We emphasize here that $\hat\beta_T$ is a preliminary estimator of $\beta^0$, and not $\theta^0$,} since it is constrained by the possibly misspecified constraint $\beta \in
\Theta _{0},$ {meaning that it cannot, in general, be a consistent estimator for $\theta^0$}. {We will only maintain that} $\hat{\beta}_{T}$ is a $\sqrt{T}$-consistent estimator of some pseudo-true value $\beta ^{0}$:
\begin{equation*}
\hat{\beta}_{T}=\left( \hat{\beta}_{T}^{1^{\prime }},\beta ^{2,0^{\prime
}}\right) ^{\prime },\beta ^{0}=(\beta ^{1,0^{\prime }},\beta ^{2,0^{\prime
}})^{\prime }.
\end{equation*}

{We can now define our pseudo-score matching estimator of $\theta^0$ as follows.}\medskip

\noindent\textbf{Definition 1:} The Approximate Maximum Likelihood (AML) estimator $\hat{\theta}_{T,H}$\ of $\theta^0 $\
is defined as the solution to the following equation:%
\begin{equation}
\Delta _{\beta }L_{T}\left( \hat{\beta}_{T}\right) =\frac{1}{H}%
\sum_{h=1}^{H}\Delta _{\beta }L_{T}^{(h)}\left( \hat{\theta}_{T,H},\hat{\beta%
}_{T}\right)  \label{AML}.
\end{equation}

{The AML estimator, (\ref{AML}), is defined as the solution of $p$ nonlinear equations, in $p$ unknown parameters,} so that we may expect existence of a
solution $\theta =\hat{\theta}_{T,H}$. However, in practice
it will be safer to minimize a squared norm of a difference between
the two terms in \eqref{AML}. The fact that the system (\ref{AML})\ is just
identified tells us that asymptotically, the behavior of the minimum should
not depend on the weighting matrix used in the squared norm, insofar as (\ref{AML}) {asymptotically defines
a unique solution, which, hopefully coincides with the
true unknown value $\theta ^{0}$.} This will be the purpose of the main
identification assumption ({given in Section \ref{sec:dist}}).

We can already state the general result.\medskip 

\noindent\textbf{Proposition 1:} If $\sqrt{T}( \hat{\beta}_{T}-\beta ^{0})
=O_{P}(1)$, under \textbf{Assumptions A1, A2}, the AML estimator, $\hat{\theta}_{T,H}$,
satisfies
\begin{equation*}
\plim_{T\rightarrow \infty }\left\{ \sqrt{T}\Delta _{\beta }L_{T}\left(
\beta ^{0}\right) -\frac{1}{H}\sum_{h=1}^{H}\sqrt{T}\Delta _{\beta
}L_{T}^{(h)}\left( \hat{\theta}_{T,H},\beta ^{0}\right) \right\} =0.
\end{equation*}
Under \textbf{Assumption A3} and other well-suited identification and
regularity conditions (see section 3 for a precise details),
\begin{eqnarray*}
	\sqrt{T}\left( \hat{\theta}_{T,H}-\theta ^{0}\right) &\rightarrow_d
	&\aleph \left( 0,\Omega _{(H)}\right), \\
	\Omega _{(H)} &=&\left( 1+\frac{1}{H}\right) \left[ J^{0}\left( \theta
	^{0},\beta ^{0}\right) \right] ^{-1}\left[ I^{0}\left( \theta ^{0},\beta
	^{0}\right) \right] \left[ J^{0}\left( \theta ^{0},\beta ^{0}\right) \right]
	^{-1},
\end{eqnarray*}
and with
$I^{0}\left( \theta ^{0},\beta ^{0}\right) =\lim_{T\rightarrow \infty
}\text{Var}\left\{ \sqrt{T}\Delta _{\beta }L_{T}\left( \beta ^{0}\right) -E\left[ 
\sqrt{T}\Delta _{\beta }L_{T}\left( \beta ^{0}\right) \big{\vert} \left\{
x_{t}\right\}_{t=1}^{T}\right] \right\}.$
\hfill\(\Box\)\medskip 

An important message of \textbf{Proposition 1} is that the probability distribution
of the AML estimator $\hat{\theta}_{T,H}$ depends on the choice of the
estimator $\hat{\beta}_{T}$ only through the pseudo-true value $\beta ^{0}$.
In other words, the AML estimator defined by (\ref{AML}) is asymptotically
equivalent to the unfeasible estimator $\breve{\theta}_{T,H}(\beta^0)$\ of $\theta^0 $ that solves
\begin{equation*}
\Delta _{\beta }L_{T}\left( \beta ^{0}\right) =\frac{1}{H}%
\sum_{h=1}^{H}\Delta _{\beta }L_{T}^{(h)}\left( \theta,\beta ^{0}\right).
\end{equation*}

\subsection{Comparison with I-I Approaches}

\subsubsection{Score Matching a la Gallant and Tauchen (1996)}

The pseudo-score that is considered by Gallant and Tauchen (1996) (GT
hereafter) is not, in general, a proxy of the structural score where the
parameter vector $\beta $ is of the same dimension as the structural parameter
vector $\theta $. On the contrary, GT consider an auxiliary model {with
likelihood function}
\begin{equation*}
Q_{T}\left( \beta \right) =\frac{1}{T}\sum_{_{1\leq t\leq T}}\log \left(
q\{y_{t}\left\vert \left( y_{\tau }\right) _{1\leq \tau \leq
	t-1},x_{t},z_{0};\beta \right\} \right) ,\;\beta \in B\subset 
\mathbb{R}
^{q}.
\end{equation*}

The function $q\{y_{t}\left\vert \left(
y_{\tau }\right) _{1\leq \tau \leq t-1},x_{t},z_{0};.\right\} $\ is not, in
general, the true transition density of the process $\left\{
y_{t}\right\} _{t=1}^{T}$. It is a pseudo-likelihood in the sense of
Gourieroux, et al. (1984), {which is precisely the reason for using} the
notations $q\{.\left\vert .\right\} $\ and $Q_{T}\left( .\right) $\ instead
of $l\{.\left\vert .\right\} $\ and $L_{T}\left( .\right) $. Then the pseudo
maximum likelihood estimator $\hat{\beta}_{T}$ satisfies
\begin{equation*}
\frac{\partial Q_{T}}{\partial \beta }\left( \hat{\beta}_{T}\right) =0.
\end{equation*}

Using $\hat{\beta}_{T}$, GT define an I-I estimator $%
\hat{\theta}_{T,H}$ of $\theta^0 $\ as the solution {of the following program}
\begin{equation}
\min_{\theta }\left\Vert \frac{1}{H}\sum_{h=1}^{H}\Delta _{\beta
}Q_{T}^{(h)}\left( \theta ,\hat{\beta}_{T}\right) \right\Vert _{W_{T}}^{2},
\label{GT}
\end{equation}{for $W_{T}$ a positive-definite\ matrix, and where $\left\Vert x\right\Vert _{W_{T}}^{2}=x^{\prime }W_{T}x$.} {While GT only consider the case $H=\infty $, the above definition is indeed
the extension of GT proposed by GMR. In GMR, the authors demonstrate that the} estimator $\hat{\theta}%
_{T,H}$ described above is asymptotically equivalent to the standard I-I estimator based on
matching estimators of $\beta $, {and which implicitly requires $q\geq p.$}

{The GT estimator $\hat{\theta}_{T,H}$ can be
equivalently viewed as the solution of}
\begin{equation*}
\min_{\theta }\left\Vert \Delta _{\beta }Q_{T}\left( \hat{\beta}_{T}\right) -%
\frac{1}{H}\sum_{h=1}^{H}\Delta _{\beta }Q_{T}^{(h)}\left( \theta ,\hat{\beta%
}_{T}\right) \right\Vert _{W_{T}}^{2}.
\end{equation*}
Therefore, if the pseudo-likelihood $Q_{T}(.)$\ would coincide with the true
likelihood $L_{T}(.)$, and $\hat{\beta}_{T}$ would not be subject to false
equality constraints, the GT I-I estimator would exactly coincide with our AML
estimator. However, it is worth keeping in mind that our philosophy for AML\
{is precisely the opposite: we are explicitly concerned with cases where, by the nature of the constraints we employ,}
\begin{equation*}
\Delta _{\beta }Q_{T}\left( \hat{\beta}_{T}\right) \neq 0.
\end{equation*}

A consequence of {this difference in estimation philosophy} is that GT underpin the
accuracy of the I-I estimator $\hat{\theta}_{T,H}$ by the asymptotic
distribution of the auxiliary estimator $\hat{\beta}_{T}$. This {point of view can be seen via}
a Taylor expansion of the first-order conditions
\begin{equation*}
\frac{\partial }{\partial \theta }\left[ \frac{1}{H}\sum_{h=1}^{H}\Delta
_{\beta }Q_{T}^{(h)}\left( \hat{\theta}_{T,H},\hat{\beta}_{T}\right)
^{\prime }\right] W_{T}\frac{\sqrt{T}}{H}\sum_{h=1}^{H}\Delta _{\beta
}Q_{T}^{(h)}\left( \hat{\theta}_{T,H},\hat{\beta}_{T}\right) =0. 
\end{equation*}
{Using} the notations in \textbf{Assumptions A2 and A3}, (and with abuse of notation as
if $L_{T}=Q_{T}$), we see that
\begin{eqnarray*}
	o_{P}\left( 1\right)  &=&J^{0}\left( \theta ^{0},\beta ^{0}\right) ^{\prime
	}W_{T}\frac{\sqrt{T}}{H}\sum_{h=1}^{H}\Delta _{\beta }Q_{T}^{(h)}\left(
	\theta ^{0},\beta ^{0}\right)  \\
	&&+J^{0}\left( \theta ^{0},\beta ^{0}\right) ^{\prime }W_{T}K^{0}\left(
	\beta ^{0}\right) \sqrt{T}\left( \hat{\beta}_{T}-\beta ^{0}\right)
	+J^{0}\left( \theta ^{0},\beta ^{0}\right) ^{\prime }W_{T}J^{0}\left( \theta
	^{0},\beta ^{0}\right) \sqrt{T}\left( \hat{\theta}_{T,H}-\theta ^{0}\right) 
\end{eqnarray*}
GMR (see the part of their Appendix 1 entitled ``The Third Version
of the Indirect Estimator") show that the above Taylor expansion allows us to
view $\sqrt{T}( \hat{\theta}_{T,H}-\theta ^{0}) $\ as an
asymptotically  linear function of the difference between $\hat{\beta}_{T}$%
\ and a similar estimator computed on simulated data. For this reason, the
asymptotic distribution of $\sqrt{T}( \hat{\theta}_{T,H}-\theta
^{0}) $ is directly determined by the asymptotic distribution of $%
\sqrt{T}( \hat{\beta}_{T}-\beta ^{0}) $, {which is in sharp contrast to}
the result of \textbf{Proposition 1} for the AML estimator.   

\subsubsection{Score Matching a la Calzolari, Fiorentini and Sentana (2004)}
Consider that the false equality constraints under which AML is implemented can be written in the implicit form
\begin{equation*}
g\left( \theta \right) =0,
\end{equation*} for some given function $g:\Theta \rightarrow \mathbb{R}^{d_{g}}$,
with $d_{g}<p$. Recall that the log-likelihood function $%
L_{T}(\theta )$ is assumed to be tractable for the set of parameters satisfying this constraint. It is then possible to estimate the parameters from
the Lagrangian function 
\begin{equation*}
\mathcal{L}_{T}(\beta ,\lambda )=L_{T}(\beta )+g(\beta )^{\prime }\lambda ,
\end{equation*}%
where $\lambda \in \mathbb{R}^{d_{g}}$ is the vector of Lagrange
multipliers. The estimator $\hat{\zeta}_{T}=(\hat{\beta}_{T}^{\prime },\hat{%
	\lambda}_{T}^{\prime })^{\prime }$ can then be defined from the
first-order conditions
\begin{eqnarray*}
0 &=&\frac{\partial \mathcal{L}_{T}(\hat{\beta}_{T},\hat{\lambda}_{T})}{%
	\partial \beta }=\Delta _{\beta }L_{T}\left( \hat{\beta}_{T}\right) +\frac{%
	\partial g(\hat{\beta}_{T})^{\prime }}{\partial \beta }\hat{\lambda}_{T},\\
0 &=&g(\hat{\beta}_{T}). 
\end{eqnarray*}

{From these conditions, Calzolari et al. (2004) argue that I-I
score matching should be corrected by the
information contained in the Lagrange multipliers.} In other words, they
propose that $\hat{\theta}_{T,H}$ solve
\begin{equation}
\frac{1}{H}\sum_{h=1}^{H}\Delta _{\beta }L_{T}^{(h)}\left( \hat{\theta}%
_{T,H},\hat{\beta}_{T}\right) +\frac{\partial g(\hat{\beta}_{T})^{\prime }}{%
	\partial \beta }\hat{\lambda}_{T}=0 , \label{scormul}
\end{equation}
which is equivalent to solving
\begin{equation*}
\frac{1}{H}\sum_{h=1}^{H}\Delta _{\beta }L_{T}^{(h)}\left( \hat{\theta}%
_{T,H},\hat{\beta}_{T}\right) -\Delta _{\beta }L_{T}\left( \hat{\beta}%
_{T}\right) =0,
\end{equation*}and coincides with our AML estimator.\footnote{It is worth knowing that Calzolari et al. (2004) also contemplate
	the I-I estimator defined by (\ref{scormul}) in the case of inequality
	constraints on the auxiliary parameters, so that \ \ $\hat{\lambda}_{T}$ is
	a vector of Kuhn-Tucker multipliers. In this case, the argument to consider
	the recentered score vector (\ref{scormul}) instead of a score vector (\ref%
	{GT}) a la Gallant and Tauchen (1996)\ is not any more to correct for a
	misspecification bias but to hedge against possible non asymptotic normality
	of estimators constrained by inequality restrictions. Then, it can be shown
	(see also Frazier and Renault (2019) for a detailed asymptotic theory in
	case of parameters near the boundary of the parameter space) that making the
	difference of the two\ score vectors as in (\ref{AML}) will restore
	asymptotic normality even though each of them is not asymptotically normal,
	due to the fact that the inequality constrained estimator $\hat{\beta}_{T}$
	is not asymptotically normal.}  

Our claim is that, even when
we have no such thing as Lagrange multipliers $\hat{\lambda}_{T}$ to
encapsulate the information about the violation of constraints (information
that should be added to the information brought by the constrained
estimators $\hat{\beta}_{T}$), it still makes sense to imagine that the full
score vector accounts for this missing information. This will be confirmed by
our general analysis in the next subsections. 

In addition, it is worth
noting that even though our AML approach is similar to the I-I
estimators proposed in Calzolari et al. (2004), it stems from a completely
different point of view. We have defined an auxiliary model with parameter
vector $\beta $ as a version of the structural model that has been
simplified. In contrast to Calzolari et al. (2004), we never contemplate
simplifying the auxiliary model, {which in their case has already chosen to be a simple approximation to the structural model.}

\subsubsection{Indirect Inference a la Calvet and Czellar}

The examples in Section \ref{sec:examps1} demonstrate that there are important cases where imposing
a simplifying constraint of the form $\theta =h(\gamma ),\gamma \in 
\mathbb{R}
^{d},d<p,$ results in an auxiliary model that is a computationally feasible version of the
structural model of interest. As explained in Calvet and Czellar (2015):
\textquotedblleft Since [under the constraints] the auxiliary and structural
models are then closely related, the resulting indirect inference estimator
is expected to have good accuracy properties.\textquotedblright 

Calvet and Czellar (2015) propose to use estimators of the auxiliary
parameters based on the observed data, say $\hat{\gamma}_{T}$, and the
simulated data, say $\tilde{\gamma}_{T}(\theta )$, to estimate the
structural parameters. However, while $\hat{\gamma}_{T}$ and  $\tilde{\gamma}%
_{T}(\theta )$ can often be obtained relatively easily, it is important to
realize that these auxiliary parameters can not generally identify the structural
parameters $\theta $, except in the unlikely case that the constraints $%
\left\{ \exists \gamma \in \Gamma ,\theta =h(\gamma )\right\} $ are
satisfied at $\theta ^{0}$ (the true value of the structural parameters).

To circumvent this identification issue, Calvet and Czellar (2015) propose
to add additional auxiliary statistics, with dimension at least as large as $%
p-d$, within the I-I procedure. Denote these statistics based on observed
data by $\hat{\eta}_{T}$ and simulated data by $\tilde{\eta}_{T}(\theta )$,
then Calvet and Czellar (2015) propose to estimate $\theta $ from the
following program: for $\hat{\beta}_{T}:=(\hat{\gamma}_{T}^{{\prime }},%
\hat{\eta}_{T}^{\prime })^{\prime }$, $\tilde{\beta}_{T}(\theta ):=(\tilde{%
	\gamma}_{T}(\theta )^{\prime },\tilde{\eta}_{T}(\theta )^{\prime })^{\prime }
$, an estimator of $\theta ^{0}$ can be obtained by 
\begin{equation}
\min_{\theta \in \Theta }\left( \hat{\beta}_{T}-\tilde{\beta}_{T}(\theta
)\right) ^{\prime }W\left( \hat{\beta}_{T}-\tilde{\beta}_{T}(\theta )\right) ,
\label{eq:czW}
\end{equation}%
where $W$ is a positive-definite weighting matrix of conformable
dimension.

In a sense, the approach of Calvet and Czellar (2015) follows the idea of
estimation under the null that is commonly encountered in testing situations
in econometrics; namely, we estimate a simpler version of the model that is
formed as a constrained version of the model we assume has actually
generated the data, and then we construct statistics about this simpler
model to determine whether or not the simpler model is appropriate to model
the observed data. Several remarks are in order.

First, it is important to keep in mind that for the minimization program (%
\ref{eq:czW}), the simulated data are obtained from the unconstrained
structural model, meaning by considering possibly any $\theta \in \Theta $ \
and not only $\theta \in \Theta ^{0}=\left\{ \theta \in \Theta ;\exists
\gamma \in \Gamma ,\theta =h(\gamma )\right\} $.

Second, since the Calvet and Czellar (2015) approach directly imposes the
constraints in explicit form within the structural model, they obtain what
they consider as an \textquotedblleft unconstrained\textquotedblright\
auxiliary model. The result is that this approach will generate simple
auxiliary estimators of $\beta $. However, the downside is that since we
have disregarded the impact of the constraints the approach can not identify
the entire vector of structural parameters without resorting to ad-hoc
statistics. While the addition of $\hat{\eta}_{T}$ to the auxiliary
estimators may result in a vector of statistics that can identify $\theta
^{0}$, the precise choice of $\hat{\eta}_{T}$ in any given example is
somewhat arbitrary and likely sub-optimal.

Third, for sake of efficient inference, one should realize that, by
definition, the estimator of the simplified structural model
(indexed by a lower dimensional parameter), while convenient, overlooks relevant information. In the following section, we demonstrate that AML can, in a sense, {account for this
information loss, and, thus, get close to the efficiency of maximum
likelihood estimation} without giving up the convenient simplification of our structural model.
\section{Asymptotic Distribution of AML Estimators}\label{sec:dist}
{In this section, we describe the asymptotic distribution of the AML estimator $\hat{\theta}_{T,H}$, which is the
solution, in $\theta$, to}
\begin{equation*}
\Delta _{\beta }L_{T}( \hat{\beta}_{T}) =\frac{1}{H}%
\sum_{h=1}^{H}\Delta _{\beta }L_{T}^{(h)}( \theta ,\hat{\beta}%
_{T}) ,
\end{equation*}
where $\hat{\beta}_{T}$ is a consistent estimator of a pseudo-true value $%
\beta ^{0}\in \Theta _{0}\subset \Theta .$ The asymptotic theory of this
estimator is not completely standard since, for each $h=1,...,H$, $%
L_{T}^{(h)}( \theta ,\hat{\beta}_{T}) $ is a sample mean of $T$
terms, each of them depending on $\hat{\beta}_{T}$, hence it is a double array. As
explained in {Section 2, in particular the result of \textbf{Proposition 1}, we set the focus on situations} where the
asymptotic distribution of the AML estimator $\hat{\theta}_{T,H}$ depends on
the estimator $\hat{\beta}_{T}$, only through its probability limit $\beta
^{0}$. 

{Therefore, to simplify the exposition, we first set the focus on the unfeasible
AML (hereafter, UAML) estimator $\breve{\theta}_{T,H}(\beta ^{0})$, defined as the solution, in $\theta$, to} 
\begin{equation*}
\Delta _{\beta }L_{T}\left( \beta ^{0}\right) =\frac{1}{H}%
\sum_{h=1}^{H}\Delta _{\beta }L_{T}^{(h)}\left( \theta ,\beta ^{0}\right) 
.
\end{equation*}{Since $\Delta _{\beta }L_{T}\left( \beta ^{0}\right) $ is a
pseudo-score, and may include components that can not be represented as partial
derivatives of $L_T(\cdot)$, we follow van der Vaart (1998) (Chapter 5) and refer to $\breve{\theta}_{T,H}(\beta ^{0})$ as a
Z-estimator of $\theta^0 $.}  Moreover, it is worth
recalling that we do not accommodate here the case where one component of the
structural parameter vector is an integer. The discussion of this case could be achieved by extending
the range of the integer parameter to the complete set of non-negative real
numbers, which is feasible by a piecewise linear extension.

\subsection{Consistency}

{For a given pseudo-true value $\beta ^{0}$, consistency of $\breve{\theta}_{T,H}(\beta ^{0})$, for $\theta^0$, follows by applying Theorem 5.9 in van der Vaart (1998), which requires the following regularity condition.} 

\medskip

\noindent\textbf{Assumption B1 (\textit{Identification given $\beta^0$})}: For any $h=1,...,H$, \ $\Delta _{\beta
}L_{T}^{(h)}\left( \theta ,\beta ^{0}\right) $ converges in probability
(as $T\rightarrow \infty $), uniformly on $\theta \in \Theta $, towards a
function $M\left( \theta ,\beta ^{0}\right) $ such that, for every $%
\varepsilon >0$,
\begin{equation*}
\inf_{\theta \in \Theta :d\left( \theta ,\theta ^{0}\right) \geq \varepsilon
}\left\Vert M\left( \theta ,\beta ^{0}\right) -M\left( \theta ^{0},\beta
^{0}\right) \right\Vert >0.
\end{equation*}

{From the i.i.d. nature of the simulation, and the definition of the simulated log-likelihood $%
L_{T}^{(h)}\left( \theta ,\beta ^{0}\right)$ in \eqref{simsco}, it} is not
restrictive to assume that $M\left( \theta ,\beta
^{0}\right) $ does not depend on $h$. Similarly, $\Delta _{\beta
}L_{T}\left( \beta ^{0}\right) $\ converges towards $M\left( \theta
^{0},\beta ^{0}\right) $. Under \textbf{Assumption B1}, we can state the following result. 

\medskip

\noindent\textbf{Proposition 2:} Under \textbf{Assumption B1}, the UAML
estimator $\breve{\theta}_{T,H}(\beta ^{0})$ is a consistent estimator of
the true unknown value $\theta ^{0}$: $\plim_{T\rightarrow \infty }\breve{\theta}_{T,H}(\beta ^{0})=\theta ^{0}$.\hfill\(\Box\)

\medskip

{We now illustrate the identification condition \textbf{Assumption B1} in two examples, and demonstrate that this condition is similar to the identification condition required by ML.} For the purpose of these illustrations, we only consider that \textbf{Assumption B1} enforces
\begin{equation*}
M\left( \theta ,\beta ^{0}\right) -M\left( \theta ^{0},\beta ^{0}\right)
\neq 0,\forall \theta \neq \theta ^{0}.
\end{equation*}{That is, we temporarily overlook the fact that the well-separated minimum of $\left\Vert M\left( \theta ,\beta ^{0}\right) -M\left( \theta
^{0},\beta ^{0}\right) \right\Vert $ generally requires additional regularity, e.g., continuity of the function $%
M(.,\beta ^{0})$ and compactness of $\Theta $.}

\subsubsection*{Example: Well-specified Models}
Assume that $\Delta _{\beta }L_{T}^{(h)}\left( \theta ,\beta \right) $ is
the score vector of a well-specified parametric model for which $\beta
^{0}=\theta ^{0}$ is the true unknown value of the parameters, i.e.,
\begin{equation*}
\Delta _{\beta }L_{T}^{(h)}\left( \theta ,\beta \right) =\frac{1}{T}%
\sum_{{t=1}}^{T}\frac{\partial \log \left[ l\{\tilde{y}%
	_{t}^{(h)}( \theta ) \vert \{\tilde{y}_{\tau
	}^{(h)}(\theta )\}_{1\leq \tau \leq t-1},x_{t};\beta \} \right] 
}{\partial \beta }.
\end{equation*}
Under standard regularity conditions
\begin{equation*}
M(\theta ,\beta )=E_{\theta }\left\{ \frac{\partial \log \left[
	l\{y_{t} | \{ y_{\tau }\}_{1\leq \tau \leq
		t-1},x_{t};\beta \} \right] }{\partial \beta }\right\},
\end{equation*}
where $E_{\theta }$ denotes expectation computed under the
probability distribution of the process $\left\{ y_{t}\right\} _{t=1}^{T}$
at the parameter value $\theta $. The standard identification condition for maximum likelihood is then
\begin{equation*}
M(\theta ,\beta )=0\Longleftrightarrow \theta =\beta.
\end{equation*}
In particular,
\begin{equation*}
M\left( \theta ,\beta ^{0}\right) -M\left( \theta ^{0},\beta ^{0}\right)
\neq 0,\forall \theta \neq \theta ^{0}=\beta ^{0}.
\end{equation*}

In other words, the identification condition in \textbf{Assumption B1} for the UAML
is tantamount to the identification condition for maximum likelihood. \hfill\(\Box\)

\subsubsection*{Example: Exponential Models}
Assume that conditionally on $%
\left\{ x_{t}\right\} _{t=1}^{T}$, the variables $y_{t}$ are
independent, for $t=1,...,T$, and the conditional distribution of $y_{t}$\ only depends on the
exogenous variable $x_{t}$ with the same index. Further, assume that this distribution has a
density $l\{y_{t}\left\vert x_{t};\theta \right\} $\ that is of the exponential form
\begin{equation*}
l\{y_{t}\left\vert x_{t};\theta \right\} =\exp \left[ c\left( x_{t},\theta
\right) +h(y_{t},x_{t})+a(x_{t},\theta )^{\prime }T(y_{t})\right],
\end{equation*}
where $c(.,.)$ and $h(.,.)$ are given functions and $%
a(x_{t},\theta )$ and $T(y_{t})$ are $r$-dimensional random vectors, {all known up to the unknown $\theta^0$}. {The extension to dynamic models, in which conditioning values would also
include lagged values of the process $y_{t}$, can also be considered at the cost of additional notations.}
From
\begin{equation*}
\frac{\partial \log \left[ l\{y_{t}\left\vert x_{t};\theta \right\} \right] 
}{\partial \theta }=\frac{\partial c\left( x_{t},\theta \right) }{\partial
	\theta }+\frac{\partial a\left( x_{t},\theta \right)^{\prime } }{\partial
	\theta }T(y_{t})
\end{equation*}
since the conditional score vector has, by definition, a
zero conditional expectation, we deduce that
\begin{equation*}
\frac{\partial L_{T}\left( \theta \right) }{\partial \theta }=\frac{1}{T}%
\sum_{t=1}^{T}\frac{\partial a^{\prime }\left( x_{t},\theta \right) }{%
	\partial \theta }\left\{ T(y_{t})-E_{\theta }[T(y_{t})\left\vert x_{t}\right]
\right\} .
\end{equation*}

Following Theorem 1 in Gourieroux et al. (1987), 
\begin{eqnarray*}
	E_{\theta }[T(y_{t})\left\vert x_{t}\right] =m\left( x_{t},\theta \right)
	,\;\;Var_{\theta }[T(y_{t})\left\vert x_{t}\right] =\Omega \left( x_{t},\theta
	\right),
\end{eqnarray*}which implies that $$\frac{\partial a\left( x_{t},\theta \right)^{\prime } }{%
\partial \theta }=\frac{\partial m^{\prime }\left( x_{t},\theta \right) }{%
\partial \theta }\Omega ^{-1}\left( x_{t},\theta \right).$$
Therefore, the maximum likelihood estimator $\hat{\theta}_{T}$\ is defined
as the solution to
\begin{equation}
\frac{\partial L_{T}\left( \theta \right) }{\partial \theta }=\frac{1}{T}%
\sum_{t=1}^{T}\frac{\partial m^{\prime }\left( x_{t},\theta \right) }{%
	\partial \theta }\Omega ^{-1}\left( x_{t},\theta \right) \left\{
T(y_{t})-m\left( x_{t},\theta \right) \right\} =0  \label{optinstm}.
\end{equation} The
first-order conditions (\ref{optinstm}) show that maximum likelihood is the
GMM estimator with optimal instruments for the conditional moment
restrictions
\begin{equation*}
E_{\theta }[T(y_{t})-m\left( x_{t},\theta \right) \left\vert x_{t}\right] =0.
\end{equation*}Under the assumptions for standard asymptotic
theory of efficient GMM (Hansen, 1982), i.e., for all $\theta \in \Theta $, the
conditional variance $\Omega \left( x_{t},\theta \right) $\ of the moment
conditions is non-singular and the Jacobian matrix $E[\partial m^{\prime
}\left( x_{t},\theta \right) /\partial \theta \left\vert x_{t}\right] $ is
full row rank, the identification condition for consistency of maximum likelihood is that
\begin{equation*}
E\left\{ \frac{\partial m^{\prime }\left( x_{t},\theta \right) }{\partial
	\theta }\Omega ^{-1}\left( x_{t},\theta \right) \left\{ T(y_{t})-m\left(
x_{t},\theta \right) \right\} \right\} =0\Longrightarrow \theta =\theta ^{0}.
\end{equation*}

We summarize the relationship between the ML identification above and the corresponding version for UAML in the following result, the details of which can be found in Appendix \ref{sec:appTs}. 
\begin{result}\label{result:expo}In the exponential model, the identification condition in \textbf{Assumption B1} can be restated as 
\begin{equation}
E\left\{ \frac{\partial m^{\prime }\left( x_{t},\beta ^{0}\right) }{\partial
	\theta }\Omega ^{-1}\left( x_{t},\beta ^{0}\right) \left\{ m\left(
x_{t},\theta \right) -m\left( x_{t},\theta ^{0}\right) \right\} \right\}
\Longrightarrow \theta =\theta ^{0}  \label{identUAMLm}. 
\end{equation}Two cases are of primary interest to demonstrate that the identification condition for UAML is tantamount to the ML identification condition. \medskip

\noindent\textbf{Case 1:} The model is a linear regression. For some known
	multivariate function $\kappa (x_{t})$ of $x_{t}$,
	\begin{equation*}
	m\left( x_{t},\theta \right) =\kappa\left( x_{t}\right)^{\prime } \theta.
	\end{equation*}	
	The identification condition (\ref{identUAMLm}) is then equivalent to
	\begin{equation*}
	E\left[ \kappa (x_{t})\Omega ^{-1}\left( x_{t},\beta ^{0}\right) \kappa(x_{t})^{\prime }\right] (\theta -\theta ^{0})=0\Longrightarrow \theta
	=\theta ^{0}.
	\end{equation*}Moreover, if $E\left[ \kappa (x_{t})\Omega ^{-1}\left( x_{t},\beta^0\right) \kappa(x_{t})^{\prime }\right]$ is full rank at $\beta^0=\theta^0$, it is full rank for any $\beta^0\in\Theta_0$. \medskip 
	
\noindent\textbf{Case 2:} The model is unconditional. In this case, a necessary identification condition is given by 
	 $$E_\theta[T(y_1)]=E_{\theta^0}[T(y_1)]\iff\theta=\theta^0.$$In this case, the AML identification condition (\ref{identUAMLm}) can be equivalently stated as 
	\begin{equation*}
	\frac{\partial m^{\prime }\left( \beta ^{0}\right) }{\partial \theta }\Omega
	^{-1}\left( \beta ^{0}\right) \left\{ E_{\theta }\left[ T(y_{1})\right]
	-E_{\theta ^{0}}\left[ T(y_{1})\right] \right\} \Longrightarrow \theta
	=\theta ^{0}.
	\end{equation*}The matrix $\partial m\left( \beta^0
	\right)^{\prime } /\partial \theta $\ is full row rank, irrespective of the value of $\beta^0$, so that if $\Omega^{}(\beta^0)$ is non-singular for any $\beta^0\in\Theta_0$, the above identification condition is implied by the identification condition $E_\theta[T(y_1)]=E_{\theta^0}[T(y_1)]\iff\theta=\theta^0. $	
\end{result}
It is also possible to extend the above analysis to the case of latent exponential models. For the sake of brevity, the details of this extension are given in Appendix \ref{sec:appTslate}. \hfill\(\Box\)\medskip 

We now return to the general case and address consistency of AML
based on a first-step consistent estimator of $\beta ^{0}$. For
this purpose, we must slightly reinforce \textbf{Assumption B1}.

\medskip

\noindent\textbf{Assumption B1$'$:} The estimator $\hat{\beta}_{T}$ satisfies $\sqrt{T}( \hat{\beta}%
_{T}-\beta ^{0}) =O_{P}(1)$. \textbf{Assumption B1} is
fulfilled, and, for any $h=1,...,H$ and any real number $\gamma >0$,
\begin{equation*}
\sup_{\theta \in \Theta }\sup_{\left\| \hat{\beta}_{T}-\beta
	^{0}\right\| \leq \frac{\gamma }{\sqrt{T}}}\left\| \Delta _{\beta
}L_{T}^{(h)}( \theta ,\hat{\beta}_{T}) -M\left( \theta ,\beta
^{0}\right) \right\| =o_{P}(1).
\end{equation*}
\medskip

\noindent\textbf{Proposition 3:} Under \textbf{Assumption B1$'$:}, the AML
estimator $\hat{\theta}_{T,H}$\ is a consistent estimator of the
true unknown value $\theta ^{0}$: $\plim_{T\rightarrow \infty }\hat{\theta}_{T,H}=\theta ^{0}$.\hfill\(\Box\)

\subsection{Asymptotic Normality and Efficiency}
{Asymptotic normality has already been demonstrated in \textbf{Proposition 1}; see Section \ref{sec:cons1}}. Ensuring the argument is rigorous only requires slightly reinforcing \textbf{Assumption A3}.

\medskip

\noindent\textbf{Assumption B2}: {For any $h=1,...,H$ and any real
number $\gamma >0$},
\begin{equation*}
\sup_{\theta \in \Theta }\sup_{\left\Vert \hat{\beta}_{T}-\beta
	^{0}\right\Vert \leq \frac{\gamma }{\sqrt{T}}}\left\Vert \frac{\partial
	\Delta _{\beta }L_{T}^{(h)}\left( \theta,\beta \right) }{\partial \theta
	^{\prime }}+J^{0}\left( \theta ,\beta \right) \right\Vert =o_{P}(1).
\end{equation*}

\noindent \textbf{Proposition 4:} Under \textbf{Assumptions A1, A2, A3} and \textbf{Assumptions B1$'$, B2,} the AML estimator $\hat{\theta}_{T,H}$\ and the UAML estimator $\breve{\theta}_{T,H}(\beta^0)$ are
asymptotically normal with zero mean and asymptotic variance
\begin{equation*}
\Omega _{(H)}=\left( 1+\frac{1}{H}\right) \left[ J^{0}\left( \theta
^{0},\beta ^{0}\right) \right] ^{-1}\left[ I^{0}\left( \theta ^{0},\beta
^{0}\right) \right] \left[ J^{0}\left( \theta ^{0},\beta ^{0}\right) \right]
^{-1}.
\end{equation*}\hfill\(\Box\)

A natural question to ask is {how close is the} asymptotic variance matrix $\Omega
_{}=\lim_{H\rightarrow\infty}\Omega_{(H)}$ {to the Cramer-Rao efficiency bound}. It is important to realize that efficiency loss can only occur if $%
\beta ^{0}\neq \theta ^{0}$\ or if the pseudo score vector $\Delta _{\beta
}L_{T}\left( \theta ^{0}\right) $\ is not the true score vector. More
precisely, we prove the following result in Appendix \ref{app:proofs}. \medskip 

\noindent\textbf{Proposition 5}: Under the assumptions of \textbf{Proposition 4}, if
\begin{equation*}
\Delta _{\beta }L_{T}\left( \theta ^{0}\right)  =\frac{1}{T}\sum_{t=2}^{T}%
\frac{\partial \log \left( l\{y_{t}\left\vert \left( y_{\tau }\right)
	_{1\leq \tau \leq t-1},x_{t},z_{0},\theta ^{0}\right\} \right) }{\partial
	\theta } =\frac{1}{T}\sum_{t=2}^{T}S\{y_{t}\left\vert \left( y_{\tau }\right)
_{1\leq \tau \leq t-1},x_{t},z_{0},\theta ^{0}\right\} ,
\end{equation*}
and if $H\rightarrow\infty$, then
asymptotic variance of the UAML estimator, $\breve{\theta}_{T,H}(\beta^0)$, 
(and that of the AML estimator $\hat{\theta}_{T,H}$) achieves the Cramer-Rao efficiency bound. \hfill\(\Box\)\medskip 

However, it is important to note that even if $\Delta _{\beta }L_{T}\left( \beta ^{0}\right)=\sum_{t=2}^{T}S\{y_{t}\left\vert \left(
y_{\tau }\right) _{1\leq \tau \leq t-1},x_{t},z_{0},\beta ^{0}\right\}/T
$, i.e., $\Delta _{\beta }L_{T}( \beta ^{0})$ is
accurately computed at the pseudo-true value $\beta ^{0}$, the
matrix
\[
I^{0}\left( \theta ^{0},\beta ^{0}\right) =\lim_{T\rightarrow \infty
}Var\left\{ \sqrt{T}\Delta _{\beta }L_{T}\left( \beta ^{0}\right) -E\left[ 
\sqrt{T}\Delta _{\beta }L_{T}\left( \beta ^{0}\right) \big{|} \left\{
x_{t}\right\} _{t=1}^{T}\right] \right\} 
\]
will coincide with the Fisher Information Matrix only if
\[
\lim_{T\rightarrow \infty }Var\left\{ E\left[ \sqrt{T}\Delta _{\beta
}L_{T}\left( \beta ^{0}\right) \left\vert \left\{ x_{t}\right\}
_{t=1}^{T}\right) \right] \right\} =0.
\]
This property is unlikely to be fulfilled in the case of a conditional model when $\beta
^{0}\neq \theta ^{0}$. However, it is automatically fulfilled in a model
that is not conditional. Moreover, it is possible to analytically calculate the proximity between the asymptotic variances of AML and genuine maximum likelihood in the, previously considered, case of exponential models.
\subsubsection*{Example: Exponential Models, Continued}

{From the first-order conditions \eqref{optinstm}, the simulated pseudo-score can be stated as}
\begin{equation*}
\Delta _{\beta }L_{T}^{(h)}\left( \theta ,\beta \right) =\frac{1}{T}%
\sum_{t=1}^{T}\frac{\partial m^{\prime }\left( x_{t},\beta \right) }{%
	\partial \theta }\Omega ^{-1}\left( x_{t},\beta \right) \left\{ T\left[ 
\tilde{y}_{t}^{(h)}\left( \theta \right) \right] -m\left( x_{t},\beta
\right) \right\}.
\end{equation*}
Recalling the definition of the UAML estimator, we see that $\breve{\theta}_{T}(\beta^0):=\lim_{H\rightarrow\infty}\breve{\theta}_{T,H}(\beta^0)$ is defined as the
solution, in $\theta $, to
\begin{equation*}
\frac{1}{T}\sum_{t=1}^{T}\frac{\partial m^{\prime }\left( x_{t},\beta
	^{0}\right) }{\partial \theta }\Omega ^{-1}\left( x_{t},\beta ^{0}\right)
\left\{ T(y_{t})-m\left( x_{t},\theta \right) \right\} =0,
\end{equation*}where we recall that $E_\theta[T(y_t)|x_t]=m(x_t,\theta)=\lim_{H\rightarrow\infty}\sum_{h=1}^{H}T[\tilde{y}_t^{(h)}(\theta)]/H$.

{Comparing the above equation with (\ref{optinstm}), the only reason why UAML may
be less efficient than ML is that the evaluation of the ``optimal instruments'' is carried out at a
pseudo-true value} of the structural parameters (i.e., $\beta ^{0}\neq \theta ^{0}$). {It is
worth revisiting the implications of this} in the two cases considered in \textbf{Result \ref{result:expo}}. 
\medskip 

\noindent\textbf{Case 1:} The model is a linear regression. For some known
multivariate function $\kappa (x_{t})$ of $x_{t}$,
\begin{equation*}
m\left( x_{t},\theta \right) =\kappa\left( x_{t}\right)^{\prime } \theta.
\end{equation*}
The equation defining the UAML estimator is then
\begin{equation*}
\frac{1}{T}\sum_{t=1}^{T}\kappa \left( x_{t}\right) \Omega ^{-1}\left(
x_{t},\beta ^{0}\right) \left\{ T(y_{t})-\kappa \left(
x_{t}\right)^{\prime } \theta \right\} =0.
\end{equation*}{From the above, we see that the presence of conditional heteroskedasticity or cross-correlation, of a parametric nature, can result in a loss of efficiency for UAML.} However, if $\Omega \left( x_{t},\beta ^{0}\right) =\sigma ^{2}\text{Id}$, UAML is
asymptotically equivalent to maximum likelihood.\medskip 

\noindent\textbf{Case 2:} The model is unconditional. The equation defining the UAML estimator is then given by
\begin{equation*}
\frac{\partial m^{\prime }\left( \beta ^{0}\right) }{\partial \theta }\Omega
^{-1}\left( \beta ^{0}\right) \frac{1}{T}\sum_{t=1}^{T}\left\{T(y_{t})-m%
\left( \theta \right) \right\} =0.
\end{equation*}{In this case, the only possible loss of efficiency will occur if the moment conditions that identify $\theta $%
\ are overidentified, i.e., when $r=\text{dim}(T)\geq p$, so that the selection
matrix $\frac{\partial m^{\prime }\left( \beta ^{0}\right) }{\partial \theta 
}\Omega ^{-1}\left( \beta ^{0}\right) $ is optimal only at $\beta
^{0}=\theta ^{0}$. An efficiency loss will then occur if, when evaluated at $\beta^0\neq\theta^0$, the vector space
spanned by the rows of the selection matrix do not coincide with the space spanned by the rows when $\beta ^{0}=\theta ^{0}$.}

\section{Examples}
In this section, we apply AML to two of the examples considered in Section \ref{sec:examps1}. First, we analyze the repeated sampling behavior of AML in the confines of the generalized Tobit model, with a pseudo-score computed under the false inequality constraint discussed in Section \ref{sec:gtm}. Next, we evaluate the performance of AML relative to ML in the MSM model, described in Section \ref{sec:msm}, and use AML to estimate the MSM model on daily S\&P500 returns. The empirical results suggest a large value of $\overline{k}$ for this data, which ensures ML can not be feasibly implemented.
\subsection{Example 1: Generalized Tobit Model}
We illustrate the performance of AML in the generalized Tobit-type model via a Monte Carlo study. We generate 1,000 replications from the structural model in equations (\ref{struct1})-(\ref{struct2}) (jointly with the logistic distribution specification for $y_{2i}^{\ast}$, as in equation \eqref{eq:logist}) for two different samples sizes $T=1000$ and $T=10,000$. We fix the true parameter values at $\theta_1=(0.1,0.2)'=\theta_2=(0.1,0.2)'$, and $\theta_3=1$, and the scale parameter for the model is $\sigma=0.5$. The explanatory variables are given by $x_i=\tilde x_i=(1,x_{1i})'$ with $x_{1i}$ generated i.i.d. from the uniform distribution on $[0,1]$. For AML, we take $H=10$ simulated samples. 

\begin{figure}[H]
	\begin{center}
		\includegraphics[width=0.9\textwidth,height=0.5\textheight]{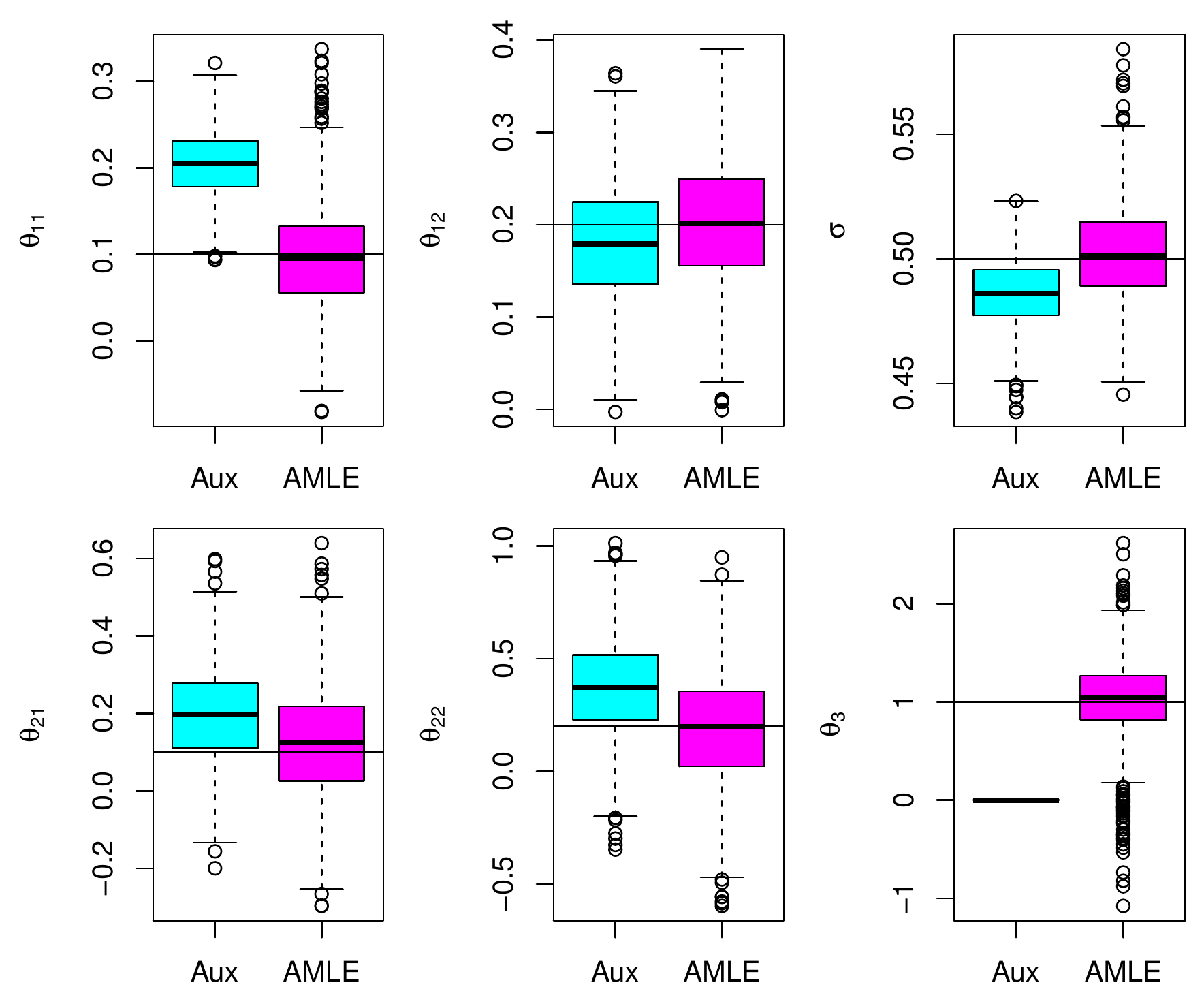}
	\end{center}
	\caption{Each boxplot reports the auxiliary (left boxplots) and AML (right boxplots) parameter estimates for the generalized Tobit model at $T=1,000$ across the Monte Carlo replications. The true parameter values are $\theta_{11}=0.1$, $\theta_{12}=0.2$, $\theta_{21}=0.1$, $\theta_{22}=0.2$, $\theta_{3}=1$, $\sigma=0.5$ and are reported as horizontal lines.}
	\label{fig:tobit1K}
\end{figure}

\begin{figure}[H]
	\begin{center}
		\includegraphics[width=0.9\textwidth,height=0.5\textheight]{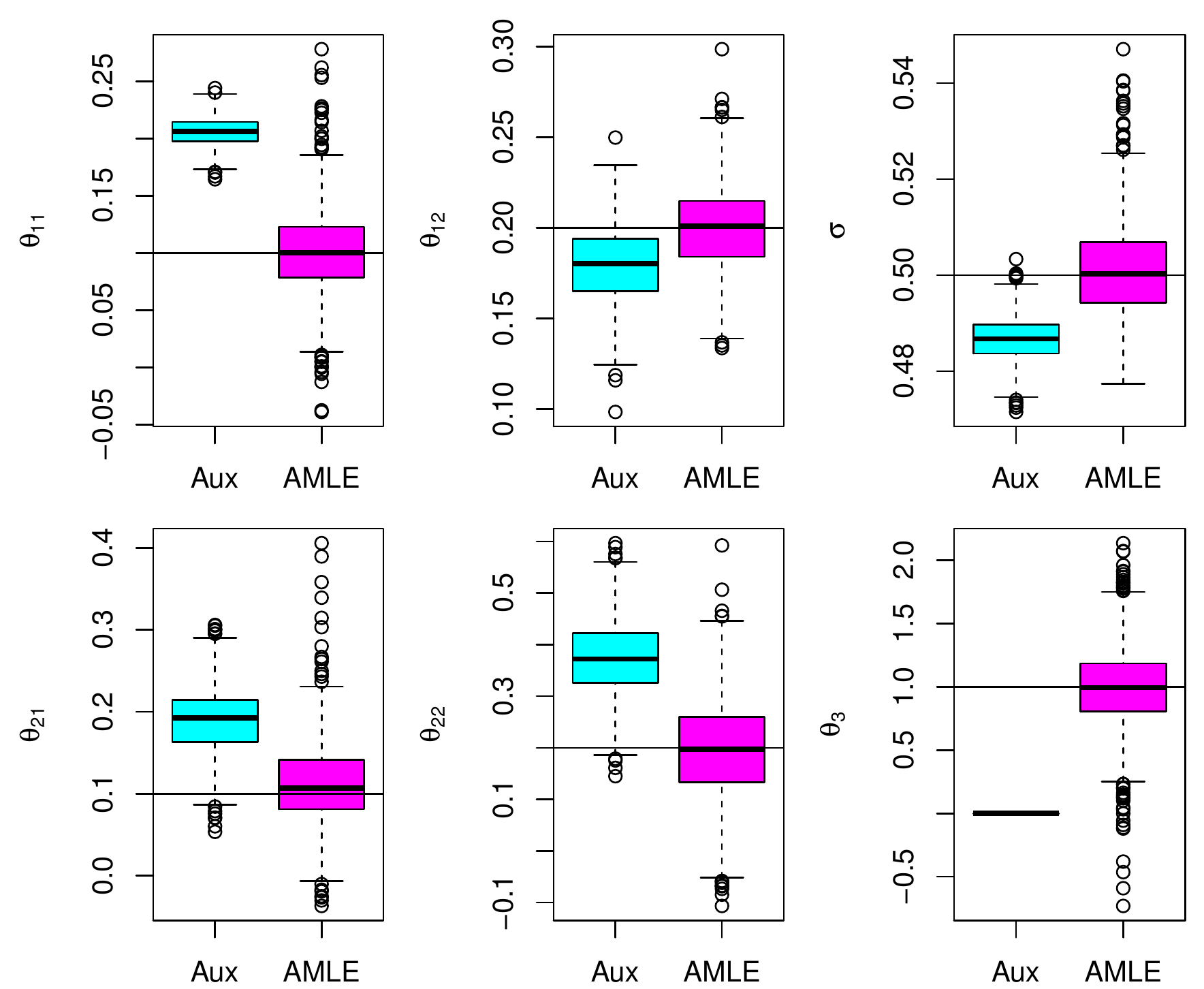}
	\end{center}
	\caption{Each boxplot reports the auxiliary (left boxplots) and AML (right boxplots) parameter estimates for the generalized Tobit model at $T=10,000$ across the Monte Carlo replications. The true parameter values are $\theta_{11}=0.1$, $\theta_{12}=0.2$, $\theta_{21}=0.1$, $\theta_{22}=0.2$, $\theta_{3}=1$, $\sigma=0.5$ and are reported as horizontal lines.}
	\label{fig:tobit10K}
\end{figure}

For each Monte Carlo replication, we calculate the constrained auxiliary estimators and the AML estimator. We compare the resulting estimates graphically in Figures \ref{fig:tobit1K} and \ref{fig:tobit10K}. For each of the parameters, the left figure represents the auxiliary estimator over the replications, and the right figure the AML estimator. The true parameter values are reported as horizontal lines. 

The results demonstrate that while the restricted model is easy to estimate, it ultimately provides biased estimators of the resulting parameters for $\theta_1$, $\theta_2$ and $\sigma$ (as well as $\theta_3$, which is fixed at a value of zero). In contrast, AML delivers point estimators that are well-centred over the true values.

Table \ref{table:tobit} compares the AML and auxiliary estimators across the two samples sizes in terms of bias (Bias), mean squared error (MSE), and Monte Carlo coverage (COV).\footnote{Monte Carlo coverage is calculated as the average number of times, across the Monte Carlo trials, that $\theta^0_j$, i.e., the true value of the $j$-th parameter, is contained in the univariate confidence interval $\hat\theta^i_j\pm\hat\sigma_{j}1.96$, where $\hat\sigma_{j}$ is the standard deviation for the $j$-th parameter over the Monte Carlo replications and $\hat\theta^i_j$ is the estimator of the $j$-th parameter in the $i$-th Monte Carlo trial. } The results demonstrate that AML delivers estimators with relatively small biases, and good Monte Carlo coverage. 

\begin{table}[H]
	\begin{center}
		\vskip10pt
		\begin{tabular}{cccccccc}
			\hline\hline
			&& $\theta_{11}$ & $\theta_{12}$ & $\theta_{21}$ & $\theta_{22}$  & $\theta_3$           & $\sigma$\\
			\hline
			\tvi && \multicolumn{6}{c}{$\underline{T=1,000}$}\\
			
			\tvi \multirow{3}{*}{\underline{Auxiliary}} &
			Bias & 0.1049 & -0.0198 &  0.0946 & 0.1747 & - & -0.0139\\
			& MSE & 0.0125 & 0.0046 & 0.0249 &  0.0815 & - & 0.0004\\
			& COV & 0.2250 & 0.9380 &  0.8930 & 0.8810 & -  & 0.8390\\
			\tvi \multirow{3}{*}{\underline{AML}} &
			Bias & -0.0038 &  0.0010 & 0.0267 & -0.0084 &  0.0218 & 0.0026\\
			& MSE & 0.0039 & 0.0048 & 0.0218 & 0.0643 & 0.1945 &  0.0004\\
			& COV &  0.9420 & 0.9500 & 0.9390 & 0.9480 & 0.9380 & 0.9490\\
			\tvi && \multicolumn{6}{c}{$\underline{T=10,000}$}\\
			
			\tvi \multirow{3}{*}{\underline{Auxiliary}} &
			Bias & 0.1062 & -0.0206 &  0.0906 & 0.1740 & - & -0.0133\\
			& MSE & 0.0114 & 0.0008 & 0.0099 & 0.0356 & - & 0.0002\\
			& COV & 0.0000 & 0.8200 & 0.3850 & 0.3400 & - & 0.1630\\
			\tvi \multirow{3}{*}{\underline{AML}} &
			Bias & 0.0008 & -0.0003 & 0.0125 & -0.0031 &  -0.0123 & 0.0012\\
			& MSE & 0.0015 & 0.0005 & 0.0028 & 0.0097 & 0.1222 & 0.0001\\
			& COV & 0.9450 & 0.9580 & 0.9450 & 0.9400 & 0.9330 & 0.9490\\
			\hline
		\end{tabular}
	\end{center}
	\vskip-15pt
	\caption{Accuracy measures for auxiliary and AML parameter estimates of the generalized Tobit model, across the sample sizes $T=1,000$ and $T=10,000$, and across the 1,000 Monte Carlo replications. The true parameter values are $\theta_{11}=0.1$, $\theta_{12}=0.2$, $\theta_{21}=0.1$, $\theta_{22}=0.2$, $\theta_{3}=1$, $\sigma=0.5$.}
			\label{table:tobit}
\end{table}

\subsection{Example 4: Markov-Switching Multifractal Model}
In this sub-section, we explore the behavior of AML and, when feasible, compare AML and ML. As discussed in Section \ref{sec:msm}, the structural parameters in the MSM model are $\theta=(\zeta',\overline{k})'$, where the parameter $\zeta=(m_0,\bar{\gamma},b,\sigma)'$ govern the behavior of the individual volatility processes, and where $\overline{k}$ denotes the (unknown) number of volatility components. The likelihood of the MSM model, $L_T(\zeta,\overline{k})$, is given in equation \eqref{MSM}, and can be optimized so long as small values of $\overline{k}$ are considered. Indeed, for fixed $\zeta$, computation of the likelihood is only feasible for values of $\overline{k}$ that are not too large: a single evaluation of the log-likelihood for a sample of size $T$ requires $O(2^{2\overline{k}}T)$ computations, and ML estimation becomes infeasible if the true value of $\overline{k}$ is large.

However, under the constraint $\overline{k}=2$, the likelihood $L_T(\zeta,\overline{k})$ requires only $O(2^{4}T)$ computations. This suggest the following constrained estimator for the purpose of AML:\footnote{The more computationally convenient constraint $\overline{k}=1$ can not be readily used as the parameter $b$ vanishes from the log-likelihood function when $\overline{k}=1$.}
\begin{equation}
\hat\beta_T=\argmax_{\beta\in\Theta} L_T(\zeta,\overline{k}),\text{ s.t }\overline{k}=2.
\end{equation}
The likelihood $L_T(\zeta,\overline{k})$ is not differentiable in $\overline{k}$, since $\overline{k}\in\{1,2,\dots,\}$, and so for the $\overline{k}$ component of the AML pseudo-score we use the difference approximation $L_T(\zeta,3)-L_T(\zeta,2)$, which yields 
\begin{equation}
\label{qtilde1}
{\Delta_\beta L_{T}}\left( \zeta ,2\right) =\left( \frac{%
	\partial L_{T}\left( \zeta ,2\right) }{\partial \zeta ^{\prime }}%
,L_{T}\left( \zeta ,3\right) -L_{T}\left( \zeta ,2\right) \right) ^{\prime }.
\end{equation}
where we note that ${\partial L_T(\zeta,2)}/{\partial{\zeta}'}$ can be reliably obtained using numerical differentiation.

To implement AML in this example, we consider $H$ i.i.d. simulated samples, from the MSM model. From these simulated samples, the AML estimator is obtained by minimizing, in the Euclidean norm, the difference between the average simulated pseudo-score $\sum_{h=1}^{H}\Delta_\beta L^{(h)}_{T}(\theta,\hat{\beta}_T)/H$ and ${\Delta_\beta L_{T}}( \hat\beta_T)$.

\subsubsection*{Monte Carlo}
We first consider data generated from the MSM model with $\mu=0$ and a relatively small value of $\overline{k}$ so that ML is computationally feasible. This allows us to compare AML and ML, and directly assess the efficiency loss of AML relative to ML. To this end, we generate 1,000 synthetic data sets from the MSM model in Section \ref{sec:msm} with $T=5,000$ observations, and where the parameter values are set as follows: $m_0=1.5$, $\overline\gamma=0.2$, $b=4$, $\sigma=0.01$ and $\overline{k}=4$. 

Numerical implementation of AML and ML require optimization over the integer parameter space for $\overline{k}$, while optimization for the $\zeta$ components can proceed via standard approaches. For both approaches, optimization over the $\zeta$ components is carried out using a quasi-Newton approach, with finite-differences used to estimate the derivatives. For the $\overline{k}$ components, the likelihood is optimized across the grid $\{1,\dots,7\}$, while AML considers a much larger grid of values.\footnote{Technically, we implement AML by extending the grid of values over which $\overline{k}$ is optimized to the entire real line. This is done by considering a piecewise linear extension of the pseudo-score for the $\overline{k}$ component, and by taking the closest integer to the resulting optimized value.}

The ability of AML to consider large values for $\overline{k}$ is possible because the computational cost required to evaluate the AML criterion function \textit{does not} increase with $\overline{k}$, and requires $O(HT)$ computations for any value of $\overline{k}$. In this Monte Carlo exercise, AML is implemented using $H=100$ pseudo-samples, as the large value of $H$ smooths the criterion function and increases the accuracy of numerical differentiation methods.\footnote{An alternative to the finite-differences considered herein would be to use the simulation-based differentiation approach in Frazier et al. (2019).}

\begin{figure}[H]
	\begin{center}
		\includegraphics[width=0.9\textwidth,height=0.5\textheight]{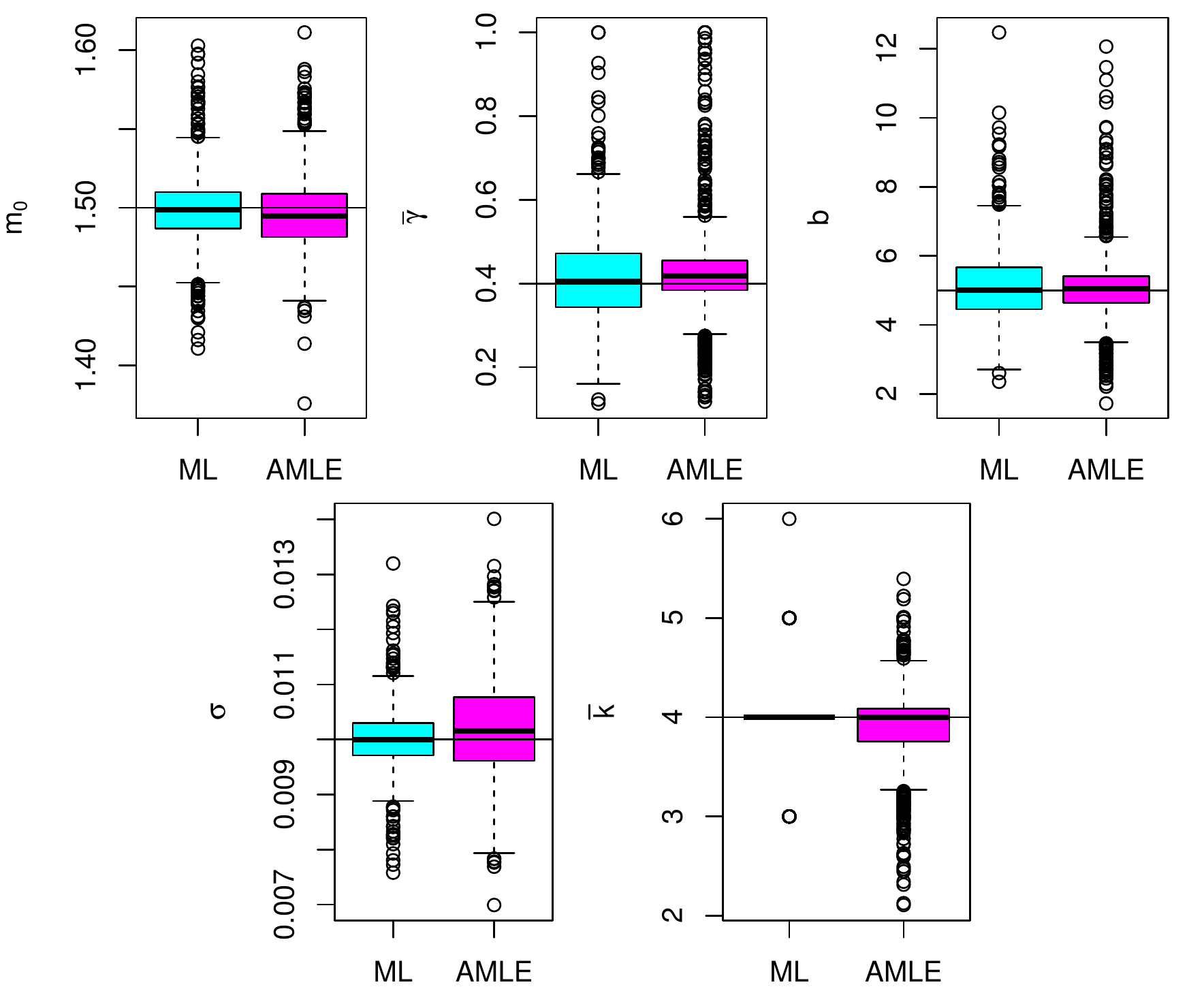}
	\end{center}
	\caption{Each boxplot reports the ML (left boxplots) and AML (right boxplots) parameter estimates for the MSM model with sample size $T=5,000$ across the Monte Carlo replications. The true parameter values are $m_0=1.5$, $\overline{\gamma}=0.4$, $b=5$, $\sigma=0.01$ and $\overline{k}=4$ and are reported as horizontal lines. }
	\label{fig:MSM4}
\end{figure}

Figure \ref{fig:MSM4} displays the results of this Monte Carlo experiment. For each sub-figure, the left plot contains the ML estimator and the right plot contains the associated AML estimator. The true parameter values are reported as horizontal lines. AML provides estimators that are well-centred over the true value of the structural parameters with, as expected, a larger variance than the ML estimator in some cases. 

Table \ref{table:MSM4} compares the bias (Bias), mean squared error (MSE) and Monte Carlo coverage (COV) of the estimators. In addition, for each replication we calculate the efficiency loss of AML with respect to ML via the average relative standard error, denoted by SE(ML)/SE(AML) in Table \ref{table:MSM4}. Using this measure, numbers below unity suggest that, on average, the ML estimator is more efficient than the AML estimator. The results in Table \ref{table:MSM4} suggest that the two estimators are comparable in terms of bias and MSE for $m_0$, $\bar{\gamma}$ and $b$, with ML yielding more accurate estimators for $\overline{k}$ and $\sigma$. Analyzing the efficiency of the two estimators, we see that, according to the SE(ML)/SE(AML) measure, AML is nearly as efficient as ML for $m_0$, $\bar{\gamma}$ and $b$, but less so for $\sigma$ and $\overline{k}$. The later is not entirely unexpected as imposing the invalid restriction $\overline{k}=2$ within the pseudo-score should lead to some efficiency loss (with respect to ML). However, this example also demonstrates that imposing this restriction only leads to a minor loss in accuracy for estimating $m_0$, $\bar{\gamma}$ and $b$.

\begin{table}[H]
	\begin{center}
		\vskip10pt
		\begin{tabular}{cccccccc}
			\hline\hline
			&& $m_0$ & $\overline{\gamma}$ & $b$ & $\sigma$  & $\overline{k}$\\
			\hline
			\tvi \multirow{3}{*}{\underline{ML}} 
			&	Bias	&	-0.0014244	&	0.0134517	&	0.1367587	&	0.0000088	&	-0.0120000	\\
			&	MSE	&	0.0004834	&	0.0121796	&	1.0688123	&	0.0000003	&	0.0900000	\\
			&	COV	&	0.9380000	&	0.9520000	&	0.9560000	&	0.9490000	&	0.9130000	\\
			\tvi \multirow{3}{*}{\underline{AML}} 
			&	Bias	&	-0.0036913	&	0.0280103	&	0.0653309	&	0.0002228	&	-0.0878051	\\
			&	MSE	&	0.0005691	&	0.0142423	&	0.9924541	&	0.0000009	&	0.1727888	\\
			&	COV	&	0.9510000	&	0.9430000	&	0.9440000	&	0.9310000	&	0.9150000	\\
			&	SE(ML)/SE(AML)	&	0.9309007	&	0.9442337	&	1.0308558	&	0.5860551	&	0.7377811	\\
			\hline
		\end{tabular}
	\end{center}
	\vskip-15pt
	\caption{Accuracy measures for ML and AML parameter estimates of the MSM for $T=5,000$, and across the 1,000 Monte Carlo replications. The true parameter values are $m_0=1.5$, $\overline{\gamma}=0.4$, $b=5$, $\sigma=0.01$ and $\overline{k}=4$. In ML estimation, $\overline{k}$ only takes values in $\{1,\dots,7\}$.}
			\label{table:MSM4}
\end{table}

\begin{figure}[H]
	\begin{center}
		\includegraphics[width=0.9\textwidth,height=0.4\textheight]{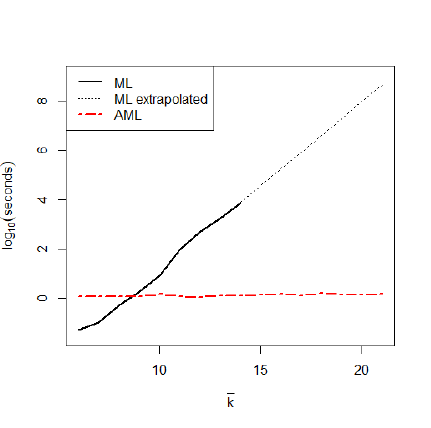}
	\end{center}
	\caption{Computation times, in $\log_{10}$ seconds, of the likelihood function (continuous line) and AML criterion function (dash-dotted line) using $H=100$. The averages presented are taken over twenty data sets simulated from the MSM model with $T=5,000$, $m_0=1.5$, $\overline\gamma=0.2$, $b=4$, $\sigma=0.01$ and $\overline{k}=6,\,7,\dots,21$. Small dotted line indicates extrapolated computation time for ML estimation for $\overline{k}\ge15$.}
	\label{fig:msmtime}
\end{figure}

While ML has an edge in terms of accuracy, due to computational cost, ML is infeasible if the true value of $\overline{k}$ is large. To illustrate this point, we compare the time, in $\log_{10}$ seconds, required to evaluate the log-likelihood function and the AML criterion function for various values of $\overline{k}$ and for a sample size of $T=5,000$. Programs were implemented in C and computation was performed on an Intel(R) Xeon(R) CPU E7-4830 v3 @ 2.10GHz. For each $\overline{k}=6,7,\dots,21$, we evaluate twenty Monte Carlo replications and report the mean computation time for the AML criterion function based on $H=100$ simulated samples. We repeat the same exercises for the log-likelihood function and for $\overline{k}=6,7,\dots,14$, with linear extrapolation used for values of $\overline{k}\ge15$. Figure~\ref{fig:msmtime} compares the mean computation times. For $\overline{k}$ small, evaluation of the likelihood is faster than the AML criterion, given the large number of simulated paths used in the AML criterion. However, when $\overline{k}$ becomes even moderately large, AML is clearly superior in terms of computational cost. For values of $\overline{k}>9$, AML is particularly attractive in terms of computation time. At a value of $\overline{k}=21$, a single evaluation of the log-likelihood would require 5459.2 days (approximately 15 years), whereas an evaluation of the AML criterion only requires 1.45 seconds. 

\begin{figure}[H]
	\begin{center}
			\includegraphics[width=0.9\textwidth,height=0.5\textheight]{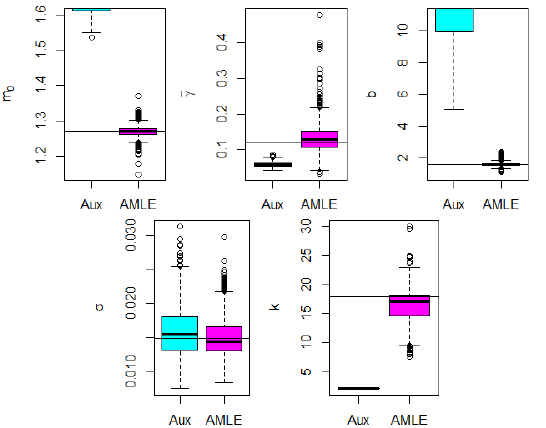}
	\end{center}
	\caption{Each boxplot reports the auxiliary (left boxplots) and AML (right boxplots) parameter estimates for the MSM model with sample size $T=23,202$ across the Monte Carlo replications.  The true parameter values are $m_0=1.2708$, $\overline{\gamma}=0.1215$, $b=1.5663$, $\sigma=0.0149$ and $\overline{k}=18$ and reported with horizontal lines.}
	\label{fig:MSM21}
\end{figure}

We now assess the performance of AML for a large value of $\overline{k}$. We choose $\overline{k}=18$ and other parameter values that resulted from the empirical example conducted later (see Table~\ref{tab:sp500est} in the following subsection). Figure~\ref{fig:MSM21} displays the estimation results over 1,000 Monte Carlo replications from the DGP associated with $T=23,202$ (as in the empirical dataset in the following subsection), and where the parameter values are $m_0=1.2708$, $\overline{\gamma}=0.1215$, $b=1.5663$, $\sigma=0.0149$ and $\overline{k}=18$. For each sample, we calculate the constrained estimator and AML estimator using $H=100$ pseudo-samples.  For each sub-figure, the left plot contains the constrained auxiliary estimates and the right plot contains the associated AML estimator. The true parameter values are reported with horizontal lines. While the restricted model is easy to estimate, it provides estimators that are significantly biased for all parameters except $\sigma$. AML corrects the resulting bias for all structural parameters and delivers estimators that are, on average, centred over the true values. Analyzing the other accuracy measures given in Table \ref{table:MSM21}, we see that AML generally yields estimators with low bias and Monte Carlo coverage close to the nominal level.  

\begin{table}[H]
	\begin{center}
		\vskip10pt
		\begin{tabular}{cccccccc}
			\hline\hline
			&& $m_0$ & $\overline{\gamma}$ & $b$ & $\sigma$  & $\overline{k}$\\
			\hline
			\tvi \multirow{3}{*}{\underline{Auxiliary}} 
			&	Bias	&	 0.363348 &   -0.061777 &  12.002244 &  0.000943 	&	-	\\
			&	MSE	&	 0.133257 &   0.003867 &   174.209123 & 0.000014 	&	-	\\
			&	COV	&	0.000000	&	0.000000	&	0.480000	&	0.939000	&	-	\\
			\tvi \multirow{3}{*}{\underline{AML}} 
			&	Bias	&	-0.001502 & 0.012439 &  0.025719 &  0.000033 & -1.558178	\\
			&	MSE	&	0.000303 &  0.002176 &  0.022416 &  0.000009 &  11.391885
			\\
			&	COV	&	 0.936000 & 0.955000 & 0.937000 & 0.945000 & 0.897000\\
			\hline
		\end{tabular}
	\end{center}
	\vskip-15pt
	\caption{Accuracy measures for auxiliary and AML estimator parameter estimates of the MSM model with $T=23,202$, and across the 1,000 Monte Carlo replications. True parameter values are $m_0=1.2708$, $\overline{\gamma}=0.1215$, $b=1.5663$, $\sigma=0.0149$ and $\overline{k}=18$.}
			\label{table:MSM21}
\end{table}

\subsubsection*{Application: S\&P500 Returns}

\begin{figure}[H]
	\begin{center}
		\includegraphics[width=0.7\textwidth,height=0.25\textheight]{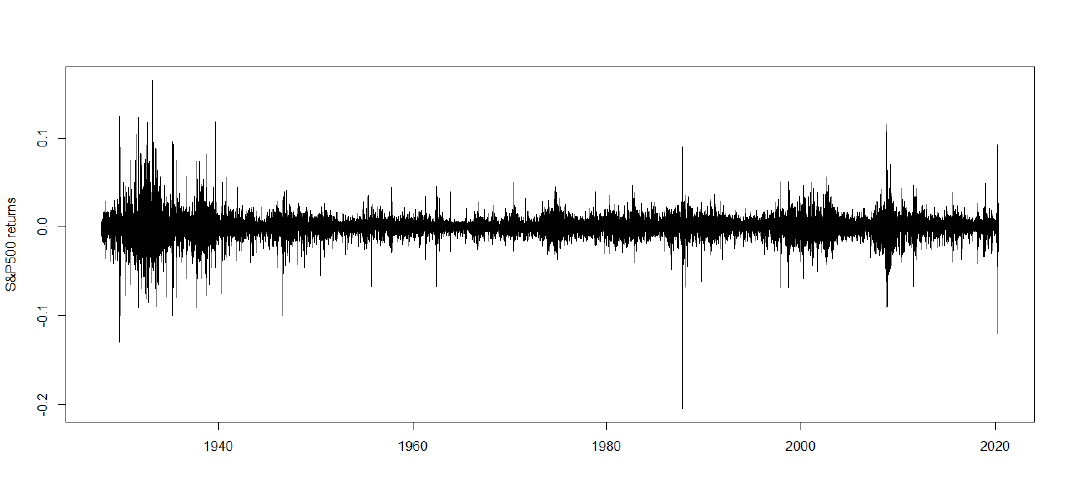}
	\end{center}
	\caption{Daily S$\&$P500 returns between January 3, 1928 and May 15, 2020.}
	\label{fig:SP500}
\end{figure}

We now estimate the Binomial MSM model (with $\mu=0$) on demeaned daily S$\&$P500 (simple) returns between January 3, 1928 and May 15, 2020\footnote{Downloaded from finance.yahoo.com on May 15, 2020.}. The sample size is $T=23,202$. The data are plotted in Figure~\ref{fig:SP500}. Using this data, Table~\ref{tab:sp500est} compares the AML estimators with those obtained from maximum likelihood for fixed values of $\overline{k}$ ranging from $\overline{k}=1$ up to $\overline{k}=10$.  The estimated value of $\overline{k}$ obtained by AML is far larger than the feasible value associated with ML. Moreover, except for $m_0$, the remaining estimated parameters are also significantly different, with the estimated values of $\bar{\gamma}$ and $b$ being markedly different across the two approaches. The standard errors for ML are calculated using the asymptotic formula, while those for AML are calculated using a parametric bootstrap based and 1,000 simulated data sets from the assumed DGP. 

In order to compare the goodness-of-fit of the eleven models enumerated in Table~\ref{tab:sp500est}, for each model we provide one-day-ahead forecasts at each in-sample date $t=1,\dots,T$ using a particle filter of size $N=10^6$. For a given model, at each date $t$, the particle filter provides $N$ simulated values from the approximate distribution of $r_t|\{r_1,\dots,r_{t-1}\}$:
\begin{equation*}
r_t^{(1)},\dots,r_t^{(N)}\,.
\end{equation*}  
At each date $t=1,\dots,T$, we calculate the $\alpha=1\%$ and $\alpha=5\%$ value-at-risk forecasts defined by
\begin{equation*}
{\text{VaR}}_{\alpha,t}=-q_\alpha(r_t^{(1)},\dots,r_t^{(N)})\,,
\end{equation*}  
where $q_\alpha(\cdot)$ indicates the $\alpha$-th sample quantile, and report the failure rate of ${\text{VaR}}_{\alpha,t}$:
\begin{equation*}
p_{\alpha}=\frac{1}{T}\sum_{t=1}^T 1_{r_t<(-{\text{VaR}}_{\alpha,t})}\,.
\end{equation*}  
The closer $p_{\alpha}$ is to $\alpha$, the better the forecasts. The left panel of Table~\ref{tab:sp500reg} reports $p_{\alpha}$ for $\alpha=0.01$ and $\alpha=0.05$ for each model specification along with asymptotic standard errors in parentheses. AML provides the only model specification for which both failure rates are not significantly different from their nominal levels. In addition, we also assess the accuracy of the $\alpha=5\%$ expected shortfall forecasts:
\begin{equation*}
ES_{\alpha,t}=\sum_{i=1}^N r_t^{(i)}1_{r_t^{(i)}<(-{\text{VaR}}_\alpha^t)}\bigg{/}\sum_{i=1}^N 1_{r_t^{(i)}<(-{\text{VaR}}_\alpha^t)}\,.
\end{equation*}  
To this end, we collect the empirical returns satisfying $r_t|r_t<-\text{Var}_{0.05,t}^{(\overline{k}=10)}$,  under the model with $\overline{k}=10$, and for each value of $\overline{k}$ in Table \ref{tab:sp500est}, we regress these returns on $ES_{\alpha,t}$, calculated under the corresponding value of $\overline{k}$ in Table \ref{tab:sp500est}. Regression intercepts, slopes, $R^2$ values and $p$-values of the Wald test associated with the joint hypothesis $({\text{intercept}},{\text{{slope}}})'=(0,1)$ are reported in the right panel of Table~\ref{tab:sp500reg}. 
The $\overline{k}=18$ specification provides the best expected shortfall forecasts, as measured by the magnitude of the corresponding $p$-values.

\begin{table}[H]
	\begin{center}
		\begin{tabular}{ccccccc}
			\hline
			& $\overline{k}$ & $m_0$ & $\overline{\gamma}$ & $b$ & $\sigma$  & Log-like.\\
			\hline																				
			\multirow{10}{*}{\underline{ML}} & 
			\tvi 1 &
			$\underset{(0.0040)}{1.8168}$ & $\underset{(0.0062)}{0.0269}$ & - & $\underset{(0.0002)}{0.0164}$ & 75476.07\\
			&  2 &
			$\underset{(0.0040)}{1.6654}$ & $\underset{(0.0062)}{0.0593}$ & $\underset{(1.1063)}{14.6239}$ & $\underset{(0.0002)}{0.0157}$ & 76409.45\\
			& 3 &
			$\underset{(0.0040)}{1.5890}$ & $\underset{(0.0062)}{0.0922}$ & $\underset{(1.1063)}{9.0988}$ & $\underset{(0.0002)}{0.0161}$ & 76779.51\\
			& 4 &
			$\underset{(0.0052)}{1.5199}$ & $\underset{(0.0861)}{0.1149}$ & $\underset{(0.3414)}{5.3760}$ & $\underset{(0.0003)}{0.0151}$ & 76874.11\\
			& 5 &
			$\underset{(0.0052)}{1.4745}$ & $\underset{(0.0861)}{0.1461}$ & $\underset{(0.3414)}{4.6768}$ & $\underset{(0.0003)}{0.0161}$ & 76940.79\\
			& 6 &
			$\underset{(0.0052)}{1.4517}$ & $\underset{(0.0861)}{0.9441}$ & $\underset{(0.3414)}{6.5357}$ & $\underset{(0.0003)}{0.0152}$ & 76978.39\\
			& 7 &
			$\underset{(0.0055)}{1.4291}$ & $\underset{(0.0929)}{0.9999}$ & $\underset{(0.2772)}{5.5954}$ & $\underset{(0.0002)}{0.0132}$ & 76994.82\\
			& 8 &
			$\underset{(0.0060)}{1.3882}$ & $\underset{(0.1093)}{1.0000}$ & $\underset{(0.1854)}{3.9099}$ & $\underset{(0.0003)}{0.0128}$ & 77001.80\\
			& 9 &
			$\underset{(0.0062)}{1.3568}$ & $\underset{(0.1224)}{1.0000}$ & $\underset{(0.1427)}{3.1657}$ & $\underset{(0.0005)}{0.0137}$ & 77006.94\\
			& 10 &
			$\underset{(0.0067)}{1.3383}$ & $\underset{(0.1305)}{1.0000}$ & $\underset{(0.1328)}{2.8090}$ & $\underset{(0.0006)}{0.0130}$ & 77009.94\\
			\hline
			\tvi 	\underline{AML} & $\underset{(2.9955)}{18}$ & $\underset{(0.0173)}{1.2708}$ & $\underset{(0.0450)}{0.1215}$ & $\underset{(0.1476)}{1.5663}$ & $\underset{(0.0030)}{0.0149}$ & - \\
			
		\end{tabular}
	\end{center}
	\vskip-15pt
	\caption{The table reports the ML estimator (ML) and AML estimator (AML) of the demeaned empirical S$\&$P500 returns (left panel). Asymptotic standard errors for the ML estimator are reported in parentheses below each value. The AML standard errors are obtained using a parametric bootstrap based on 1,000 simulated samples (of length $T=23,202$) generated from the MSM model at the AML point estimates.}
	\label{tab:sp500est}
	
\end{table}

\begin{table}[H]
	\caption{Goodness-of-fit comparisons of AML and ML with various $\overline{k}$.}
	\begin{center}
		\begin{tabular}{cc|cc|cccc}
			\hline
			\tvi & \multicolumn{3}{c}{{\phantom{space}}\underline{VaR failure rates}} & \multicolumn{4}{c}{\underline{${\text{ES}}_{0.05}$ regressions}}\\
			\tvi & $\overline{k}$ & $p_{0.05}$ & $p_{0.01}$ & Intercept & Slope & $R^2$ & Wald\\
			\hline
			\multirow{10}{*}{\underline{ML}}
			& \tvi 1 & $\underset{(0.0013)}{{0.0427}}$ & $\underset{(0.0006)}{{0.0081}}$ &  $\underset{(0.0008)}{0.0007}$ & $\underset{(0.0277)}{0.9112}$ & 0.4771 & ${3\cdot 10^{-19}}$\\	 
			& 2 & $\underset{(0.0014)}{{0.0463}}$ & $\underset{(0.0006)}{{0.0082}}$ &  $\underset{(0.0007)}{{0.0024}}$ & $\underset{(0.0244)}{1.0322}$ & 0.6015 & ${5\cdot 10^{-7}}$\\	 
			& 3 & $\underset{(0.0014)}{{0.0463}}$ & $\underset{(0.0006)}{{0.0082}}$ &  $\underset{(0.0006)}{0.0009}$ & $\underset{(0.0220)}{0.9947}$ & 0.6331 & {{0.0013}}\\	 
			& 4 & $\underset{(0.0014)}{0.0486}$ & $\underset{(0.0006)}{{0.0082}}$ &  $\underset{(0.0006)}{0.0005}$ & $\underset{(0.0216)}{0.9975}$ & 0.6420 & 0.1256\\
			& 5 & $\underset{(0.0014)}{0.0479}$ & $\underset{(0.0006)}{{0.0085}}$ &  $\underset{(0.0006)}{-0.0003}$ & $\underset{(0.0209)}{0.9622}$ & 0.6420 & {{0.0151}}\\
			& 6 & $\underset{(0.0014)}{{0.0461}}$ & $\underset{(0.0006)}{{0.0075}}$ &  $\underset{(0.0006)}{0.0004}$ & $\underset{(0.0222)}{0.9666}$ & 0.6149 & ${7\cdot 10^{-5}}$\\
			& 7 & $\underset{(0.0014)}{0.0477}$ & $\underset{(0.0006)}{{0.0078}}$ &  $\underset{(0.0006)}{0.0006}$ & $\underset{(0.0228)}{0.9873}$ & 0.6127 & {{0.0115}}\\
			& 8 & $\underset{(0.0014)}{0.0489}$ & $\underset{(0.0006)}{{0.0080}}$ &  $\underset{(0.0006)}{0.0008}$ & $\underset{(0.0228)}{1.0071}$ & 0.6215 & 0.0741\\
			&  9 & $\underset{(0.0014)}{0.0486}$ & $\underset{(0.0006)}{{0.0081}}$ &  $\underset{(0.0006)}{0.0005}$ & $\underset{(0.0224)}{0.9944}$ & 0.6236 & 0.0725\\
			&  10 & $\underset{(0.0014)}{0.0488}$ & $\underset{(0.0006)}{{0.0082}}$ &  $\underset{(0.0006)}{0.0006}$ & $\underset{(0.0225)}{1.0054}$ & 0.6271 & 0.1981\\
			\hline
			\underline{AML} &	\tvi 18 & $\underset{(0.0015)}{0.0522}$ & $\underset{(0.0007)}{0.0106}$ &  $\underset{(0.0006)}{-0.0005}$ & $\underset{(0.0217)}{1.0010}$ & 0.6412 & 0.2022 \\
		\end{tabular}
	\end{center}
	\vskip-15pt
	\caption{The table reports accuracies of the 1$\%$ and 5$\%$ value-at-risk (left panel) and 5$\%$ expected shortfall forecasts (right panel) using a particle filter with $10^6$ particles. In the left panel, failure rates of the 1$\%$ and 5$\%$ value-at-risk are reported with asymptotic standard errors in parentheses. In the right panel, for each $\overline{k}$, the empirical returns satisfying $\{r_t|r_t<-{\text{VaR}}_{0.05,t}^{(\overline{k}=10)}\}$ are regressed on $\{{\text{ES}}_{0.05,t}^{\overline{k}}|r_t<-{\text{VaR}}_{0.05,t}^{(\overline{k})}\}$, where $\text{VaR}_{0.05,t}^{(\overline{k})}$ corresponds to the 5$\%$ value-at-risk at date $t$ forecasted with $\overline{k}$ and $\text{ES}_{0.05,t}^{(\overline{k})}$ corresponds to the 5$\%$ expected shortfall at date $t$ forecasted with $\overline{k}$. For each regression, the intercepts and slopes are reported with standard errors in parentheses along with the $R^2$ values and the p-values of the Wald test $H_0:\,({\text{intercept}},{\text{slope}})=(0,1)$.}
		\label{tab:sp500reg}
\end{table}

\section{Conclusion}

{In this paper, we provide an alternative to indirect inference (hereafter, I-I) estimation that simultaneously allows us to circumvent the intractability of maximum likelihood estimation (as with standard I-I), but which, in contrast to naive I-I, respects the goal of obtaining asymptotically efficient inference in the context of a fully parametric model.} Although close in spirit to I-I, the approximate maximum likelihood (hereafter, AML) method developed in this paper {does \textit{not} belong to the realm} of I-I for two reasons:
First, the asymptotic distribution of the AML estimator only depends on the probability limit of the estimated auxiliary parameters and not on its asymptotic distribution. Second, while the AML estimator is obtained by matching two sample moments, one computed on observed data, and one computed on simulated data, both {sample moments depend} on the observed data through the value of the preliminary estimator of the auxiliary parameters. {Interestingly, the sampling} uncertainty carried by this preliminary estimator has no impact on the asymptotic distribution of the AML estimator because it is erased through the matching procedure.

The message of our {paper is threefold.
First, we demonstrate that} the idea of matching proxies of the score for the structural model seems productive to reach near efficiency for inference on the structural parameters. We show theoretically that, at least for exponential models or transformation of them, the efficiency loss should be {manageable} since it is mainly due to the effect of a misspecification bias created by our simplification of the structural model. 

{Second, there are many non-linear time series models, which are popular in financial econometrics and dynamic/nonlinear microeconometrics, where a natural simplification of the structural model yields a convenient proxy for the score of the structural model.} Since the misspecification bias created by this simplification is only due to imposing some possible false equality constraints, or to numerical approximations {for certain elements of the gradient vector}, one may reasonably hope that the {resulting efficiency loss is minimal}. {While our general results (and theoretical examples) suggest that this finding is valid in many examples, including dynamic discrete choice and stochastic volatility models,} we provide numerical evidence in three specific examples: generalized Tobit, Markov-switching multifractal models and stable distributions. The numerical results largely confirm our intuitions. {Our method can alleviate the computational cost of maximum likelihood associated with complex models, at the cost of a limited loss in efficiency.} Moreover, we confirm that even in finite-samples, the Wald confidence intervals associated to AML estimators {display excellent coverage, since, thanks to matching the misspecification bias, the preliminary estimators have no impact on the central tendency of the AML estimator.
}

{A third and even more general} message is that the matching principle put forward by I-I estimation can be extended to situations where the two empirical moments to match, one based on observed data, one based on simulated data may both depend on the observed data through a convenient summary of them. {While we have used this idea to aim for (nearly) efficient inference, Gospodinov, et al., (2017) employ a similar approach to hedge against} misspecification bias due to the use of a misspecified simulator. Even though they have not derived the asymptotic distribution theory in their case, the two methods are essentially similar and could be nested within a general asymptotic theory where {both the moments to match and the simulator} depend on observed data.

\appendix

\appendix 
\section{Proofs of Main Results}\label{app:proofs}
\subsection{Proof of Proposition 1}

With standard abuse of notation, a Taylor expansion gives:%
\begin{eqnarray*}
	\sqrt{T}\Delta _{\beta }L_{T}\left( \hat{\beta}_{T}\right) &=&\sqrt{T}\Delta
	_{\beta }L_{T}\left( \beta ^{0}\right) -K^{0}\left[ \tilde{\beta}_{T}\right] 
	\sqrt{T}\left[ \hat{\beta}_{T}-\beta ^{0}\right] \\
	\frac{\sqrt{T}}{H}\sum_{h=1}^{H}\Delta _{\beta }L_{T}^{(h)}\left( \theta ,%
	\hat{\beta}_{T}\right) &=&\frac{1}{H}\sum_{h=1}^{H}\sqrt{T}\Delta _{\beta
	}L_{T}^{(h)}\left( \theta ,\beta ^{0}\right) -\left\{ \frac{1}{H}%
	\sum_{h=1}^{H}K^{0}\left( \tilde{\beta}_{T}^{(h)}(\theta )\right) \right\} 
	\sqrt{T}\left[ \hat{\beta}_{T}-\beta ^{0}\right]
\end{eqnarray*}

where $\tilde{\beta}_{T}$\ and $\tilde{\beta}_{T}^{(h)}(\theta ),h=1,...,H$
are all in the interval $\left[ \beta ^{0},\hat{\beta}_{T}\right] .$Hence:%
\begin{equation*}
\sqrt{T}\Delta _{\beta }L_{T}\left( \beta ^{0}\right) -\frac{1}{H}%
\sum_{h=1}^{H}\sqrt{T}\Delta _{\beta }L_{T}^{(h)}\left( \theta ,\beta
^{0}\right) =\left\{ K^{0}\left[ \tilde{\beta}_{T}\right] -\frac{1}{H}%
\sum_{h=1}^{H}K^{0}\left( \tilde{\beta}_{T}^{(h)}(\theta )\right) \right\} 
\sqrt{T}\left[ \hat{\beta}_{T}-\beta ^{0}\right]
\end{equation*}

with, thanks to assumptions A1 and A2, and the fact that

$\sqrt{T}\left[ \hat{\beta}_{T}-\beta ^{0}\right] =O_{P}(1)$\ implies that
our AML estimator is such that:%
\begin{equation*}
\sqrt{T}\Delta _{\beta }L_{T}\left( \beta ^{0}\right) -\frac{1}{H}%
\sum_{h=1}^{H}\sqrt{T}\Delta _{\beta }L_{T}^{(h)}\left( \hat{\theta}%
_{T},\beta ^{0}\right) =o_{P}(1)
\end{equation*}

Under assumption A3, an additional Taylor expansion gives%
\begin{equation*}
\sqrt{T}\Delta _{\beta }L_{T}\left( \beta ^{0}\right) -\frac{1}{H}%
\sum_{h=1}^{H}\sqrt{T}\Delta _{\beta }L_{T}^{(h)}\left( \theta ^{0},\beta
^{0}\right) +o_{P}(1)=-\left( \frac{1}{H}\sum_{h=1}^{H}J^{0}\left( \tilde{%
	\theta}_{T}^{(h)},\beta ^{0}\right) \right) \sqrt{T}\left( \hat{\theta}%
_{T}-\theta ^{0}\right)
\end{equation*}

where $\tilde{\theta}_{T}^{(h)},h=1,...,H$ are all in the interval $\left[
\theta ^{0},\hat{\theta}_{T}\right] .$Hence:%
\begin{equation*}
\sqrt{T}\left( \hat{\theta}_{T}-\theta ^{0}\right) =\left[ J^{0}\left(
\theta ^{0},\beta ^{0}\right) \right] ^{-1}\left\{ \sqrt{T}\Delta _{\beta
}L_{T}\left( \beta ^{0}\right) -\frac{1}{H}\sum_{h=1}^{H}\sqrt{T}\Delta
_{\beta }L_{T}^{(h)}\left( \theta ^{0},\beta ^{0}\right) \right\} +o_{P}(1)
\end{equation*}

We know from Gourieroux, Monfort and Renault (1993) (see their proposition 3
and its proof) that:%
\begin{equation*}
\left\{ \sqrt{T}\Delta _{\beta }L_{T}\left( \beta ^{0}\right) -\frac{1}{H}%
\sum_{h=1}^{H}\sqrt{T}\Delta _{\beta }L_{T}^{(h)}\left( \theta ^{0},\beta
^{0}\right) \right\} \rightarrow _{d}\aleph \left( 0,\left( 1+\frac{1}{H}%
\right) I^{0}\left( \theta ^{0},\beta ^{0}\right) \right)
\end{equation*}
which completes the proof of Proposition 1. \hfill\(\Box\)

\subsection{Proof of Proposition 5}
By virtue of \textbf{Proposition 4}, we only need to prove that the asymptotic
variance $\Omega _{(H)}$ of the UAML estimator $\breve{\theta}_{T,H}(\theta
^{0})$ coincides with the Cramer-Rao efficiency bound when $H\rightarrow
\infty $. When $H\rightarrow \infty $, this estimator, denoted $\breve{\theta%
}_{T}$, can be seen as the solution in $\theta $ of the system of equations:  
\[
\Delta _{\beta }L_{T}\left( \theta ^{0}\right) =E_{\theta }[\Delta _{\beta
}L_{T}\left( \theta ^{0}\right) | \left\{ x_{t}\right\} _{t=1}^{T}]. 
\]

If we define
\[
m_{T}\left( \beta ,\theta \right) =\Delta _{\beta }L_{T}\left( \beta \right)
-E_{\theta }[\Delta _{\beta }L_{T}\left( \beta \right) \big{|} \left\{
x_{t}\right\} _{t=1}^{T}] ,
\]
we have, by definition,
\begin{eqnarray*}
	0 &=&\sqrt{T}m_{T}\left( \theta ^{0},\breve{\theta}_{T}\right)  \\
	&=&\sqrt{T}m_{T}\left( \theta ^{0},\theta ^{0}\right) +\frac{\partial
		m_{T}\left( \theta ^{0},\theta ^{0}\right) }{\partial \theta ^{\prime }}%
	\sqrt{T}\left( \breve{\theta}_{T}-\theta ^{0}\right) +o_{P}(1).
\end{eqnarray*}
Recall the definition of $\Delta _{\beta }L_{T}\left( \theta ^{0}\right)$,
\begin{eqnarray}
\Delta _{\beta }L_{T}\left( \theta ^{0}\right)  &=&\frac{1}{T}\sum_{t=1}^{T}%
\frac{\partial \log \left( l\{y_{t}\left\vert \{ y_{\tau }\}_{\tau=1}^{t-1},x_{t},z_{0},\theta ^{0}\right\} \right) }{\partial
	\theta }  \label{exactscore} \\
&=&\frac{1}{T}\sum_{t=1}^{T}S\{y_{t}\left\vert \{ y_{\tau }\}_{\tau=1}^{t-1},x_{t},z_{0},\theta ^{0}\right\}\nonumber ,
\end{eqnarray} and note that, by virtue of \eqref{exactscore}, 
\[
\sqrt{T}m_{T}\left( \theta ^{0},\theta ^{0}\right) =\sqrt{T}\Delta _{\beta
}L_{T}\left( \theta ^{0}\right) =\frac{1}{\sqrt{T}}\sum_{t=1}^{T}\frac{%
	\partial \log \left( l\{y_{t}\left\vert \{ y_{\tau }\}_{\tau=1}^{t-1},x_{t},z_{0},\theta ^{0}\right\} \right) }{\partial \theta }
\]
converges in distribution to a $\aleph \left( 0,I^{0}\right)$ random variable, where $%
I^{0}=I^{0}\left( \theta ^{0},\theta ^{0}\right) $ is the Fisher information
matrix.

Moreover,
\[
\plim_{T\rightarrow \infty }\frac{\partial m_{T}\left( \theta ^{0},\theta
	^{0}\right) }{\partial \theta ^{\prime }}=\plim_{T\rightarrow \infty }\frac{1%
}{T}\sum_{t=1}^{T }\frac{\partial }{\partial \theta ^{\prime }}%
E_{\theta }\left\{ \frac{\partial \log \left( l\{y_{t}\left\vert \{ y_{\tau }\}_{\tau=1}^{t-1},x_{t},z_{0},\theta ^{0}\right\}
	\right) }{\partial \theta }\right\} _{\theta =\theta ^{0}},
\]
and we have
\begin{eqnarray*}
	&&\frac{\partial }{\partial \theta ^{\prime }}E_{\theta }\left\{ \frac{%
		\partial \log \left( l\{y_{t}\left\vert \{ y_{\tau }\}_{\tau=1}^{t-1},x_{t},z_{0},\theta ^{0}\right\} \right) }{\partial \theta }%
	\right\}  \\
	&=&\frac{\partial }{\partial \theta ^{\prime }}\int \frac{\partial \log
		\left( l\{y_{t}\left\vert \{ y_{\tau }\}_{\tau=1}^{t-1},x_{t},z_{0},\theta ^{0}\right\} \right) }{\partial \theta }%
	l\{y_{t}\left\vert \{ y_{\tau }\}_{\tau=1}^{t-1},x_{t},z_{0},\theta \right\} d\nu \left( y_{t}\big{|} \{ y_{\tau }\}_{\tau=1}^{t-1},x_{t} \right) ,
\end{eqnarray*}
where $\nu $ denotes some dominating measure. Thus,
\begin{flalign*}
	&\frac{\partial }{\partial \theta ^{\prime }}E_{\theta }\left\{ \frac{%
		\partial \log \left( l\{y_{t}\left\vert \{ y_{\tau }\}_{\tau=1}^{t-1},x_{t},z_{0},\theta ^{0}\right\} \right) }{\partial \theta }%
	\right\}  \\
	&=\int S\{y_{t}\left\vert \{ y_{\tau }\}_{\tau=1}^{t-1},x_{t},z_{0},\theta ^{0}\right\} S\{y_{t}\left\vert\{ y_{\tau }\}_{\tau=1}^{t-1},x_{t},z_{0},\theta \right\}'
	l\{y_{t}\left\vert\{ y_{\tau }\}_{\tau=1}^{t-1},x_{t},z_{0},\theta \right\} d\nu \left( y_{t}\big{|} \{ y_{\tau }\}_{\tau=1}^{t-1},x_{t} \right) .
\end{flalign*}

Therefore,
\[
\plim_{T\rightarrow \infty }\frac{\partial m_{T}\left( \theta ^{0},\theta
	^{0}\right) }{\partial \theta ^{\prime }}=E\left[ S\{y_{t}\mid\{ y_{\tau }\}_{\tau=1}^{t-1},x_{t},z_{0},\theta ^{0}\}
S^{\prime }\{y_{t}\mid \{ y_{\tau }\}_{\tau=1}^{t-1},x_{t},z_{0},\theta ^{0}\} \mid \left\{ x_{\tau }\right\}
_{\tau =1}^{t}\right] 
\]
is the Fisher information matrix\ $I^{0}$. Consequently,
\[
\sqrt{T}\left( \breve{\theta}_{T}-\theta ^{0}\right) =-\left( I^{0}\right)
^{-1}\sqrt{T}m_{T}\left( \theta ^{0},\theta ^{0}\right) +o_{P}\left(
1\right) \longrightarrow _{d}\aleph \left( 0,\left( I^{0}\right)
^{-1}\right) .
\]

\section{GARCH-like Stochastic Volatility Models: Pseudo-Score}\label{sec:glsv}
In this section, we give the necessary details required to obtain \textbf{Result \ref{result:g-sv}} in Section \ref{sec:psv}. 

To this end, we first compute the latent score, and then use this to interpret the score in terms of generalized
residuals, it is worth computing the latent score. We first decompose the
latent log-likelihood as follows:
\begin{eqnarray*}
	L_{T}^{\ast }\left( \zeta ,0\right) &=&L_{1,T}^{\ast }\left( \mu ,\omega
	,\alpha \right) +L_{2,T}^{\ast }\left( \varpi \right) ,\\
	L_{1,T}^{\ast }\left( \mu ,\omega ,\alpha \right) &=&\frac{1}{T}%
	\sum_{t=1}^{T}\left\{ -\frac{1}{2}\left[ \log (2\pi )+\log \left( \left[
	\omega +\alpha \varepsilon _{t}^{2}+\eta _{t}\right] \right) \right]
	\right\} -\frac{1}{2T}\sum_{t=1}^{T}\frac{\left( r_{t+1}-\mu \right) ^{2}}{%
		\omega +\alpha \varepsilon _{t}^{2}+\eta _{t}}, \\
	L_{2,T}^{\ast }\left( \varpi \right) &=&-\log \left( \varpi \right) +\frac{1%
	}{T}\sum_{t=1}^{T}\log f_{\chi }\left( \frac{\eta _{t}}{\varpi }\right).
\end{eqnarray*}

Computations very similar to the case of Gaussian QMLE of ARCH models give:%
\begin{eqnarray*}
	\frac{\partial L_{T}^{\ast }\left( \zeta ,0\right) }{\partial \mu } &=&\frac{%
		1}{T}\sum_{t=1}^{T}\frac{r_{t+1}-\mu }{\sigma _{t}^{2}} ,\\
	\frac{\partial L_{T}^{\ast }\left( \zeta ,0\right) }{\partial \omega } &=&%
	\frac{1}{2T}\sum_{t=1}^{T}\frac{1}{\sigma _{t}^{2}}-\frac{1}{2T}%
	\sum_{t=1}^{T}\frac{\left( r_{t+1}-\mu \right) ^{2}}{\sigma _{t}^{4}} ,\\
	\frac{\partial L_{T}^{\ast }\left( \zeta ,0\right) }{\partial \alpha } &=&%
	\frac{1}{2T}\sum_{t=1}^{T}\frac{\varepsilon _{t}^{2}}{\sigma _{t}^{2}}-\frac{%
		1}{2T}\sum_{t=1}^{T}\frac{\left( r_{t+1}-\mu \right) ^{2}}{\sigma _{t}^{4}}%
	\varepsilon _{t}^{2},
\end{eqnarray*}
while
\begin{equation*}
\frac{\partial L_{T}^{\ast }\left( \zeta ,0\right) }{\partial \varpi }=-%
\frac{1}{\varpi }-\frac{1}{T\varpi ^{2}}\sum_{t=1}^{T}\frac{f_{\chi
	}^{\prime }\left( \frac{\eta _{t}}{\varpi }\right) }{f_{\chi }\left( \frac{%
		\eta _{t}}{\varpi }\right) }\eta _{t},
\end{equation*}
where $f_{\chi }^{\prime }$ is the derivative of the probability density
function $f_{\chi }$. Note that for sake of non-negativity of variance, we
expect the probability distribution of $\chi _{t}$\ to have a lower bounded
support, like for instance a demeaned log-normal distribution. However, it
is a reasonable hypothesis to see $\chi _{t}$\ as a Gaussian variable if we
consider that the correction term is small enough such that a Gaussian
approximation is accurate enough. We would then get a proxy of the latent
score by: 
\begin{eqnarray*}
	\frac{\partial \tilde{L}_{T}^{\ast }\left( \zeta ,0\right) }{\partial \varpi 
	} &=&-\frac{1}{\varpi }+\frac{1}{\varpi ^{3}}\frac{1}{T}\sum_{t=1}^{T}\eta
	_{t}^{2} \\
	&=&-\frac{1}{\varpi }+\frac{1}{\varpi ^{3}}\frac{1}{T}\sum_{t=1}^{T}\left[
	\sigma _{t}^{2}-\omega -\alpha \varepsilon _{t}^{2}\right] .
\end{eqnarray*}

The message from (\ref{obssco}) is that we will go from latent
score vector to observable one by replacing all functions of latent
volatility by its optimal filter. Let us define these filters:%
\begin{eqnarray}
\left[ \sigma _{t}^{2}\right] _{F,t} &=&E[\sigma _{t}^{2}\left\vert r_{\tau
},\tau \leq t\right]   \label{filter}, \\
\left[ \frac{1}{\sigma _{t}^{2}}\right] _{F,t} &=&E[\frac{1}{\sigma _{t}^{2}}%
\left\vert r_{\tau },\tau \leq t\right] ,  \notag \\
\left[ \frac{1}{\sigma _{t}^{4}}\right] _{F,t} &=&E[\frac{1}{\sigma _{t}^{4}}%
\left\vert r_{\tau },\tau \leq t\right].   \notag
\end{eqnarray}
Then, we have
\begin{eqnarray*}
	\frac{\partial \tilde{L}_{T}\left( \zeta ,0\right) }{\partial \mu } &=&\frac{%
		1}{T}\sum_{t=1}^{T}\left[ \frac{1}{\sigma _{t}^{2}}\right] _{F,t}\left(
	r_{t+1}-\mu \right),  \\
	\frac{\partial \tilde{L}_{T}\left( \zeta ,0\right) }{\partial \omega } &=&%
	\frac{1}{2T}\sum_{t=1}^{T}\left[ \frac{1}{\sigma _{t}^{2}}\right] _{F,t}-%
	\frac{1}{2T}\sum_{t=1}^{T}\left[ \frac{1}{\sigma _{t}^{4}}\right]
	_{F,t}\left( r_{t+1}-\mu \right) ^{2} ,\\
	\frac{\partial \tilde{L}_{T}\left( \zeta ,0\right) }{\partial \alpha } &=&%
	\frac{1}{2T}\sum_{t=1}^{T}\left[ \frac{1}{\sigma _{t}^{2}}\right]
	_{F,t}\varepsilon _{t}^{2}-\frac{1}{2T}\sum_{t=1}^{T}\left[ \frac{1}{\sigma
		_{t}^{4}}\right] _{F,t}\left( r_{t+1}-\mu \right) ^{2}\varepsilon _{t}^{2} ,\\
	\frac{\partial \tilde{L}_{T}\left( \zeta ,0\right) }{\partial \varpi } &=&-%
	\frac{1}{\varpi }+\frac{1}{\varpi ^{3}}\frac{1}{T}\sum_{t=1}^{T}\left[ \left[
	\sigma _{t}^{2}\right] _{F,t}-\omega -\alpha \varepsilon _{t}^{2}\right] .
\end{eqnarray*}

We recall that we denote these pseudo-score components with
notation $\tilde{L}$\ to stress that they are only approximations. They have
been computed with filtering formulas (\ref{filter}) that are only
approximations since doing as if $\rho =0.$ The filtered values (\ref{filter}%
) allow us to compute "generalized residuals" similar to the one computed in
the dynamic Probit example. However, by contrast with this example, we do
not have in general closed form formulas for these filters. Any filtering
strategy may be worth applying in this context.  At least, a very simple one
is to use the $ARCH(1)$ approximation as a convenient filter, meaning that
we replace in all filtering formulas , the latent quantity $\sigma _{t}^{2}$%
\ by the observed one $\hat{\sigma}_{t}^{2}$ (erasing then the conditional
expectation operator) that comes from fitting an $ARCH(1)$ model to our data
set $\left\{ r_{t+1}\right\} _{t=1}^{T}$.   

We now address the computation of the partial derivative $\partial \tilde{L}%
_{T}\left( \zeta ,0\right) /\partial \rho $ of the observed log-likelihood
with respect to the parameter $\rho $.

Using the definition of the latent likelihood, see Section \ref{sec:g-sv}, we can write:%
\begin{equation*}
L_{T}(\theta )=\frac{1}{T}\log \left( \int ...\int G_{T}\left( \mu ,\omega
,\alpha \right) \dprod\limits_{t=1}^{T}\frac{1}{\varpi }f_{\chi }\left( 
\frac{\eta _{t}-\rho \eta _{t-1}}{\varpi }\right) \right) d\eta _{1}...d\eta
_{T},
\end{equation*}
where
\begin{equation*}
G_{T}\left( \mu ,\omega ,\alpha \right) =\dprod\limits_{t=1}^{T}\frac{1}{%
	\sqrt{2\pi }}\frac{1}{\left[ \omega +\alpha \varepsilon _{t}^{2}+\eta _{t}%
	\right] ^{1/2}}\exp \left( -\frac{(r_{t+1}-\mu )^{2}}{2\left[ \omega +\alpha
	\varepsilon _{t}^{2}+\eta _{t}\right] }\right) .
\end{equation*}
Then,
\begin{eqnarray*}
	\frac{\partial L_{T}\left( \theta \right) }{\partial \rho } &=&\left[
	Tl_{T}\left( \theta \right) \right] ^{-1}\int_{-\infty }^{+\infty
	}...\int_{-\infty }^{+\infty }G_{T}\left( \mu ,\omega ,\alpha \right) \frac{1%
	}{\varpi ^{T}}\frac{\partial }{\delta \rho }\dprod\limits_{t=1}^{T}f_{\chi
	}\left( \frac{\eta _{t}-\rho \eta _{t-1}}{\varpi }\right) d\eta _{1}...d\eta
	_{T}, \\
	l_{T}\left( \theta \right)  &=&\int_{-\infty }^{+\infty }...\int_{-\infty
	}^{+\infty }G_{T}\left( \mu ,\omega ,\alpha \right) \frac{1}{\varpi ^{T}}%
	\dprod\limits_{t=1}^{T}f_{\chi }\left( \frac{\eta _{t}-\rho \eta _{t-1}}{%
		\varpi }\right) d\eta _{1}...d\eta _{T}.
\end{eqnarray*}
With an innovation process $\chi _{t}$ that is a standard Gaussian, this
leads (by computing the derivative of the product as a sum of products with
one term differentiated in each) to:%
\begin{eqnarray}
&&l_{T}\left( \zeta ,0\right) \frac{\partial L_{T}\left( \zeta ,0\right) }{%
	\partial \rho }  \label{scorerho} \\
&=&\int_{-\infty }^{+\infty }...\int_{-\infty }^{+\infty }G_{T}\left( \mu
,\omega ,\alpha \right) \left[ \dprod\limits_{t=1}^{T}\frac{1}{\varpi }%
f_{\chi }\left( \frac{\eta _{t}}{\varpi }\right) \right] \frac{\gamma _{\eta
		,T}}{\varpi ^{2}}d\eta _{1}...d\eta _{T} \notag\\
&=&\int_{-\infty }^{+\infty }...\int_{-\infty }^{+\infty }l^{\ast }[\left\{
r_{t+1},\eta _{t}\right\} _{t=1}^{T}\left\vert \left( \zeta ,0\right) \right]
\frac{\gamma _{\eta ,T}}{\varpi ^{2}}d\eta _{1}...d\eta _{T},  \notag
\end{eqnarray}
where $\gamma _{\eta ,T}$\ is the sample autocovariance of order $1$ of the
latent process
\begin{equation*}
\gamma _{\eta ,T}=\frac{1}{T}\sum_{t=1}^{T}\eta _{t}\eta _{t-1}.
\end{equation*}

We note that
\begin{equation*}
l_{T}\left( \zeta ,0\right) =\int_{-\infty }^{+\infty }...\int_{-\infty
}^{+\infty }l^{\ast }[\left\{ r_{t+1},\eta _{t}\right\} _{t=1}^{T}\left\vert
\left( \zeta ,0\right) \right] d\eta _{1}...d\eta _{T}=l^{\ast }[\left\{
r_{t+1}\right\} _{t=1}^{T}\left\vert \left( \zeta ,0\right) \right] 
\end{equation*}
so that (\ref{scorerho}) gives
\begin{equation}
\frac{\partial \tilde{L}_{T}\left( \zeta ,0\right) }{\partial \rho }=\frac{1%
}{\varpi ^{2}}E[\gamma _{\eta ,T}\left\vert \left\{ r_{t+1}\right\}
_{t=1}^{T}\right]   \label{smoothing}.
\end{equation}
Again, the computation of the observed score component is germane to the
computation of generalized residuals. However, it is worth noting that (\ref{smoothing}) is a smoothing formula instead of a filtering formula. The pseudo-score $\partial \tilde{L}_{T}\left( \zeta ,0\right) /{\partial \rho }$ can then be based on the approximation 
\begin{equation*}
\frac{1}{\varpi ^{2}}\frac{1}{T}\sum_{t=2}^{T}\left( [{\sigma}%
_{t}^{2}]_{F,t}-\omega -\alpha \varepsilon _{t}^{2}\right) \left( [{\sigma}%
_{t-1}^{2}]_{F,t-1}-\omega -\alpha \varepsilon _{t-1}^{2}\right) .
\end{equation*}

\section{Details for Examples in Section \ref{sec:dist}}\label{sec:appTs}
In this section, we give the details required to obtain Result \ref{result:expo} in Section \ref{sec:dist}. In addition, we also extend this example to consider latent exponential models. 
\subsection{Example: Exponential Models}
For the sake of exposition, we assume that conditionally on $%
\left\{ x_{t}\right\} _{t=1}^{T}$, the variables $y_{t},t=1,...,T$ are
independent and the conditional distribution of $y_{t}$\ only depends on the
exogenous variable $x_{t}$ with the same index. This distribution has a
density $l\{y_{t}\left\vert x_{t};\theta \right\} $\ that is assumed to be
exponential:%
\begin{equation*}
l\{y_{t}\left\vert x_{t};\theta \right\} =\exp \left[ c\left( x_{t},\theta
\right) +h(y_{t},x_{t})+a^{\prime }(x_{t},\theta )T(y_{t})\right]
\end{equation*}
where $c(.,.)$ and $h(.,.)$ are given numerical functions and $%
a(x_{t},\theta )$ and $T(y_{t})$ are $r$-dimensional random vectors. Note
that the extension to dynamic models in which conditioning values would also
include some lagged values of the process $y_{t}$\ would be easy to devise.
From:%
\begin{equation*}
\frac{\partial \log \left[ l\{y_{t}\left\vert x_{t};\theta \right\} \right] 
}{\partial \theta }=\frac{\partial c\left( x_{t},\theta \right) }{\partial
	\theta }+\frac{\partial a^{\prime }\left( x_{t},\theta \right) }{\partial
	\theta }T(y_{t})
\end{equation*}
 we deduce, since the conditional score vector has by definition a
zero conditional expectation, that:%
\begin{equation*}
\frac{\partial L_{T}\left( \theta \right) }{\partial \theta }=\frac{1}{T}%
\sum_{t=1}^{T}\frac{\partial a^{\prime }\left( x_{t},\theta \right) }{%
	\partial \theta }\left\{ T(y_{t})-E_{\theta }[T(y_{t})\left\vert x_{t}\right]
\right\} 
\end{equation*}

Following Theorem 1 in Gourieroux et al. (1987),
\begin{eqnarray*}
	E_{\theta }[T(y_{t})\left\vert x_{t}\right] &=&m\left( x_{t},\theta \right)
	,Var_{\theta }[T(y_{t})\left\vert x_{t}\right] =\Omega \left( x_{t},\theta
	\right) \\
	&\Longrightarrow &\frac{\partial a^{\prime }\left( x_{t},\theta \right) }{%
		\partial \theta }=\frac{\partial m^{\prime }\left( x_{t},\theta \right) }{%
		\partial \theta }\Omega ^{-1}\left( x_{t},\theta \right)
\end{eqnarray*}

Therefore, the maximum likelihood estimator $\hat{\theta}_{T}$\ is defined
as solution of:%
\begin{equation}
\frac{\partial L_{T}\left( \theta \right) }{\partial \theta }=\frac{1}{T}%
\sum_{t=1}^{T}\frac{\partial m^{\prime }\left( x_{t},\theta \right) }{%
	\partial \theta }\Omega ^{-1}\left( x_{t},\theta \right) \left\{
T(y_{t})-m\left( x_{t},\theta \right) \right\} =0  \label{optinst}
\end{equation}

We actually generalize the remark of van der Vaart (1998), Section 4.2.,
noting that "the maximum likelihood estimators are moment estimators" based
on the (conditional) expectation of the sufficient statistic $T(y)$. The
first-order conditions (\ref{optinst}) show that maximum likelihood is the
GMM estimator with optimal instruments for the conditional moment
restrictions:%
\begin{equation*}
E_{\theta }[T(y_{t})-m\left( x_{t},\theta \right) \left\vert x_{t}\right] =0.
\end{equation*}

Note that we implicitly maintain the assumptions for standard asymptotic
theory of efficient GMM (Hansen, 1982): for all $\theta \in \Theta $, the
conditional variance $\Omega \left( x_{t},\theta \right) $\ of the moment
conditions is non-singular and the Jacobian matrix $E[\partial m^{\prime
}\left( x_{t},\theta \right) /\partial \theta \left\vert x_{t}\right] $ is
full row rank.

The identification condition for consistency of maximum likelihood is then
that:%
\begin{equation*}
E\left\{ \frac{\partial m^{\prime }\left( x_{t},\theta \right) }{\partial
	\theta }\Omega ^{-1}\left( x_{t},\theta \right) \left\{ T(y_{t})-m\left(
x_{t},\theta \right) \right\} \right\} =0\Longrightarrow \theta =\theta ^{0}.
\end{equation*}

In terms of GMM, it means that optimal instruments are assumed to identify
the true unknown value $\theta ^{0}$\ of the parameter vector $\theta $, by
contrast with cases put forward by Dominguez and Lobato (2004). By the Law
of Iterated Expectations, this can be rewritten:%
\begin{equation*}
E\left\{ \frac{\partial m^{\prime }\left( x_{t},\theta \right) }{\partial
	\theta }\Omega ^{-1}\left( x_{t},\theta \right) \left\{ m\left( x_{t},\theta
^{0}\right) -m\left( x_{t},\theta \right) \right\} \right\}
=0\Longrightarrow \theta =\theta ^{0}
\end{equation*}

or equivalently (by symmetry):

\begin{equation}
E\left\{ \frac{\partial m^{\prime }\left( x_{t},\theta ^{0}\right) }{%
	\partial \theta }\Omega ^{-1}\left( x_{t},\theta ^{0}\right) \left\{ m\left(
x_{t},\theta \right) -m\left( x_{t},\theta ^{0}\right) \right\} \right\}
=0\Longrightarrow \theta =\theta ^{0}  \label{MLEid}.
\end{equation}

By extension of (\ref{optinst}), we have:

\begin{equation}
\Delta _{\beta }L_{T}^{(h)}\left( \theta ,\beta \right) =\frac{1}{T}%
\sum_{t=1}^{T}\frac{\partial m^{\prime }\left( x_{t},\beta \right) }{%
	\partial \theta }\Omega ^{-1}\left( x_{t},\beta \right) \left\{ T\left[ 
\tilde{y}_{t}^{(h)}\left( \theta \right) \right] -m\left( x_{t},\beta
\right) \right\}  \label{simoptinst}
\end{equation}

so that: 
\begin{equation*}
M\left( \theta ,\beta ^{0}\right) =E\left\{ \frac{\partial m^{\prime }\left(
	x_{t},\beta ^{0}\right) }{\partial \theta }\Omega ^{-1}\left( x_{t},\beta
^{0}\right) \left\{ T\left[ \tilde{y}_{t}^{(h)}\left( \theta \right) \right]
-m\left( x_{t},\beta ^{0}\right) \right\} \right\}.
\end{equation*}

 Hence:%
\begin{equation*}
M\left( \theta ,\beta ^{0}\right) -M\left( \theta ^{0},\beta ^{0}\right)
=E\left\{ \frac{\partial m^{\prime }\left( x_{t},\beta ^{0}\right) }{%
	\partial \theta }\Omega ^{-1}\left( x_{t},\beta ^{0}\right) \left\{ T\left[ 
\tilde{y}_{t}^{(h)}\left( \theta \right) \right] -T\left[ \tilde{y}%
_{t}^{(h)}\left( \theta ^{0}\right) \right] \right\} \right\}.
\end{equation*}

By the Law of Iterated Expectations:%
\begin{equation*}
M\left( \theta ,\beta ^{0}\right) -M\left( \theta ^{0},\beta ^{0}\right)
=E\left\{ \frac{\partial m^{\prime }\left( x_{t},\beta ^{0}\right) }{%
	\partial \theta }\Omega ^{-1}\left( x_{t},\beta ^{0}\right) \left\{ m\left(
x_{t},\theta \right) -m\left( x_{t},\theta ^{0}\right) \right\} \right\},
\end{equation*}
so that the identification Assumption B1 amounts to:%
\begin{equation}
E\left\{ \frac{\partial m^{\prime }\left( x_{t},\beta ^{0}\right) }{\partial
	\theta }\Omega ^{-1}\left( x_{t},\beta ^{0}\right) \left\{ m\left(
x_{t},\theta \right) -m\left( x_{t},\theta ^{0}\right) \right\} \right\}
\Longrightarrow \theta =\theta ^{0}  \label{identUAML}.
\end{equation}

When $\beta ^{0}=\theta ^{0}$, we are back to the well-specified example  and (\ref{identUAML}%
) is obviously identical to the identification condition (\ref{MLEid}) for
consistency of maximum likelihood.

Moreover, the identification assumption (\ref{identUAML}) for consistency of
the UAML estimator $\breve{\theta}_{T,H}(\beta ^{0})$ is clearly likely
implied by the standard condition (\ref{MLEid}) for consistency of maximum
likelihood, at least in two particular cases:

\textbf{1st case:} The model is a linear regression model w.r.t. some known
multivariate function $\kappa (x_{t})$ of $x_{t}$:%
\begin{equation*}
m\left( x_{t},\theta \right) =\kappa ^{\prime }\left( x_{t}\right) \theta.
\end{equation*}

In this case, the identification condition (\ref{identUAML}) is akin to:%
\begin{equation*}
E\left[ \kappa (x_{t})\Omega ^{-1}\left( x_{t},\beta ^{0}\right) \kappa
^{\prime }(x_{t})\right] (\theta -\theta ^{0})=0\Longrightarrow \theta
=\theta ^{0}.
\end{equation*}

Obviously, when the matrix:%
\begin{equation*}
E\left[ \kappa (x_{t})\Omega ^{-1}\left( x_{t},\beta ^{0}\right) \kappa
^{\prime }(x_{t})\right]
\end{equation*}
is positive definite for $\beta ^{0}=\theta ^{0}$, it is positive definite
for any possible value of $\beta ^{0}$.

\textbf{2nd case:} The model is not conditional. In this case, a necessary
condition for identification condition is:%
\begin{equation}
E_{\theta }\left\{ T(y_{1})\right\} =E_{\theta ^{0}}\left\{ T(y_{1})\right\}
\Longleftrightarrow \theta =\theta ^{0}  \label{stat}.
\end{equation}

This is basically the case considered by van der Vaart (1998) when noting
that "the maximum likelihood estimators are moment estimators" based on the
expectation of the sufficient statistic $T(y)$. This identification
condition should be maintained when picking $p$ linear independent equations
out of possibly overidentified equations (\ref{stat}). More precisely, the
identification condition for UAML, written as:%
\begin{equation*}
\frac{\partial m^{\prime }\left( \beta ^{0}\right) }{\partial \theta }\Omega
^{-1}\left( \beta ^{0}\right) \left\{ E_{\theta }\left\{ T(y_{1})\right\}
-E_{\theta ^{0}}\left\{ T(y_{1})\right\} \right\} \Longrightarrow \theta
=\theta ^{0}
\end{equation*}

should generically be implied by (\ref{stat}), since, irrespective of the
value of $\beta ^{0}$, the matrix $\partial m^{\prime }\left( \beta
^{0}\right) /\partial \theta $\ is full row rank.

More generally, one may expect that the identification condition (\ref%
{identUAML}), when fulfilled for $\beta ^{0}=\theta ^{0}$, should be more
often than not fulfilled for any value of $\beta ^{0}$.

\subsection{Example: Latent Exponential Model}\label{sec:appTslate}
{We now extend the exponential model example to incorporate a sequence of latent
	variables $\left\{ y_{t}^{\ast }\right\} _{t=1}^{T}$, such that, conditionally on $%
	\{x_{t}\}_{t=1}^{T}$, the variables $y_{t}^{\ast }$
	are independent, for all $t=1,\dots,T$, and the conditional distribution of $y_{t}^{\ast }$ only
	depends on the exogenous variable $x_{t}$ with the same index.} This
distribution has a density $l\{y_{t}^{\ast }\left\vert x_{t};\theta \right\} 
$, with respect to the dominating measure $\nu(dy_t^{\ast})$, that is assumed to be exponential:%
\begin{equation*}
l\{y_{t}^{\ast }\left\vert x_{t};\theta \right\} =\exp \left[ c\left(
x_{t},\theta \right) +h(y_{t}^{\ast },x_{t})+a^{\prime }(x_{t},\theta
)T(y_{t}^{\ast })\right]
\end{equation*}

Let $g$ be a known vector function that defines the observed endogenous
variable $y_{t}$ as:%
\begin{equation*}
y_{t}=g\left( y_{t}^{\ast },x_{t}\right).
\end{equation*}Then, conditionally on $\left\{ x_{t}\right\} _{t=1}^{T}$, the variables $%
y_{t},t=1,...,T$ are independent and the conditional distribution of $y_{t}$ only depends on the exogenous variables $x_t$ with the same index. {This conditional
	distribution has a density $l\{y_{t}\left\vert x_{t};\theta \right\} $, with
	respect to the measure $\nu^g(dy)$, which is the transformation of the original measure $\nu(dy^\ast_t)$ by $g$, and where we recall that $\nu(dy^\ast_t)$ was the
	dominating measure used to define the latent density $l\{y_{t}^{\ast
	}\left\vert x_{t};\theta \right\} $.} The observable log-likelihood can then be stated as
\begin{equation*}
L_{T}(\theta )=\frac{1}{T}\sum_{t=1}^{T}\log \left[ l\{y_{t}\left\vert
x_{t};\theta \right\} \right].
\end{equation*}

{In general, the observable density is not of an exponential form, see Gourieroux et al.
	(1987) for the
	particular case where $y_{t}=g\left( y_{t}^{\ast }\right) $ and for examples of Probit, bivariate Probit, Tobit,
	generalized Tobit, disequilibrium and Gompit models.} As already mentioned in
Section \ref{sec:psv}, Gourieroux et al. (1987), extending a result of Louis
(1982), give a method to compute the observable score as a conditional
expectation of the latent score
\begin{equation*}
\frac{\partial L_{T}\left( \theta \right) }{\partial \theta }=\frac{1}{T}%
\sum_{t=1}^{T}E_{\theta }\left[ \frac{\partial \log \left[ l\{y_{t}^{\ast
	}\left\vert x_{t};\theta \right\} \right] }{\partial \theta }\left\vert
y_{t},x_{t}\right] \right] .
\end{equation*}
Then, by applying (\ref{optinst}) we get
\begin{equation}
\frac{\partial L_{T}\left( \theta \right) }{\partial \theta }=\frac{1}{T}%
\sum_{t=1}^{T}\frac{\partial m^{\prime }\left( x_{t},\theta \right) }{%
	\partial \theta }\Omega ^{-1}\left( x_{t},\theta \right) \left\{ E_{\theta
}[T(y_{t}^{\ast })\left\vert y_{t},x_{t}\right] -m\left( x_{t},\theta
\right) \right\}  \label{MLEres}.
\end{equation}

As exemplified by Gourieroux et al. (1987) for many limited dependent
variable models, we can define and compute a generalized error as:%
\begin{eqnarray*}
	u\left( y_{t},x_{t},\theta \right) &=&\tilde{T}\left( y_{t},x_{t},\theta
	\right) -m\left( x_{t},\theta \right) \\
	\tilde{T}\left( y_{t},x_{t},\theta \right) &=&E_{\theta }[T(y_{t}^{\ast
	})\left\vert y_{t},x_{t}\right].
\end{eqnarray*}
Then, the maximum likelihood estimator $\hat{\theta}_{T}$\ is defined as
solution of
\begin{equation}
\frac{\partial L_{T}\left( \theta \right) }{\partial \theta }=\frac{1}{T}%
\sum_{t=1}^{T}\frac{\partial m^{\prime }\left( x_{t},\theta \right) }{%
	\partial \theta }\Omega ^{-1}\left( x_{t},\theta \right) u\left(
y_{t},x_{t},\theta \right) =0  \label{normal}.
\end{equation}
Hence, the identification condition for consistency of maximum likelihood
can be written:%
\begin{equation}
E\left[ \frac{\partial m^{\prime }\left( x_{t},\theta \right) }{\partial
	\theta }\Omega ^{-1}\left( x_{t},\theta \right) u\left( y_{t},x_{t},\theta
\right) \right] =0\Longleftrightarrow \theta =\theta ^{0}  \label{identgen}.
\end{equation}

We also note that MLE is not any more a moment estimator with optimal
instruments (confirming that the model is not exponential any more) since:%
\begin{equation*}
Var[u\left( y_{t},x_{t},\theta ^{0}\right) \left\vert x_{t}\right] =Var\left[
E_{\theta ^{0}}[T(y_{t}^{\ast })\left\vert y_{t},x_{t}\right] \left\vert
x_{t}\right] \right] \neq \Omega \left( x_{t},\theta ^{0}\right)
=Var[T(y_{t}^{\ast })\left\vert x_{t}\right].
\end{equation*}
More generally, by extension of (\ref{normal}) we have:
\begin{equation*}
\Delta _{\beta }L_{T}^{(h)}\left( \theta ,\beta \right) =\frac{1}{T}%
\sum_{t=1}^{T}\frac{\partial m^{\prime }\left( x_{t},\beta \right) }{%
	\partial \theta }\Omega ^{-1}\left( x_{t},\beta \right) u\left[ \tilde{y}%
_{t}^{(h)}\left( \theta \right) ,x_{t},\beta \right].
\end{equation*}
Hence,
\begin{equation*}
M\left( \theta ,\beta ^{0}\right) =E\left\{ \frac{\partial m^{\prime }\left(
	x_{t},\beta ^{0}\right) }{\partial \theta }\Omega ^{-1}\left( x_{t},\beta
^{0}\right) u\left[ \tilde{y}_{t}^{(h)}\left( \theta \right) ,x_{t},\beta
^{0}\right] \right\}.
\end{equation*}
so that
\begin{eqnarray*}
	&&M\left( \theta ,\beta ^{0}\right) -M\left( \theta ^{0},\beta ^{0}\right) \\
	&=&E\left\{ \frac{\partial m^{\prime }\left( x_{t},\beta ^{0}\right) }{%
		\partial \theta }\Omega ^{-1}\left( x_{t},\beta ^{0}\right) \left[ u\left[ 
	\tilde{y}_{t}^{(h)}\left( \theta \right) ,x_{t},\beta ^{0}\right] -u\left[ 
	\tilde{y}_{t}^{(h)}\left( \theta ^{0}\right) ,x_{t},\beta ^{0}\right] \right].
	\right\}
\end{eqnarray*}

When $\beta ^{0}=\theta ^{0}$, we are back to the well-specified example and we note that by
definition:%
\begin{eqnarray*}
	E\{u\left[ \tilde{y}_{t}^{(h)}\left( \theta ^{0}\right) ,x_{t},\theta ^{0}%
	\right] \left\vert x_{t}\right\}  &=&0\Longrightarrow \forall h \\
	E\left\{ h(x_{t})u\left[ \tilde{y}_{t}^{(h)}\left( \theta ^{0}\right)
	,x_{t},\theta ^{0}\right] \right\}  &=&0\Longrightarrow  \\
	M\left( \theta ,\beta ^{0}\right) -M\left( \theta ^{0},\beta ^{0}\right) 
	&=&E\left\{ \frac{\partial m^{\prime }\left( x_{t},\theta ^{0}\right) }{%
		\partial \theta }\Omega ^{-1}\left( x_{t},\theta ^{0}\right) u\left[ \tilde{y%
	}_{t}^{(h)}\left( \theta \right) ,x_{t},\theta ^{0}\right] \right\} =0.
\end{eqnarray*}
 so that the identification condition
\begin{equation*}
M\left( \theta ,\beta ^{0}\right) -M\left( \theta ^{0},\beta ^{0}\right)
\Longleftrightarrow \theta =\theta ^{0},
\end{equation*}
can be written
\begin{equation}
E\left\{ \frac{\partial m^{\prime }\left( x_{t},\theta ^{0}\right) }{%
	\partial \theta }\Omega ^{-1}\left( x_{t},\theta ^{0}\right) u\left[ \tilde{y%
}_{t}^{(h)}\left( \theta \right) ,x_{t},\theta ^{0}\right] \right\}
=0\Longleftrightarrow \theta =\theta ^{0}  \label{unbiasMLE}.
\end{equation}

By commuting the roles of $\theta $\ and $\theta ^{0}$, this is clearly
tantamount to the identification condition (\ref{identgen}) for maximum
likelihood. In the general case, the identification condition B1($\beta ^{0})$\ for UAML
can be written:%
\begin{equation*}
E\left\{ \frac{\partial m^{\prime }\left( x_{t},\beta ^{0}\right) }{\partial
	\theta }\Omega ^{-1}\left( x_{t},\beta ^{0}\right) \left[ u\left[ \tilde{y}%
_{t}^{(h)}\left( \theta \right) ,x_{t},\beta ^{0}\right] -u\left[ \tilde{y}%
_{t}^{(h)}\left( \theta ^{0}\right) ,x_{t},\beta ^{0}\right] \right]
\right\} =0\Longleftrightarrow \theta =\theta ^{0}.
\end{equation*}

Note that by the Law of Iterated Expectations, this can be written:%
\begin{equation*}
E\left\{ \frac{\partial m^{\prime }\left( x_{t},\beta ^{0}\right) }{\partial
	\theta }\Omega ^{-1}\left( x_{t},\beta ^{0}\right) \left[ \tilde{m}%
(x_{t},\theta ,\beta ^{0})-\tilde{m}(x_{t},\theta ^{0},\beta ^{0})\right]
\right\} =0\Longleftrightarrow \theta =\theta ^{0},
\end{equation*}
where
\begin{equation*}
\tilde{m}(x_{t},\theta ,\beta ^{0})=E[u\left( \tilde{y}_{t}^{(h)}\left(
\theta \right) ,x_{t},\beta ^{0}\right) \left\vert x_{t}\right].
\end{equation*}

By comparison with (\ref{unbiasMLE}), we see that while both generalized
errors\ $u\left[ \tilde{y}_{t}^{(h)}\left( \theta \right) ,x_{t},\beta ^{0}%
\right] $\ and $u\left[ \tilde{y}_{t}^{(h)}\left( \theta ^{0}\right)
,x_{t},\beta ^{0}\right] $ will in general have a non-zero conditional
expectation given $x_{t}$ (when $\beta ^{0}\notin \left\{ \theta ,\theta
^{0}\right\} $), identification means that when $\theta \neq \theta ^{0}$,
their difference cannot be orthogonal to the $p$\ specific functions of $%
x_{t}$ that define the rows of the selection matrix:%
\begin{equation*}
\frac{\partial m^{\prime }\left( x_{t},\beta ^{0}\right) }{\partial \theta }%
\Omega ^{-1}\left( x_{t},\beta ^{0}\right).
\end{equation*}

 This condition is similar to the condition (\ref{identUAML}) of
identification for UAML in the exponential model example, except that, due to the
transformation $y_{t}=g\left( y_{t}^{\ast },x_{t}\right) $, the conditional
expectation given $x_{t}$ along simulated paths still depend on $\beta ^{0}$. In the particular case of a latent model defined by a univariate linear
and homoskedastic regression equation:%
\begin{equation*}
m\left( x_{t},\theta \right) =x_{t}^{\prime }\theta ,\Omega \left(
x_{t},\theta \right) =\sigma ^{2},
\end{equation*}
the identification condition in \textbf{Assumption B1} for UAML becomes:%
\begin{equation*}
E\left\{ x_{t}\left[ \tilde{m}(x_{t},\theta ,\beta ^{0})-\tilde{m}%
(x_{t},\theta ^{0},\beta ^{0})\right] \right\} =0\Longleftrightarrow \theta
=\theta ^{0}.
\end{equation*}

For instance, in the case of a Probit model ($\sigma ^{2}=1$):%
\begin{equation*}
E\left\{ x_{t}\frac{\varphi (x_{t}^{\prime }\beta ^{0})}{\Phi (x_{t}^{\prime
	}\beta ^{0})\left[ 1-\Phi (x_{t}^{\prime }\beta ^{0})\right] }\left[ \Phi
(x_{t}^{\prime }\theta )-\Phi (x_{t}^{\prime }\theta ^{0})\right] \right\}
=0\Longleftrightarrow \theta =\theta ^{0},
\end{equation*}
which we can compare to the standard identification condition for a Probit
model
\begin{equation*}
E\left\{ x_{t}\frac{\varphi (x_{t}^{\prime }\theta )}{\Phi (x_{t}^{\prime
	}\theta )\left[ 1-\Phi (x_{t}^{\prime }\theta )\right] }\left[ \Phi
(x_{t}^{\prime }\theta )-\Phi (x_{t}^{\prime }\theta ^{0})\right] \right\}
=0\Longleftrightarrow \theta =\theta ^{0}.
\end{equation*}
These conditions appear to be quite reasonable.

\section{Example 5: (\textit{Stable Distribution})}

Consider i.i.d. observations $y_1,\dots,y_T$ generated from a stable distribution with stability parameter $a\in(0,2]$, skewness parameter $b\in[-1,1]$, scale parameter $c>0$ and location parameter $\mu\in\mathbb{R}$. The structural parameter vector is given by
\begin{equation*}
\theta =\left(a,b,\zeta ^{\prime }\right) ^{\prime },\zeta =\left( c,\mu
\right) ^{\prime }.
\end{equation*}

We consider this model under the false equality constraint:%
\begin{equation*}
(a,b)'=(1,0)'
\end{equation*}
corresponding to a Cauchy distribution with location $\mu $\
and scale $c$, which gives the log-likelihood:%
\begin{equation*}
L_{T}\left(1,0,\zeta \right) =-\log \left[ \pi c\right] -\frac{1}{T}%
\sum_{t=1}^{T}\log \left[ 1+\left( \frac{y_{t}-\mu }{c}\right) ^{2}\right]
\end{equation*}
We can define the pseudo-score vector as:%
\begin{equation*}
{\Delta_\theta L_{T}}\left( 1,0,\zeta \right) =\left( \frac{%
	\partial L_{T}\left( 1,0,\zeta 
	\right) }{\partial \zeta ^{\prime }}%
,L_{T}\left( 2,0,\zeta\right) -L_{T}\left(1,0, \zeta\right) ,\tilde{L}%
_{T}\left(1,1, \zeta\right) -L_{T}\left(1,0, \zeta\right) \right) ^{\prime
}.
\end{equation*}

Note that, the finite difference $\left[ L_{T}\left(2,0, \zeta\right)
-L_{T}\left(1,0 ,\zeta\right) \right] $ is a convenient approximation of
the partial derivative $\partial L_{T}\left(1,0, \zeta\right) /\partial a$
since the log-likelihood function $L_{T}\left(2,0, \zeta\right) $ is
computed as the likelihood for i.i.d. draws in a Normal distribution with
mean $\mu $ and variance $2c^{2}$. Second, the finite difference $\left[ L_{T}\left(1,1, \zeta\right)
-L_{T}\left( 1,0,\zeta\right) \right] $ is a convenient approximation of
the partial derivative $\partial L_{T}\left( 1,0,\zeta\right) /\partial b$
since the log-likelihood function $L_{T}\left(1,1, \zeta\right) $ could be
computed as the likelihood for i.i.d. draws in a Landau distribution with
location parameter $\mu $ and scale parameter $c$
$$
L_T(1,1,\zeta)=\sum_{t=1}^{T}\log(f(y_t)),\text{ where }f(y)=\frac{1}{\pi c} \int_{0}^{\infty} e^{-x} \cos \left[x\left(\frac{y-\mu}{e}\right)+\frac{2 x}{\pi} \log \left(\frac{x}{c}\right)\right] d x.
$$To speed up the computation, we use the following approximation to $f(y)$ given by Behrens and Melissinos (1981).\footnote{Similar results were obtained whether or not the approximation was employed. Given the similarity of the results, and the drastic speed difference, the approximation approach is more reasonable to apply in practice. } 
$$
f(y)\approxeq\frac{1}{\sqrt{2\pi}c}\exp\left\{-(y-\mu)/(2c) -\exp\left[-\left(\{y-\mu\}/c\right)\right]/{2}\right\}.
$$

\subsection{Monte Carlo}
We now compare the behavior of AML using the above pseudo-score, and $H=10$ simulations, against two alternative approaches: one based on sample quantiles, due to McCullough (1986), and one based on an auxiliary regression model, due to Koutrouvelis (1981). To this end, we generate 1,000 synthetic datasets from the alpha stable models, each with $T=10,000$ observations, and under $\theta=(1.8,-0.1,1,0)'$. 

\begin{figure}[h!]
	\begin{center}
		\includegraphics[height=0.6\textheight, width=0.95\textwidth]{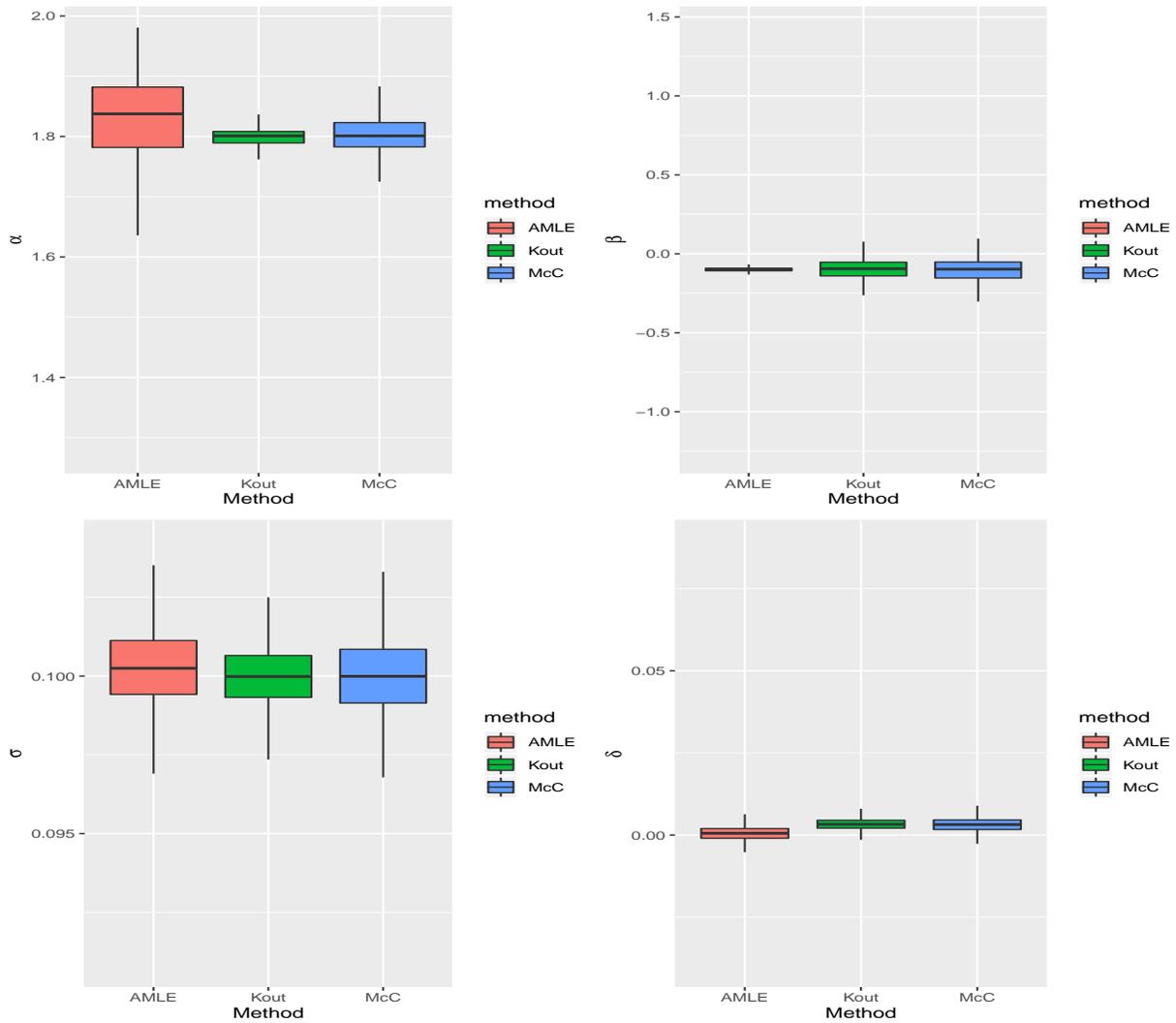}
	\end{center}
	\caption{Boxplots of estimators across 1000 Monte Carlo replications from the stable distribution. The true values used to generate the data are $\theta=(a,b,c,\mu)=(1.8,-0.1,0.1,0)'$. AML- approximate maximum likelihood estimator, Kout- Koutrouvelis (1981) regression approach, McC- McCullough (1986) quantile approach.}
	\label{fig:stable}
\end{figure}

We display the resulting estimators across the replications in Figure \ref{fig:stable}.\footnote{We remark that while ML estimation is feasible in the $\alpha$-stable model for small numbers of observations, given the sample size considered herein, obtaining the MLE proved to be computationally infeasible.} Analyzing the results, we see that the three procedures perform similarly for $\sigma$, but display different behavior for $\alpha,\beta,\delta$, although all estimators seem quite reliable, and are well-centred over the true values.

Table \ref{tab:stable} records the Monte Carlo bias (Bias), root mean squared error (RMSE), and Monte Carlo coverage (COV), based on individual 95\% Wald interval, across the replications. The results demonstrate that the methods all yield accurate estimators of the corresponding true values. However, we note that the simpler methods do outperform AML in terms of bias and RMSE, but display worse coverage than AML in almost all cases.  

\begin{table}[htbp]
  \centering
  \caption{Summary accuracy measures for stable example. Acronyms are as described in Figure \ref{fig:stable}, while Aux refers to the auxiliary estimator estimated under the restriction $(a,b)=(1,0)$. To aid readability of the table, the reported bias has been multiplied by 1000, and reported RMSE has been multiplied by 100.  }
    \begin{tabular}{lrrrrrlrrrr}
          & \multicolumn{1}{l}{$a$ } &       &       &       &       &       & \multicolumn{1}{l}{$b$} &       &       &  \\
          & \multicolumn{1}{l}{AML} & \multicolumn{1}{l}{Aux} & \multicolumn{1}{l}{Kout} & \multicolumn{1}{l}{McC} &       &       & \multicolumn{1}{l}{AML} & \multicolumn{1}{l}{Aux} & \multicolumn{1}{l}{Kout} & \multicolumn{1}{l}{McC} \\
    Mean  & 1.8190 & 1.0000 & 1.7994 & 1.8031 &       & Mean  & -0.0948 & 0.0000 & -0.0966 & -0.1039 \\
    Bias  & 19.0315 & -800.0000 & -0.6072 & 3.1120 &       & Bias  & 5.1846 & 100.0000 & 3.4381 & -3.8520 \\
    RMSE  & 9.6607 & 80.0000 & 1.4561 & 2.9785 &       & RMSE  & 13.6959 & 10.0000 & 6.5433 & 7.9542 \\
    COV   & 0.9600 & 0.0000 & 0.9410 & 0.9540 &       & COV   & 0.9650 & 0.0000 & 0.9540 & 0.9440 \\
          &       &       &       &       &       &       &       &       &       &  \\
          & \multicolumn{1}{l}{$c$} &       &       &       &       &       & \multicolumn{1}{l}{$\mu$} &       &       &  \\
          & \multicolumn{1}{l}{AML} & \multicolumn{1}{l}{Aux} & \multicolumn{1}{l}{Kout} & \multicolumn{1}{l}{McC} &       &       & \multicolumn{1}{l}{AML} & \multicolumn{1}{l}{Aux} & \multicolumn{1}{l}{Kout} & \multicolumn{1}{l}{McC} \\
    Mean  & 0.1002 & 0.0881 & 0.1000 & 0.1000 &       & Mean  & 0.0007 & 0.0025 & 0.0032 & 0.0031 \\
    Bias  & 0.1636 & -11.8524 & 0.0016 & -0.0074 &       & Bias  & 0.6945 & 2.4720 & 3.2351 & 3.1469 \\
    RMSE  & 0.1488 & 1.1884 & 0.0978 & 0.1251 &       & RMSE  & 0.6313 & 0.2986 & 0.3737 & 0.3781 \\
    COV   & 0.9480 & 0.0000 & 0.9480 & 0.9510 &       & COV   & 0.9810 & 0.6650 & 0.6030 & 0.6640 \\
    \end{tabular}%
  \label{tab:stable}%
\end{table}%

\end{document}